\documentclass[11pt]{article}
\usepackage{amsfonts,amsmath,amssymb,amsthm}
\allowdisplaybreaks
\usepackage[pdftex]{graphicx}
\usepackage[hmargin=1in,vmargin=1in]{geometry}
\usepackage{natbib}
\usepackage[colorlinks,citecolor=blue,urlcolor=blue]{hyperref}
\usepackage{enumerate}
\usepackage{caption}
\usepackage[usenames,dvipsnames]{xcolor}
\usepackage{footmisc,fancyvrb}
\usepackage{xfrac}
\usepackage{cleveref}
\usepackage{multirow}
\usepackage{threeparttable}
\usepackage{mathtools}
\usepackage{breqn}
\usepackage{array}
\newcommand{\PreserveBackslash}[1]{\let\temp=\\#1\let\\=\temp}
\newcolumntype{C}[1]{>{\PreserveBackslash\centering}p{#1}}
\newcolumntype{L}[1]{>{\raggedright\let\newline\\\arraybackslash\hspace{0pt}}m{#1}}
\newcolumntype{R}[1]{>{\raggedleft\let\newline\\\arraybackslash\hspace{0pt}}m{#1}}
\usepackage{standalone}
\usepackage{booktabs}
\usepackage{soul}
\setuldepth{Berlin}
\usepackage[position=below]{subcaption}
\usepackage{footmisc}

\usepackage{xr}
\VerbatimFootnotes
\def\qed{\rule{2mm}{2mm}}

\DeclareMathOperator*{\argmax}{arg\,max}

\usepackage{tikz,graphicx,pgfplots,pgfkeys,subcaption} 

\def\addlegendimage{\csname pgfplots@addlegendimage\endcsname}
\usetikzlibrary{decorations.pathreplacing,angles,quotes}

\usetikzlibrary{external}
\usepackage[pagewise,mathlines]{lineno}
\mathchardef\dash="2D

\theoremstyle{definition}

\newtheorem{proposition}{Proposition}

\newtheorem{assump}{Assumption}
\newenvironment{assumption2}[2][]
{\renewcommand\theassump{#2}\begin{assump}[#1]}
{\end{assump}}
\newtheorem{defi}{Definition}
\newenvironment{definition2}[2][]
{\renewcommand\thedefi{#2}\begin{defi}[#1]}
{\end{defi}}

\usepackage{etoolbox}
\AtEndEnvironment{remark}{~\qed}%
\AtEndEnvironment{example}{~\qed}%

\newcommand{\overbar}[1]{\mkern 1.5mu\overline{\mkern-1.5mu#1\mkern-1.5mu}\mkern 1.5mu}

\begin{document}

\author{
Vishal Kamat \\
Toulouse School of Economics\\
University of Toulouse Capitole\\
\url{vishal.kamat@tse-fr.eu}
\and
Samuel Norris\\
Department of Economics\\
University of British Columbia\\
\url{sam.norris@ubc.ca} \vspace{0.5cm}
}

\title{Estimating Welfare Effects in a Nonparametric Choice Model: The Case of School Vouchers\thanks{We thank the editor, three anonymous referees, Ivan Canay, Isis Durrmeyer, Dmitri Koustas, Thierry Magnac, Elena Manresa, Adam Rosen, Max Tabord-Meehan, Alex Torgovitsky, and participants at several seminars and conferences for useful comments. Vishal Kamat gratefully acknowledges funding from ANR under grant ANR-17-EURE-0010 (Investissements d'Avenir program). First version: arXiv:2002.00103 dated January 31, 2020.}}

\maketitle

\vspace{-0.5cm}
\begin{abstract}
We develop new robust discrete choice tools to learn about the average willingness to pay for a price subsidy and its effects on demand given exogenous, discrete variation in prices. Our starting point is a nonparametric, nonseparable model of choice. We exploit the insight that our welfare parameters in this model can be expressed as functions of demand for the different alternatives. However, while the variation in the data reveals the value of demand at the observed prices, the parameters generally depend on its values beyond these prices. We show how to sharply characterize what we can learn when demand is specified to be entirely nonparametric or to be parameterized in a flexible manner, both of which imply that the parameters are not necessarily point identified. We use our tools to analyze the welfare effects of price subsidies provided by school vouchers in the DC Opportunity Scholarship Program.  We find that the provision of the status quo voucher and a wide range of counterfactual vouchers of different amounts can have positive and potentially large benefits net of costs. The positive effect can be explained by the popularity of low-tuition schools in the program; removing them from the program can result in a negative net benefit. We also find that various standard logit specifications, in comparison, limit attention to demand functions with low demand for the voucher, which do not capture the large magnitudes of benefits credibly consistent with the data.
\end{abstract}

\thispagestyle{empty}



\newpage
\setcounter{page}{1}

\section{Introduction}\label{sec:intro}

Price subsidies are a common feature of many social programs that aim to encourage the use of certain alternatives or make them more affordable to disadvantaged populations. Policy relevant examples include school vouchers that subsidize tuition for eligible private schools \citep{epple/etal:17}, subsidies on health insurance \citep{finkelsein/etal:19}, and price subsidies for various essential goods in developing countries \citep{dupas:14}. Quantifying individuals' willingness to pay for a price subsidy and its effects on demand are key inputs in performing cost benefit analyses of implemented subsidies and in their counterfactual design.

In this paper, our first contribution is to develop new discrete choice tools that show how to robustly learn about such welfare effects of a price subsidy given data with exogenous, discrete variation in prices. The starting point of our analysis is a nonparametric, nonseparable model of choice. In this model, we exploit the fact that our welfare parameters of interest can be expressed in terms of the demand for the various alternatives. Exogenous, discrete variation in prices---which arises naturally in randomized evaluations of price changes---reveals the value of demand at the prices observed in the data. But, our parameters generally depend on values of demand beyond those observed in the data, which introduces an identification problem.

The traditional approach to this problem is to consider parameterizations of demand through various models such as logit and probit \citep[e.g.,][]{berry/etal:95, mcfadden:74, train:09}. These parameterizations are carefully chosen such that they imply a unique demand function consistent with the data, and hence that the welfare parameters are point identified. However, a natural concern with this approach is that it may limit attention to only specific demand functions that can potentially drive the welfare estimates and resulting policy conclusions.

To this end, our main methodological contribution is to show how to characterize what we can learn about welfare under more flexible specifications of demand. Our baseline specification leaves demand to be entirely nonparametric and only imposes a fundamental shape restriction that takes demand for each alternative to be increasing in the prices of other alternatives. In this case, there exists a space of infinite-dimensional demand functions consistent with the data and the parameters are generally only partially identified. The key complication is how to compute the sharp identified sets for the parameters generated by this space of functions. Our arguments show how to carefully exploit the geometric structure of the parameters as well as the information provided by the data and shape restrictions, such that the identified sets can be computed using finite-dimensional optimization problems.

We also consider several extensions. First, we propose dimension reduction methods to ensure tractability in cases where our baseline optimization problems can be large and potentially impractical. Specifically, we show how to obtain outer sets by considering sub-programs of the baseline ones as well as how to get sharp sets under additional separability assumptions on demand that reduce the dimensions of the baseline programs. Second, we show how to extend our baseline result to accommodate additional parametric restrictions on demand. This is in the spirit of traditional methods, but we do not solely restrict attention to point identified demand functions and allow for a range of parameterized demand functions to be consistent with the data.

Our final extension is motivated by the fact that the above arguments are essentially limited to the case where individuals are assumed to be able to afford all alternatives. Such an assumption, however, can be particularly suspect in our empirical setting as discussed below. We therefore also show how to extend our analysis to learn about welfare under a latent liquidity constraint model which limits the maximum amount individuals can pay for an alternative and where a price subsidy is allowed to relax these constraints.

Our second contribution is to use the developed tools to perform a welfare analysis of the price subsidy for eligible private schools provided by school vouchers. A large empirical literature has estimated the effects of vouchers on various outcomes using data from programs that randomly allocate vouchers \citep[e.g.,][]{abdulkadiroglu/etal:18, angrist:02, dynarski/etal:18, howell/etal:00, krueger/zhu:04, mayer/etal:02, mills/wolf:17, murali/sundar:15, wolf/etal:10}. However, as surveyed in \cite{epple/etal:17}, the evidence from these studies is mixed: some find positive effects, while others find null or even negative effects. Our motivation arises from the fact that, despite this mixed evidence on the effects on outcomes, the data across these studies indicate that a non-trivial proportion of recipients choose to use the voucher. Revealed preference arguments then suggest that recipients in general value vouchers and hence that vouchers may be welfare-enhancing. Yet, little empirical work has attempted to quantify these welfare benefits and analyze whether they can justify the costs of providing vouchers.

We apply our tools to data from the DC Opportunity Scholarship Program (OSP), a voucher program in Washington, DC. The program randomly allocated a voucher worth up to \$7,500 to participants, inducing exogeneous binary variation in prices, namely the prices of schools with and without the voucher.  Our estimated bounds reveal that provision of the status quo amount as well as a range of counterfactual amounts can have a positive and potentially large welfare benefit net of the costs the government faces to provide them. We find that the positive effect can be explained by the fact that there are a large number of popular, low-tuition schools in the program; counterfactual exercises indicate that if these schools were removed from the program, the overall net benefit might be negative. 

In our setting, it is likely that some students are liquidity constrained as the program was geared towards those from low-income families. We find that accounting for the fact that the voucher can relax liquidity constraints can be important in regards to the conclusion one draws on the magnitude of the benefits of the voucher. Specifically, in this case, the upper bound becomes arbitrarily large unless one assumes a maximum willingness to pay for schools in the program. In this sense, the magnitude could be significantly downward biased if one doesn't account for liquidity constraints. Such potential biases are consistent with a growing number of recent work that highlight the importance of accounting for unobserved heterogeneity in the fact that individuals may not be choosing from the set of all alternatives in various settings \citep[e.g.,][]{abaluck/adams:21,barseghyan/etal:21,goeree:08}.

When interpreting the above findings, it is worth highlighting certain features of our analysis. We equate the welfare effect of receiving the voucher with the willingness to pay for it of parents, who often make schooling decisions for their child. While this is a natural money metric for parents' valuation for the voucher, it may not fully capture students welfare due to internal inefficiencies in parental decisions resulting from either imperfect information on school effects or misaligned objectives  \citep[e.g.,][]{abdulkadiroglu/etal:18,hastings/weinstein:08}, or capture social welfare effects due to externalities \citep[e.g.,][]{acemoglu/angirst:00}. In this sense, our analysis complements the traditional one that directly evaluates effects on test scores or other measures of broader social gains, but leaves open the question of quantifying welfare based on underlying preferences. Moreover, our analysis only captures the effects of vouchers through the decrease in school prices it induces, and not through any potential general equilibrium effects on the school system they can have \citep{friedman:62}. Consequently, it is directly informative about the effect of marginal policies that increase the supply of vouchers, rather than of those that scale up the voucher program. It is therefore important to emphasize that our results provide a partial picture on the overall welfare effects of vouchers, and one should be cautious when drawing broader policy conclusions based on them.

We also compare our empirical results to those under various standard logit parameterizations. In general, these parameterizations imply demand estimates that match well the variation in enrollment shares induced by the receipt of the voucher. However, we find that they do not capture the range of demand functions that are credibly consistent with these shares, but limit attention to those that have low demand for the voucher. Our main takeaway is that as a result, the logit estimates do not capture the large benefits consistent with the data. For example, absent liquidity constraints, our upper bounds reveal that benefits for the status quo voucher can be close to half its value, but all the logit estimates allow them to equal only a fourth of its value. The difference is even starker when accounting for liquidity constraints. The logit estimates in this case stay almost the same as those without liquidity constraints, and do not capture the potentially large biases from not accounting for such constraints as revealed by our bounds. Our comparison contributes to a growing set of results that document similar attenuation of logit based estimates and highlight the importance of nonparametric methods in various empirical settings \citep[e.g.,][]{compiani:19,ho/pakes:14,tebaldi/etal:19}.

In the following subsection we describe the relation of our analysis to the literature, after which the remainder of the paper is organized as follows. Section \ref{sec:baseline} develops our baseline identification analysis. Section \ref{sec:extensions} presents the extensions. Section \ref{sec:empirical} applies our tools to analyze the welfare effects of school vouchers in the OSP. Section \ref{sec:comparison} compares our empirical results to those using traditional parametric methods. Section \ref{sec:conclusion} concludes. Proofs and additional details pertinent to the analysis are presented in the Supplementary Appendix. A \texttt{Python} package to implement our developed tools is available at \url{https://github.com/vishalkamat/npdemand}.

\subsection{Related Literature}\label{sec:rel_literature}

A growing literature studies nonparametric identification of various quantities in discrete choice settings. One approach pursued in this literature is to argue point identification, which is often based on requiring large amounts of exogenous variation in the data \citep[e.g.,][]{berry/haile:09,berry/haile:14, briesch/etal:10, chiappori/etal:09, matzkin:93}. However, in many applications such as the ones we focus on, there exists only discrete variation, which generally gives rise to the case of partial identification. A number of recent papers have developed tools to evaluate various questions---such as estimating the effect of different prices and choice sets on demand, characterizing the underlying utility functions, and testing the premise of utility maximization---in setups that permit partial identification \citep[e.g.,][]{chesher/etal:13, kamat:19, kitamura/stoye:18, manski:07,tebaldi/etal:19}. As in our analysis, these papers carefully exploit the specific structure of their models and parameters to show how to construct the sharp identified set. But, as our setup and the parameters of interest are different from theirs, the developed arguments are distinct and complementary.

Our analysis is most closely related to recent work in the literature on nonparametric welfare analysis.  A building block of our analysis is the fact that we can express the average willingness to pay in terms of demand. To show this, we apply results from \cite{bhattacharya:15,bhattacharya:18} who formally derived such expressions for the class of nonparametric choice models we consider. If demand was point-identified, we could directly use these results to identify the welfare effects of interest. Our novelty is to show how to exploit these results when demand might be only partially identified. Recently, \cite{bhattacharya:19} derives analytic nonparametric bounds for welfare effects in such cases for a binary choice problem with a single price dimension. As in our approach, the paper's arguments are based on demand functions that are constant over a carefully constructed partition of the space of prices. However, as highlighted in Section \ref{sec:partition}, the arguments behind the construction of this partition rely on the unidimensionality of the space and require novel extensions to generalize to the case of multiple alternatives and prices we consider in our setup.

Our analysis is also conceptually related to the work of \cite{mogstad/etal:18} in an alternative setting of a binary treatment model. Their identification problem shares a similar structure where the parameters of interest can be expressed in terms of primitive functions---marginal treatment effects in their setup---that are only partially identified by the data. Indeed, our approach to incorporate parametric restrictions follows that in their paper. In contrast, as highlighted in Section \ref{sec:partition}, their arguments to compute nonparametric bounds rely on the unidimensionality of their primitive functions, which arises due to the focus on a binary treatment. In this sense, our arguments which allow for multidimensional functions can provide insights on how to obtain nonparametric bounds in settings with multiple treatments \citep[e.g.,][]{kamat/etal:22}.

Our empirical analysis contributes to the literature on the evaluation of school voucher programs, and particularly a smaller group of papers that uses choice models to study various school choice-related questions of interest \citep[e.g.,][]{allende:19, arcidiacono/etal:16, carneiro/etal:19, gazmuri:19, neilson:13}. While these papers consider richer models that allow studying various effects of vouchers that go beyond the scope of our analysis, such as general equilibrium effects through school competition or peer quality, they do so using fully parameterized models. Our analysis complements these studies by evaluating a narrower, yet relevant question, but doing so using robust nonparametric tools.

\section{Baseline Identification Analysis}\label{sec:baseline}

\subsection{Model}\label{sec:model}

Let $\mathcal{J}$ be a discrete set of choice alternatives such that $|\mathcal{J}| \geq 2$. For each individual $i$, suppose that we observe $(D_i,P_i)$, where $D_i$ denotes the chosen alternative from $\mathcal{J}$, and $P_i = (P_{ij} : j \in \mathcal{J})$ denotes a vector of prices for each alternative that the individual faces. Let $\mathcal{P}_{\text{obs}}$ denote the support of the observed price vector, which we assume to be discrete. Certain alternatives potentially may not exhibit any price variation in which case we normalize their prices to 0. We assume the observed choice is the product of an underlying utility maximizing decision. Specifically, denoting by $Y_i$ the individual's (unobserved) disposable income and by $U_{ij}:\mathbf{R} \to \mathbf{R}$ their (unobserved) utility function for alternative $j \in \mathcal{J}$, we take the observed choice to be given by
\begin{align}\label{eq:utlitymax}
  D_i = \argmax_{j \in \mathcal{J}} U_{ij}(Y_{i} - P_{ij})
\end{align}
i.e. the alternative maximizing the utility of the disposable income net of its price.

Apart from the utility maximizing structure, we highlight that our choice model is nonparametric and nonseparable, and allows for completely general unobserved heterogeneity. This is in contrast to traditional models employed in the literature that impose a combination of additional restrictions such as functional forms on the utility and parametric distributions on the unobserved heterogeneity---see Section \ref{sec:comparison} for details. A limitation of our setup, however, is that we do not model a supply side that generates prices as well as other factors beyond prices that may affect choice.  As noted in Section \ref{sec:intro} in terms of our empirical findings, this has potential implications on the interpretation and scope of our counterfactuals.

Given the above structure, our analysis exploits the fact that our parameters of interest can be expressed in terms of demand functions. In turn, we frame our problem in terms of these functions and consider various assumptions directly on them. The demand functions correspond to the distribution of choices across individuals at a given value of the price vector. More formally, let $\mathcal{P} = \prod_{j \in \mathcal{J}} \mathcal{P}_j$ denote the domain of price vectors, where $\mathcal{P}_j = \{0\}$ if prices of the $j$th alternative are normalized to 0 and $\mathcal{P}_j = \mathbf{R}$ otherwise, and let $D_i(p) = \argmax_{j \in \mathcal{J}} U_{ij}(Y_{i} - p_{j})$ denote the individual's choice had the price vector been set to $p \in \mathcal{P}$. Using this additional notation, we can respectively define the unconditional and conditional on $P_i = p' \in \mathcal{P}_{\text{obs}}$ demand by
\begin{align}
q_j(p) &= \text{Prob}\{D_i(p) = j\}~, \label{eq:q_def} \\
q_j(p|p') &= \text{Prob}\{D_i(p) = j|P_i = p'\} \label{eq:q_def_cond}
\end{align}
for each  $j \in \mathcal{J}$ and $p \in \mathcal{P}$. Our analysis is primarily based on the unconditional demand functions. But we also define conditional demand functions as they allow us to formally state the fact that our analysis throughout takes the observed variation in prices to be exogenous. In particular, we do so by assuming the following relation between the conditional and unconditional demand functions:

\begin{assumption2}{E}{(Exogeneity)}\label{ass:E}~
For each $j \in \mathcal{J}$, $q_j(p) = q_j(p|p')$ for all $p \in \mathcal{P}$ and $p' \in \mathcal{P}_{\text{obs}}$.
\end{assumption2}

\noindent Assumption \ref{ass:E} states that demand is invariant to values of the observed price vector, and captures that the observed price is exogenous of the remaining underlying variables affecting choices. This implies that conditional and unconditional demand are equal, and hence that the underlying demand functions can be uniquely captured by the vector $q \equiv (q_j : j \in \mathcal{J})$ of unconditional demand functions. As a result, in the remainder of our analysis, we focus solely on unconditional demand; whenever we refer to demand, it is understood we are referring to unconditional demand.

In our analysis, we also consider various additional assumptions on demand that restrict $q$ to lie in some space of functions. Let $\mathbf{F}$ generically denote this restricted space of functions. We postpone the description of these assumptions until after we present our parameters of interest and the objective of our analysis, as they will better motivate their purpose.

\subsection{Parameters of Interest}\label{sec:parameters}

We are interesting in evaluating the welfare effect of a price subsidy that decreases prices between two pre-specified price vectors.\footnote{The welfare effects for general price changes cannot be simply expressed in terms of demand defined in \eqref{eq:q_def}, but require defining demand at counterfactual prices as well as disposable income---see \cite{bhattacharya:15,bhattacharya:18}. Identification in this case therefore not only requires variation and assumptions along the price dimension but also along that of disposable income, which we leave for future work. We note, however, that our analysis straightforwardly applies to evaluate the effects of general price changes solely on demand, i.e. \eqref{eq:theta_q} with $g^{a,b} = 0$. \label{foot:1}} Let $p^a,p^b \in \mathcal{P}$ respectively denote the larger and smaller pre-specified vectors in this price decrease in the sense that $p^b_j \leq p^a_j$ for $j \in \mathcal{J}$.

We measure the welfare effect of the price subsidy by the willingness to pay for it. It provides a natural money metric for the gains and, equivalently, corresponds to the negative of the compensating variation for the price decrease induced by the subsidy. Formally, an individual's willingness to pay for the subsidy can be defined by the variable $B_i^{a,b}$ that solves
\begin{align}\label{eq:B_t}
\max_{j \in \mathcal{J}}U_{ij}(Y_{i} - p^a_j) = \max_{j \in \mathcal{J}}U_{ij}(Y_{i} - p^b_j - B_i^{a,b})~,
\end{align}
i.e. the amount of money to be subtracted from the individual's income under the lower price so that they are indifferent and obtain the same utility as that under the higher price. Our analysis focuses on the average willingness to pay which is defined by
\begin{align}\label{eq:E_B}
E[B_i^{a,b}]~.
\end{align}

As mentioned, our analysis exploits the fact that our parameters can be expressed as functions of the demand functions. In order to show this for \eqref{eq:E_B}, we exploit results from \cite{bhattacharya:15,bhattacharya:18} who precisely showed this in the context of a nonparametric, nonseparable model of choice as that in \eqref{eq:utlitymax}. In the following proposition, we reproduce this result in terms of our notation. To this end, it is useful to first introduce some additional notation. Let $\Delta_1^{a,b} \leq \ldots \leq \Delta_{|\mathcal{J}|}^{a,b}$ denote the ordered values of $p^a_j - p^{b}_j$ across $j \in \mathcal{J}$, i.e. the price decrements for the different alternatives, and let $\mathcal{J}^{a,b}_{l} = \{j \in \mathcal{J} : p^a_j - p^{b}_j \geq \Delta_{l}^{a,b}\}$ denote the alternatives whose price decrease is at least greater than the $l$th ordered price decrement. Moreover, with some abuse of notation, let $\text{min}\{p^a,p^b+t\} = (\text{min}\{p_j^a,p_j^b+t\} : j \in \mathcal{J})$ for $t \in \mathbf{R}$ denote the element wise minimum. Using this notation, we can then formally state the result as follows.

\begin{proposition}\label{prop:EB}
For each individual $i$, suppose $U_{ij}$ is continuous and strictly increasing for each $j \in \mathcal{J}$. Then we have that $B_i^{a,b}$ defined in \eqref{eq:B_t} exists and is unique, and that
\begin{align}\label{eq:AB_t_q}
  E[B_i^{a,b}] = \Delta_1^{a,b} + \sum_{l=1}^{|\mathcal{J}|-1} \sum\limits_{j \in \mathcal{J}_{l+1}^{a,b}} ~\int\limits_{\Delta_{l}^{a,b}}^{\Delta_{l+1}^{a,b}} q_j\left(\text{min}\{p^a,p^b+t\}\right) dt~.
\end{align}
\end{proposition}

Proposition \ref{prop:EB} requires utility to be increasing, which is captured by our restrictions on demand. Moreover, it requires them to be continuous. Importantly, this means that the utilities may not be able to implicitly capture a scenario where an individual is unable to afford an alternative at a higher price but can do so at a lower price, since this implies that the price change could discontinuously affect utility. See Section \ref{sec:liquidity} where we make this point more precisely and show how to extend our analysis under an explicit model of liquidity constraints.

To intuitively understand the expression in \eqref{eq:AB_t_q}, observe for the $l$th price decrement that the price decrease for the alternatives in $\mathcal{J}^{a,b}_{l+1}$ jointly goes from $\Delta_{l}^{a,b}$ to $\Delta_{l+1}^{a,b}$. As this price decrease can simply be viewed as a cash transfer conditional on choosing alternatives in $\mathcal{J}^{a,b}_{l+1}$, the willingness to pay for it can potentially be only between the minimum and maximum value of the transfer, namely $\Delta_{l}^{a,b}$ and $\Delta_{l+1}^{a,b}$. The expression in turn states that the average willingness to pay for the $l$th decrement corresponds to the area under the demand curve for the alternatives in $\mathcal{J}^{a,b}_{l+1}$ as prices jointly vary between the minimum and maximum values in the presence of the transfer, and the total average willingness to pay is the sum across all the decrements. See \citet[][Section 2.1]{bhattacharya:15} for more discussion on the intuition.

In addition to the above, we are also interested in parameters that evaluate the effect of the price subsidy on demand. Moreover, we are interested in those that measure the difference in the welfare effect and a weighted change in demand, which can for example allow us to compare the benefits and costs of the subsidy as we do in our application. For the purposes of our analysis, these various parameters all share a common underlying structure, which can be captured by a general parameter that can be expressed as a function of $q$ as follows
\begin{align}\label{eq:theta_q}
\theta(q) = g^{a,b} \left(\Delta_1^{a,b} + \sum_{l=1}^{|\mathcal{J}|-1} \sum\limits_{j \in \mathcal{J}_{l+1}^{a,b}} \int\limits_{\Delta_{l}^{a,b}}^{\Delta_{l+1}^{a,b}} q_j\left(\text{min}\{p^a,p^b+t\}\right) dt \right) + \sum_{j \in J} g_j^a q_j(p^a) + g_j^b q_j(p^b)~,
\end{align}
where $g^{a,b}$, $\{g_j^a : j \in \mathcal{J}\}$ and $\{g_j^b : j \in \mathcal{J}\}$ are pre-specified values that depend on the parameter of interest, i.e. the various parameters all correspond to linear combinations of the expression in \eqref{eq:AB_t_q} and demand evaluated at $p^a$ and $p^b$. Indeed, taking $g^{a,b} = 1$ and $g^a_j = g^b_j = 0$ for $j \in \mathcal{J}$, we have the parameter in \eqref{eq:AB_t_q}. Alternatively, changing $g_j^a$ and $g_j^b$ to be the costs one may associate with the demand for alternative $j \in \mathcal{J}$ under prices $p^a$ and $p^b$, we can also analyze welfare effect net of the costs associated with the price decrease. More generally, by specifying various other values for $g^{a,b}$, $\{g_j^a : j \in \mathcal{J}\}$ and $\{g_j^b : j \in \mathcal{J}\}$, we can analyze a range of parameters that capture the welfare and demand effects of the price subsidy---see our application in Section \ref{sec:empirical} for more concreteness.

\subsection{Identified Set}\label{sec:identifiedset}

The goal of the analysis is to learn about a pre-specified parameter of interest $\theta(q)$ given by \eqref{eq:theta_q}. Given the function $\theta$ is known, what we can learn about the parameter translates to what we know about $q$ through the data and imposed assumptions. From the data, we observe the distribution of $(D_i,P_i)$, which for the purposes of the identification analysis is assumed to be perfectly known without uncertainty---we discuss estimation and inference in Section \ref{sec:empirical}. Given the structure in \eqref{eq:utlitymax}, the definition of demand in \eqref{eq:q_def}-\eqref{eq:q_def_cond} and Assumption \ref{ass:E}, it follows that demand must satisfy
\begin{align}
q_j(p) &= \text{Prob}[D_i = j | P_i = p] \label{eq:q_data}
\end{align}
for $j \in \mathcal{J}$ and $p \in \mathcal{P}_{\text{obs}}$, i.e. the random variation in prices reveals the value of demand at prices observed in the data. From the assumptions, we have that demand is restricted to lie in a space of functions $\mathbf{F}$. The admissible space of demand that satisfies the data and assumptions can be defined by
\begin{align}\label{eq:Q}
\mathbf{Q} = \{q \in \mathbf{F} : q \text{ satisfies } \eqref{eq:q_data}\}~.
\end{align}

What can be learned about the parameter of interest can then be formally captured by the identified set, which is defined by
\begin{align}\label{eq:theta_identifiedset}
\theta(\mathbf{Q}) = \left\{\theta_0 \in \mathbf{R} : \theta(q) = \theta_0 \text{ for some } q \in \mathbf{Q} \right\} \equiv \Theta~,
\end{align}
i.e. the image of the space of admissible functions $\mathbf{Q}$ under the function $\theta$. Our goal is to compute the identified set under the assumptions we impose on demand, which we describe next.

\subsection{Demand Specification}\label{sec:assumptions}

Given our nonparametric model, the demand functions remain entirely unrestricted apart from the logical ones arising from the fact that they are distributions, namely
\begin{align}
 q_j(p) &\geq 0~\text{for}~j \in \mathcal{J}~, \label{eq:q_logical1} \\
 \sum\limits_{j \in \mathcal{J}} q_j(p) &= 1~, \label{eq:q_logical2}
\end{align}
for $p \in \mathcal{P}$, i.e. demand is positive and sum to one. On the other hand, observe that while the data restrictions in \eqref{eq:q_data} reveal the value of demand at certain prices, the parameters of interest generally depend on values beyond these prices. To reach informative conclusions, our analysis therefore considers additional assumptions that restrict how demand varies with prices.

In our baseline analysis, we consider the following nonparametric shape restriction, which then defines $\mathbf{Q}$ in \eqref{eq:Q}---see Sections \ref{sec:dim_red} and \ref{sec:parametric} for extensions to additional restrictions.

\begin{assumption2}{B}{(Baseline)}\label{ass:B}~
For each $j \in \mathcal{J}$, $q_j$ is weakly increasing in $p_m$ for each $m \in \mathcal{J} \setminus \{j\}$.
\end{assumption2}

\noindent Assumption \ref{ass:B}, referred to as weak substitutes in \cite{berry/etal:13}, is a fundamental shape restriction present in the majority of discrete choice models and is implied by taking $U_{ij}$ to be increasing for each $j \in \mathcal{J}$. It specifically imposes that for each $p,p' \in \mathcal{P}$ such that $p_j > p'_j$ for $j \in \mathcal{J}' \subseteq \mathcal{J}$ and $p_j = p'_j$ for $j \in \mathcal{J} \setminus \mathcal{J}'$, we have that
\begin{align}\label{eq:q_shape}
 q_j(p) \geq q_j(p')
\end{align}
for each $j \in \mathcal{J} \setminus \mathcal{J}'$. Under this specification, observe that the restricted space for demand is given by $\mathbf{F}_B = \{q \in \overbar{\mathbf{F}} : q \text{ satisfies } \eqref{eq:q_logical1}-\eqref{eq:q_shape} \}$, and, in turn, the admissible space of functions in \eqref{eq:Q} by
\begin{align}\label{eq:Q_B}
\mathbf{Q}_B = \left\{q \in \overbar{\mathbf{F}} : q \text{ satisfies } \eqref{eq:q_logical1}-\eqref{eq:q_shape} \text{ and } \eqref{eq:q_data} \right\}~,
\end{align}
where $\overbar{\mathbf{F}}$ denotes the set of all functions from $\mathcal{P}$ to $\mathbf{R}^{|\mathcal{J}|}$.

\subsection{Computing the Identified Set}\label{sec:compute_identifiedset}

In principle, the identified set in \eqref{eq:theta_identifiedset} can be computed by searching over $q$ in $\mathbf{Q}$ and taking their image under the function $\theta$. However, under our baseline specification above, doing so is infeasible as $\mathbf{Q}_B$ is an infinite-dimensional space. We conclude this section by showing how to proceed in this case. The main idea is to show how to replace $\mathbf{Q}_B$ by a finite-dimensional space $\mathbf{Q}_B^{\text{fd}}$ such that there is no loss of information in the sense that $\theta(\mathbf{Q}_B) = \theta(\mathbf{Q}^{\text{fd}}_B)$. This allows translating the problem to a tractable finite-dimensional one of simply searching through $\mathbf{Q}_B^{\text{fd}}$. In particular, taking $\mathcal{W}$ to be a finite partition of $\mathcal{P}$, we consider a finite dimensional space of functions given by
\begin{align}\label{eq:Q_B_fd}
\mathbf{Q}_B^{\text{fd}} = \left\{q \in \mathbf{Q}_B : q_j(p) = \sum_{w \in \mathcal{W}} 1_w(p) \cdot  \beta_{j}(w) \text{ for some } \{\beta_{j}(w)\}_{w \in \mathcal{W}} \text{ for each } j \in \mathcal{J} \right\}~,
\end{align}
where $1_w(p) \equiv 1\{p \in w\}$, and $\{\beta_j(w) : w \in \mathcal{W},~j \in \mathcal{J}\}$ are unknown parameters, i.e. a subset of $\mathbf{Q}_B$ such that $q$ is parameterized to be constant over the elements of $\mathcal{W}$. The main challenge is how to choose the partition $\mathcal{W}$ such that $\theta(\mathbf{Q}_B) = \theta(\mathbf{Q}^{\text{fd}}_B)$. As shown below, we carefully do so such that the resulting $q$ is sufficiently rich to define the parameter of interest and preserve the information provided by the data and shape restrictions.

\subsubsection{Partitioning the Space of Prices}\label{sec:partition}

Denoting by $\mathcal{P}_l^{a,b} = \{ p \in \mathcal{P} : p_j = \min \{p_j^a, p_j^b + t\} \text{ for } t \in (\Delta_l^{a,b},\Delta_{l+1}^{a,b}),~ j \in \mathcal{J}  \}$ for $1 \leq l \leq |\mathcal{J}|-1$ and $\{p\}$ for $p \in \{p^a,p^b\} \cup \mathcal{P}_{\text{obs}}$ the various sets of prices that underlie the parameter in \eqref{eq:theta_q} and data restrictions in \eqref{eq:q_data}, let
\begin{align}\label{eq:set_P}
\left\{\mathcal{P}_l^{a,b} : 1 \leq l \leq |\mathcal{J}|-1 \right\} \cup \left\{\{p\} : p \in \left\{p^a,p^b\right\} \cup \mathcal{P}_{\text{obs}} \right\} \equiv \{\mathcal{P}^*_1 , \ldots , \mathcal{P}^*_L\}
\end{align}
denote the collection of these price sets. Given these sets of prices, we define a collection of sets that is the key building block in our construction of $\mathcal{W}$. In this definition and in what follows, for a set $v \subseteq \mathbf{R}^{|\mathcal{J}|}$, we take $v_{[j]} = \left\{ t \in \mathbf{R} : p_j = t \text{ for some } p \in v \right\}$ to the denote the set of values of $v$ in the $j$th coordinate.

\begin{definition2}{V}{(Partition $\mathcal{V}$)}\label{def:V}
Let $\mathcal{V}$ denote a finite partition of $\mathcal{P}^* = \bigcup_{l=1}^L \mathcal{P}^*_l$ such that: (i) for each $l \in \{1,\ldots,L\}$, there exists $\mathcal{V}_l \subseteq \mathcal{V}$ such that $\mathcal{P}_l^* = \bigcup_{v \in \mathcal{V}_l} v$; (ii) for each $v \in \mathcal{V}$ and $j \in \mathcal{J}$, $v_{[j]}$ is an interval; and (iii) for all $v,v' \in \mathcal{V}$ and each $j \in \mathcal{J}$, either $v_{[j]} = v'_{[j]} ~\text{ or }~ v_{[j]} \cap v'_{[j]} = \emptyset$.
\end{definition2}

\noindent Definition \ref{def:V} states that $\mathcal{V}$ is a finite partition of the union of the sets in \eqref{eq:set_P} such that its elements satisfy certain properties: (i) their unions allow building the sets in \eqref{eq:set_P}; (ii) they are connected in each coordinate; and (iii) each pair of elements either completely overlap or are disjoint in each coordinate. Intuitively, we highlight that the first property, as the sets in \eqref{eq:set_P} underlie the parameter of interest and data restrictions, is what ensures that our finite-dimensional $q$ will be sufficiently rich to define the parameter and data restrictions. On the other hand, the latter two properties, which implies that the sets can be ordered and pairwise compared across each coordinate, is what ensures that our $q$ will preserve the information provided by the shape restrictions in \eqref{eq:q_shape}, which we can observe are based on pairwise comparisons of prices.

\begin{figure}
\centering
\caption{Various sets of prices for an example with $\mathcal{J}=\{1,2\}$ and $\mathcal{P}_{\text{obs}} = \{p’\}$, where $p^a_2 - p^b_2 > p^a_1 - p^b_1 \equiv 0$ and $p'$ is such that $p'_1 < p^a_1 = p^b_1$ and $p^b_2 < p'_2 < p^a_2$}
\makebox[\textwidth]{ \begin{tabular}{c c c}
\subcaptionbox{Sets underlying $\mathcal{P}^*$}{~~~~~~\includegraphics[width = 2in]{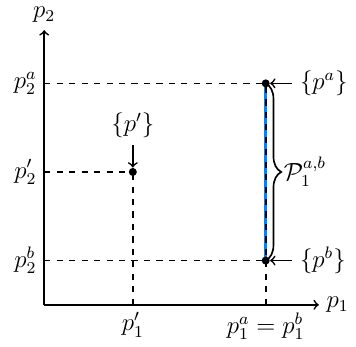}~~~~~~}& ~~~~ &
\subcaptionbox{$\mathcal{V} \equiv \{v_1 , \ldots, v_{5}\}$ satisfying Definition \ref{def:V}}{~~~~~~\includegraphics[width = 2in]{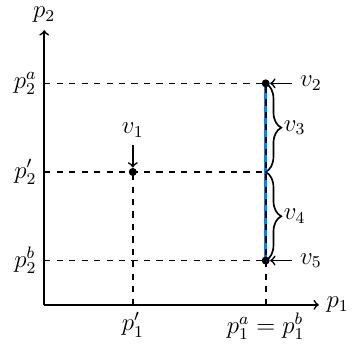}~~~~~~}
\end{tabular}}
\label{fig:sets}
\end{figure}

To better understand the various sets of prices, Figure \ref{fig:sets}(a) first graphically illustrates those in \eqref{eq:set_P} in the context of a stylized example with two alternatives. They are given by $\{\mathcal{P}^{a,b}_1 \} \cup \{\{p\} : p \in \{p^a,p^b,p'\}\}$, where we can explicitly write $\mathcal{P}^{a,b}_1$ as
\begin{align*}
 \mathcal{P}^{a,b}_1 &= \{p \in \mathbf{R}_+^2 : p_1 = p^a_1,~p_2 = p^b_2 + t,~t \in (\Delta^{a,b}_1,\Delta^{a,b}_2)\}~,
\end{align*}
with $\Delta^{a,b}_1 = 0$ and $\Delta^{a,b}_2 = p^a_2 - p^b_2$ as $p^a_2 - p^b_2 > p^a_1 - p^b_1 \equiv 0$. Figure \ref{fig:sets}(b) then illustrates a partition of the union of these sets satisfying Definition \ref{def:V}. Specifically, while these sets satisfy Definition \ref{def:V}(i) by construction and Definition \ref{def:V}(ii) as they are connected, Figure \ref{fig:sets}(a) reveals that $\{p'\}$ partially overlaps with $\mathcal{P}^{a,b}_1$ in the second coordinate and hence that they do not satisfy Definition \ref{def:V}(iii). Figure \ref{fig:sets}(b) in turn partitions $\mathcal{P}^{a,b}_1$ where it intersects with $\{p'\}$ to obtain a finer collection of sets that also satisfies Definition \ref{def:V}(iii). Moreover, observe that by construction it is the coarsest collection of such sets. In Appendix \ref{S-sec:procedure_V}, we describe how the idea behind this construction can be extended to generally obtain a partition $\mathcal{V}$ from the sets in \eqref{eq:set_P} and argue that it corresponds to the coarsest possible partition.

Using $\mathcal{V}$, we construct $\mathcal{W}$ as follows. For each $j \in \mathcal{J}$, observe that the collection of sets determined by the prices in $v \in \mathcal{V}$ in the $j$th coordinate, i.e. $\{v_{[j]} : v \in \mathcal{V} \}$, generates a partition of $[p_j^b+\Delta^{a,b}_1, p_j^a] \cup \{p_j : \mathcal{P}_{\text{obs}}\}$.\footnote{Note that stating $\{v_{[j]} : v \in \mathcal{V}\}$ is a partition of $[p_j^b+\Delta^{a,b}_1,p_j^a]$ is not formally correct as $\mathcal{P}^{a,b}_l$ for $1 \leq l \leq |\mathcal{J}|-1$ are open and hence their end points are not necessarily contained in the partition. To ensure it is a partition, we need to carefully alter the boundaries of certain $v \in \mathcal{V}$ to be either closed or open. However, for expositional ease, we abstract away from doing so as this distinction is not practically important for our analysis as our parameter of interest in \eqref{eq:theta_q} only takes Lebesgue integrals over these sets.}  Moreover, denoting by $\underline{p}_{1,j} \leq \ldots \leq \underline{p}_{\underline{M}_j,j}$ and $\bar{p}_{1,j} \leq \ldots \leq \bar{p}_{\overline{M}_j,j}$ the ordered values of $\{p_j : p \in \mathcal{P}_{\text{obs}},~p_j \leq p^b_j+\Delta^{a,b}_1\} \cup \{p^b_j+\Delta^{a,b}_1\}$ and $\{p_j : p \in \mathcal{P}_{\text{obs}},~p_j \geq p^a_j\} \cup \{p^a_j\}$, respectively, consider the collection of sets determined by the intervals between these values $\{ (-\infty, \underline{p}_{1,j} ) \} \cup \{(\underline{p}_{m-1,j},\underline{p}_{m,j})\}_{m=2}^{\underline{M}_j}  \cup \{(\bar{p}_{m-1,j},\bar{p}_{m,j})\}_{m=2}^{\overline{M}_j} \cup \{(\bar{p}_{\bar{M}_j,j},\infty ) \}$, which partitions the remaining space $\mathbf{R} \setminus ( [p_j^b+\Delta^{a,b}_1, p_j^a] \cup \{p_j : \mathcal{P}_{\text{obs}}\} )$. We together have that
\begin{align}\label{eq:set_Vj}
\mathcal{V}_j = &\{(-\infty, \underline{p}_{1,j})\} \cup\{(\underline{p}_{m-1,j},\underline{p}_{m,j})\}_{m=2}^{\underline{M}_j} \cup\{ v_{[j]}\}_{v \in \mathcal{V}} \cup \{(\bar{p}_{m-1,j},\bar{p}_{m,j})\}_{m=2}^{\overline{M}_j} \cup\{(\bar{p}_{\bar{M}_j,j},\infty)\}
\end{align}
generates a partition for the domain of prices for the $j$th coordinate when its price is not normalized to 0, while $\mathcal{V}_j$ simply equals $\{\{0\}\}$ when its price is normalized. For example, for that in Figure \ref{fig:sets}, these partitions given $\mathcal{V}$ in Figure \ref{fig:sets}(b) correspond to
\begin{align*}
 \mathcal{V}_1 &= \left\{ (-\infty,p'_1),~\{p'_1\},~ (p'_1,p^a_1),~\{p^a_1\},~(p^a_1,\infty) \right\}~, \\
 \mathcal{V}_2 &= \left\{ (-\infty,p^b_2),~\{p^b_2\},~ (p^b_2,p'_2),~\{p'_2\},~(p'_2,p^a_2),~\{p^a_2\},~(p^a_2,\infty) \right\}~.
\end{align*}
Given these partitions in each dimension, we then take $\mathcal{W}$ to be their Cartesian product, i.e.
\begin{align}\label{eq:set_W}
\mathcal{W} = \prod_{j \in \mathcal{J}} \mathcal{V}_j \equiv \left\{ w_1 , \ldots , w_{M}\right\}~.
\end{align}

It is useful to highlight that our construction of $\mathcal{W}$ simplifies in the case where demand is unidimensional---essentially arising when all alternatives except one have prices normalized to 0. As noted in Section \ref{sec:rel_literature}, this special case is similar to the identification problem studied in \cite{bhattacharya:19} and \cite{mogstad/etal:18}, who propose a finite dimensional space comparable to that in \eqref{eq:Q_B_fd} as a solution. In this case, we need not impose that $\mathcal{V}$  satisfies Definition \ref{def:V}(iii) as it is automatically implied by the fact that $\mathcal{V}$ is a partition. In the multidimensional case, this is not so and we need to explicitly introduce it to ensure that the information provided by the shape restrictions in \eqref{eq:q_shape} is preserved. Moreover, given $\mathcal{V}$, the construction of $\mathcal{W}$ follows more straightforwardly in the unidimensional case as the sets in $\mathcal{V}$ and those outside it, i.e. those in \eqref{eq:set_Vj}, directly generate a partition of the space of prices for a single coordinate. For the multidimensional case, an additional complication remains of how to combine these one dimensional partitions to partition the entire space of prices, which we propose to solve by taking their Cartesian product as in \eqref{eq:set_W}.

\subsubsection{Equivalent Finite Dimensional Characterization}\label{sec:equi_charac}

Taking our constructed $\mathcal{W}$ and finite-dimensional space $\mathbf{Q}_B^{\text{fd}}$ in \eqref{eq:Q_B_fd}, we can transform our problem in terms of $\beta \equiv (\beta'_j : j \in \mathcal{J})'$, where $\beta_j = (\beta_{j}(w_1), \ldots, \beta_{j}(w_{M}) )$, i.e. the variable parameterizing $q \in \mathbf{Q}_B^{\text{fd}}$. As $\theta$ is continuous in $q$ and $q$ is continuous in $\beta$, let $\theta_B$ be the continuous function of $\beta$ such that $\theta(q) = \theta_B(\beta)$. Moreover, $\mathbf{Q}_B^{\text{fd}}$ can be written in terms of $\beta$ by
\begin{align}\label{eq:beta_rest}
\mathbf{B} = \left\{ \beta \in \mathbf{R}^{d_{\beta}} : \left( \sum_{w \in \mathcal{W}} 1_{w} \cdot \beta_{j}(w) : j \in \mathcal{J} \right) \in \mathbf{Q}_B \right\}~,
\end{align}
where $d_{\beta}$ denotes the dimension of $\beta$, i.e. the set of values of $\beta$ that ensure that the corresponding $q$ is in $\mathbf{Q}_B$. We then have that
\begin{align}\label{eq:b_identifiedset}
\theta\left(\mathbf{Q}^{\text{fd}}_B\right) = \left\{\theta_0 \in \mathbf{R} : \theta_B(\beta) = \theta_0 \text{ for some } \beta \in \mathbf{B} \right\} \equiv \Theta_B~.
\end{align}
For example, for the setup in Figure \ref{fig:sets}, we have
\begin{align}
 \theta_B(\beta) =& g^{a,b} \Big((p_2'-p^b_2) \beta_2(w') + (p^a_2-p'_2) \beta_2(w'') \Big) +  \sum_{j \in \{1,2\}} g^{a}_j \beta_j(\{p^a\}) + g^{b}_j \beta_j(\{p^b\}) ~, \label{eq:beta_ex_theta}
\end{align}
where $w' = \{p^a_1\} \times [p^b_2,p'_2]$ and $w'' = \{p^a_1\} \times [p'_2,p^a_2]$, while $\mathbf{B}$ is given by the set of all $\beta$ satisfying
\begin{align}
 \beta_j(\{p\}) &= \text{Prob}[D_j = j|P_i = p] \text{ for } p \in \{p'\},~j \in \mathcal{J}~, \label{eq:beta_ex_data} \\
 \beta_j(w) &\geq 0 \text{ for } w \in \mathcal{W},~j \in \mathcal{J}~, \label{eq:beta_ex_logical1}\\
 \sum_{j \in \mathcal{J}}\beta_j(w) &= 1 \text{ for } w \in \mathcal{W}~, \label{eq:beta_ex_logical2}
\end{align}
which capture the restrictions in \eqref{eq:q_data}, \eqref{eq:q_logical1} and \eqref{eq:q_logical2}, respectively, and
\begin{align}\label{eq:beta_ex_shape}
 \beta_j(w) \geq \beta_j(w')
\end{align}
for all $w,w' \in \mathcal{W}$ and $j \in \mathcal{J} \setminus \mathcal{J}'$, where $\mathcal{J}' \subseteq \mathcal{J}$, such that $t > t'$ for all $t \in w_{[j]},~t' \in w'_{[j]}$ for $j \in \mathcal{J}'$ and $w_{[j]} = w_{[j']}$ for $j \in \mathcal{J} \setminus \mathcal{J}'$, which captures the restriction in \eqref{eq:q_shape}.

Indeed, we can observe that computing $\theta(\mathbf{Q}^{\text{fd}}_B)$ is a finite-dimensional problem corresponding to searching through $\beta$. However, as $\mathbf{Q}^{\text{fd}}_B \subset \mathbf{Q}_B$, a concern is that we may only have $\theta(\mathbf{Q}^{\text{fd}}_B) \subseteq \theta(\mathbf{Q}_B)$ rather than $\theta(\mathbf{Q}^{\text{fd}}_B) = \theta(\mathbf{Q}_B)$. The following proposition shows this is not the case and the finite-dimensional problem in fact preserves all the information.

\begin{proposition}\label{prop:identifiedset_base_1}
Suppose that $\mathbf{Q} = \mathbf{Q}_B$. Then, the identified set in \eqref{eq:theta_identifiedset} is equal to that in \eqref{eq:b_identifiedset}, i.e. $\Theta = \Theta_B$. In addition, if $\mathbf{B}$ is empty then by definition $\Theta_B$ is empty; whereas, if $\mathbf{B}$ is non-empty then $\Theta_B = [\underline{\theta}_{B}, \bar{\theta}_{B}]$, where
\begin{align}\label{eq:bounds_opt}
\underline{\theta}_{B} = \min_{\beta \in \mathbf{B}}~ \theta_B(\beta) ~\text{ and }~ \bar{\theta}_{B} = \max_{\beta \in \mathbf{B}} ~\theta_B(\beta)~.
\end{align}
\end{proposition}

Proposition \ref{prop:identifiedset_base_1} exploits the idea that as the parameter and data restrictions are defined as integrals over certain sets, directly taking $q$ to be a constant over a partition with values equal to these integrals is sufficient to learn about the parameter of interest rather than working with the underlying $q$. The proof formally illustrates this by showing that for each $q \in \mathbf{Q}_B$ we can construct a version of the finite-dimensional variable $\beta \in \mathbf{B}$ that generates the same parameter value as $q$, i.e. $\theta(q) = \theta_B(\beta)$. This construction corresponds to taking
\begin{align}\label{eq:beta_constr}
 \beta_j(w) = \int_0^1 q_j(p(t,w)) dt~,
\end{align}
for $w \in \mathcal{W}$ and $j \in \mathcal{J}$, where $p(t,w) = (p_j(t,w) : j \in \mathcal{J})$ such that $p_j(t,w) = \underline{w}_{[j]} + (\bar{w}_{[j]} - \underline{w}_{[j]})~t$ with  $\underline{w}_{[j]} = \text{inf}~ w_{[j]}$ and $\bar{w}_{[j]} = \text{sup} ~w_{[j]}$, i.e. taking the constant value over $w$ to be equal to a specific integral. As $\mathcal{W}$ is constructed based on $\mathcal{V}$ that satisfies Definition \ref{def:V}, these integrals can be shown to equal those underlying the parameter and data restriction by construction given Definition \ref{def:V}(i) and to preserve the shape restrictions as they can be ordered given Definition \ref{def:V}(ii)-(iii), which together ensure that this constructed $\beta$ falls in $\mathbf{B}$ and $\theta(q)=\theta_B(\beta)$. 

We conclude this section by highlighting that Proposition \ref{prop:identifiedset_base_1} also shows that the finite-dimensional problem boils down to solely solving two optimization problems in \eqref{eq:bounds_opt}, which produce the endpoints of the identified set. Moreover, as in \eqref{eq:beta_ex_theta} and \eqref{eq:beta_ex_logical1}-\eqref{eq:beta_ex_shape}, the proof of the proposition explicitly derives $\theta_B$ and $\mathbf{B}$, the objective and constraint set of these optimization problems, respectively, which are revealed to be all linear in $\beta$. This implies that the optimization problems are linear programs, a useful observation in their implementation.

\section{Extensions}\label{sec:extensions}

In this section, we briefly consider several extensions of our baseline analysis, whose details are provided in the appendix for brevity.

\subsection{Dimension Reduction}\label{sec:dim_red}

While the optimization problems in Proposition \ref{prop:identifiedset_base_1} are linear programs, they can nonetheless be computationally expensive when the dimension of the optimizing variable $\beta$ is large. As the dimension of the optimizing variable equals $|\mathcal{J}| |\mathcal{W}|$, where $|\mathcal{W}|$ is given by $\prod_{j \in \mathcal{J}} |\mathcal{V}_j|$, observe that such a case arises especially when $|\mathcal{J}|$ is large, as in our empirical application. To ensure tractability in such cases, we propose below two lower-dimensional linear programs that are easier to compute and can continue to allow us to learn about our parameter.

Our first proposal considers sub-programs of \eqref{eq:bounds_opt} that obtain outer sets containing $\Theta_B$. In particular, given how $\mathcal{V}$ how was constructed, observe that the following subset of $\mathcal{W}$
\begin{align*}
\mathcal{W}^{\text{r}} = \left\{ w \in \mathcal{W} : w = \prod_{j \in \mathcal{J}} v_{[j]} \text{ for some } v \in \mathcal{V} \right\} \equiv \left\{w^{\text{r}}_1, \ldots, w^{\text{r}}_{M^{\text{r}}}\right\}~,
\end{align*}
captures the collection of price sets that play a role in the definition of the parameter, and in turn the subvector of $\beta$ defined over these sets given by $\beta^{\text{r}} =  (\beta^{\text{r}'}_j : j \in \mathcal{J} )' \equiv \phi(\beta)$, where $\beta^{\text{r}}_j = (\beta_{j}(w^{\text{r}}_1), \ldots, \beta_{j}(w^{\text{r}}_{M^{\text{r}}}) )$, is sufficient in determining $\theta_B$ in the sense that there exists a linear function $\theta_B^{\text{r}}$ such that $\theta_B^{\text{r}}(\beta^{\text{r}}) = \theta_B(\beta)$. Moreover, for the purposes of dimension reduction, observe that its dimension is given by $|\mathcal{J}| |\mathcal{V}|$, which can be substantially smaller than that of $\beta$. The lower-dimensional linear programs we then consider are those in terms of $\beta^{\text{r}}$ given by
\begin{align}\label{eq:b_r_opt}
\underline{\theta}_{B}^{\text{r}} = \min_{\beta^{\text{r}} \in \mathbf{B}^{\text{r}}}~ \theta^{\text{r}}_B(\beta^{\text{r}}) ~\text{ and }~ \bar{\theta}_{B}^{\text{r}} = \max_{\beta^{\text{r}} \in \mathbf{B}^{\text{r}}}~ \theta^{\text{r}}_B(\beta^{\text{r}})~,
\end{align}
where $\mathbf{B}^{\text{r}}$ denotes a set of $\beta^{\text{r}}$ determined by linear constraints. Indeed, if $\mathbf{B}^{\text{r}} = \phi(\mathbf{B})$, we have by construction that these programs are equivalent to those in \eqref{eq:bounds_opt}. In turn, taking $\mathbf{B}^{\text{r}}$ to be such that $\phi(\mathbf{B}) \subseteq \mathbf{B}^{\text{r}}$ implies $\underline{\theta}_{B}^{\text{r}} \leq \underline{\theta}_{B}$ and $\bar{\theta}_{B}^{\text{r}} \geq \bar{\theta}_{B}$, and hence provides an outer set for $\Theta_B$, i.e. $\Theta_B \in \left[ \underline{\theta}_{B}^{\text{r}}, \bar{\theta}_{B}^{\text{r}} \right]$. In Appendix \ref{S-sec:B_r}, we provide a natural choice of such a $\mathbf{B}^{\text{r}}$ determined by restrictions on $\beta^{\text{r}}$ implied by those in $\mathbf{B}$, which we implement in our empirical analysis, and find to be tractable and result in informative conclusions.

Our second proposal is to additionally impose separability on the demand functions given which we can sharply compute the identified set in a tractable manner. In our empirical analysis, we specifically consider the following separability assumption that imposes that demand to be a sum of lower-dimensional functions:

\begin{assumption2}{S}{(Separability)}\label{ass:AS}~
For each $j \in \mathcal{J}$, $q_j(p) = \sum\limits_{m \in \mathcal{J}} h_{jm}(p_m)$ for some unknown functions $\{h_{jm} : m \in \mathcal{J}\}$.
\end{assumption2}

\noindent Assumption \ref{ass:AS} imposes demand to be additively separable in prices of all the alternatives. In Appendix \ref{S-sec:separability}, we consider a more general separability assumption and show how Proposition \ref{prop:identifiedset_base_1} can be extended such that we can similarly use two linear programs as in \eqref{eq:bounds_opt} to compute the identified set under these additional assumptions. For the purposes of dimension reduction, we note here that the dimensions of these corresponding programs are strictly smaller than that of $\beta$. For example, in the case of Assumption \ref{ass:AS}, it is given by $|\mathcal{J}| \sum_{j \in \mathcal{J}} |\mathcal{V}_j|$, which we can observe to be substantially smaller than that of $\beta$.

\subsection{Parametric Specifications}\label{sec:parametric}

Our baseline analysis allowed demand to be solely nonparametric. Below, in the spirit of traditional methods discussed in Section \ref{sec:comparison}, we show how our analysis can be extended to allow for flexible functional form restrictions on demand. Our analysis straightforwardly allows for the following general class of functional forms on $q$.

\begin{assumption2}{P}{(Parametric)}\label{ass:A}~
For each $j \in \mathcal{J}$,
\begin{align}\label{eq:q_para}
  q_j(p) = \sum_{k=0}^{K_j} \alpha_{jk} \cdot b_{jk}(p)~
\end{align}
for some unknown parameters $\{\alpha_{jk} : 0 \leq k \leq K_j\}$ and known functions $\{ b_{jk} : 0 \leq k \leq K_j\}$.
\end{assumption2}

\noindent Assumption \ref{ass:A} imposes that demand is a linear function of some basis of prices, where the variable $\alpha \equiv (\alpha_j': j \in \mathcal{J} )'$ with $\alpha_j = (\alpha_{j0} , \ldots, \alpha_{jK_j} )$ parameterizes demand. As discussed below, we focus on linear functions for their computational benefits. The assumption allows for a range of flexibility through the choice of $b_{jk}$ and $K_j$. For example, in our application in Section \ref{sec:empirical}, we consider parsimonious polynomial specifications that are parameterizations of the separable functions in Assumption \ref{ass:AS} given by
\begin{align}\label{eq:q_A_AS}
 q_j(p) = \sum\limits_{m \in \mathcal{J}} \sum_{k=0}^K \alpha_{jmk} \cdot p_m^k
\end{align}
for each $j \in \mathcal{J}$ and some unknown parameters $\left\{ \alpha_{jmk} : m \in \mathcal{J},~ 0 \leq k \leq K \right\}$. Analogous to traditional methods, Assumption \ref{ass:A} allows demand to be point identified in special cases, loosely when the number of unknown parameters in $\alpha_j$ for each $j \in \mathcal{J}$ is taken to be equal to the cardinality of the support of observed price variation. However, it also allows for more general cases, where these functions may not be point-identified.

In Appendix \ref{S-sec:parametric_ext}, we show how to compute the identified set when $q$ satisfies Assumption \ref{ass:A} in addition to the restrictions in $\mathbf{Q}_B$. In contrast to the nonparametric specification, as the admissible space of demand is a finite-dimensional parameterized space, say $\mathbf{Q}_P$, we can directly compute the identified set here by searching over $q \in \mathbf{Q}_P$. The arguments to do so are symmetric to those in Section \ref{sec:equi_charac} to compute $\theta(\mathbf{Q}_B^{\text{fd}})$. Specifically, we first transform $\theta(\mathbf{Q}_P)$ in terms of $\alpha$, i.e. the variable parameterizing $q \in \mathbf{Q}_P$ through \eqref{eq:q_para}. We then show that the linear structure of the basis functions in \eqref{eq:q_para} as well as the function $\theta$ and restrictions determining $\mathbf{Q}_B$ can be exploited to compute bounds as in \eqref{eq:bounds_opt} by solving two linear programs.

\subsection{Liquidity Constraints}\label{sec:liquidity}

For our final extension, recall our baseline analysis relied on the expression in Proposition \ref{prop:EB}, which required the utilities to be continuous. However, when individuals' liquidity constraints restrict the maximum price they can pay for an alternative as in our application in Section \ref{sec:empirical}, this assumption might be suspect. To see this, let $\mathcal{J} = \mathcal{J}_0 \cup \mathcal{J}_1$, where $\mathcal{J}_0$ denotes the set of alternatives with prices normalized to 0 and $\mathcal{J}_1$ denotes those for whom prices vary. Moreover, for each individual, let $E_i$ denote the (unobserved) maximum price they can afford to pay for those in $\mathcal{J}_1$, which implies $C_{1i}(p) = \{j \in \mathcal{J}_1 : p_j \leq E_i\}$ is the underlying subset of affordable alternatives in $\mathcal{J}_1$ when prices equal $p \in \mathcal{P}$. In this case, denoting by $\tilde{U}_{ij}: \mathbf{R} \to \mathbf{R}$ the utility function for alternative $j \in \mathcal{J}$, the counterfactual choice at price $p \in \mathcal{P}$ is given by
\begin{align}\label{eq:utlitymax_const}
D_i(p) = \argmax_{j \in C_i(p)} \tilde{U}_{ij}(Y_i - p_j)~,
\end{align}
where $C_i(p) = \mathcal{J}_{0} \cup C_{1i}(p)$, i.e. the utility maximizing choice that accounts for the affordability of alternatives in $\mathcal{J}_1$. Observe that this falls in our baseline setup in Section \ref{sec:model} by taking $U_{ij}$ that does not separate the role of affordability of the alternatives in $\mathcal{J}_1$ to be related to $\tilde{U}_{ij}$ and $C_i(p)$ such that $U_{ij}(Y_i - p_j) = \tilde{U}_{ij}(Y_i - p_j)$ if $j \in C_i(p)$ and $U_{ij}(Y_i - p_j) = \max_{j_0 \in \mathcal{J}_0}\tilde{U}_{ij_0}(Y_i) - \epsilon$ otherwise for some $\epsilon > 0$, i.e. it equals $\tilde{U}_{ij}$ if the alternative is affordable, and sufficiently small so that it is never preferred to an alternative in $\mathcal{J}_0$ otherwise. In turn, even if $\tilde{U}_{ij}$ is continuous, $U_{ij}$ for $j \in \mathcal{J}_1$ can be discontinuous at $Y_i-E_i$, i.e. when the price just relaxes the liquidity constraint.\footnote{In contrast, if additionally only a non-empty subset $C_{0i} \subseteq \mathcal{J}_0$ was affordable, $U_{ij}$ for $j \in \mathcal{J}_0$ remains continuous when similarly taking $U_{ij}(Y_i) = \tilde{U}_{ij}(Y_i)$ if $j \in C_{0i}$ and $U_{ij}(Y_i) = \max_{j_0 \in C_{0i}}\tilde{U}_{ij_0}(Y_i) - \epsilon$ otherwise as prices don't vary for these alternatives. In turn, note that the assumption of continuous utilities is not suspect for those in $\mathcal{J}_0$ even if only a subset of them are affordable and we hence don't explicitly introduce their affordability for ease of exposition.}

Below, we show how to extend our analysis using an alternative expression for the average value of $B_i^{a,b}$ that solves \eqref{eq:B_t} in this case defined in terms of $\tilde{U}_{ij}$ and $C_i(p)$ by
\begin{align}\label{eq:B_t_const_LC}
\max_{j \in C_i(p^a)}\tilde{U}_{ij}(Y_{i} - p^a_j) = \max_{j \in C_i(p^b)}\tilde{U}_{ij}(Y_{i} - p^b_j - B_i^{a,b})~.
\end{align}
To do so, we impose the following additional assumption.

\begin{assumption2}{LC}{(Liquidity Constraint)}\label{ass:LC}~
For each individual $i$, there exists some known $r \in \mathbf{R}$ such that $\max\limits_{j_0 \in \mathcal{J}_0}\tilde{U}_{ij_0}(Y_i) > \tilde{U}_{ij}(Y_i - p_j)$ for $p_j \geq r$ and $j \in \mathcal{J}_1$.
\end{assumption2}

\noindent Assumption \ref{ass:LC} states that there exists an alternative with prices normalized to 0 that is always preferred to alternatives with varying prices when their prices are large enough to be above some given $r$. Alternatively, it imposes that the willingness to pay for the alternatives in $\mathcal{J}_1$ must be bounded by $r$. Its strength is hence governed by the choice of $r$ and can be made arbitrarily weak by choosing $r$ to be sufficiently large. For our purposes, following a common approach in the literature \citep[e.g.,][Section 2.2]{bhattacharya:18}, this assumption allows one to model that an alternative is not in the choice set in terms of prices. This is done by taking its price to be equal to $r$, so that it is never preferred, through the following relationship
\begin{align}\label{eq:relation_choiceset_price}
 \max_{j \in C_i(p)}\tilde{U}_{ij}(Y_{i} - p_j)  = \max_{j \in \mathcal{J}}\tilde{U}_{ij}(Y_{i} - \tilde{p}_j(p,E_i))~,
\end{align}
where, for each $j \in \mathcal{J}$, $\tilde{p}_j$ is a function such that $\tilde{p}_j(p,e) = p_j$ if $p_j \leq e$ or $j \in \mathcal{J}_0$ and $\tilde{p}_j(p,e) = r$ otherwise, i.e. it takes the price to be equal to $p_j$ if the $j$th alternative is affordable or has normalized prices, and equal to $r$ if not.

In the following proposition, we show that this relation allows us to express $E[B_i^{a,b}]$ in term of a richer definition of demand than that in \eqref{eq:q_def} that accounts for the individual's maximum affordable price. Specifically, taking the support of $E_i$ to be contained in a known discrete set $\mathcal{E}$, let
\begin{align*}
 \tilde{q}_j(p,e) = \text{Prob}\Big\{\argmax_{j \in \mathcal{J}} \tilde{U}_{ij}(Y_i - p_j) = j,~E_i = e \Big\}
\end{align*}
for $p\in \mathcal{P}$ and $e \in \mathcal{E}$ be this richer version of demand that corresponds to the joint probability of choosing $j \in \mathcal{J}$ when all alternatives are affordable and the maximum affordable price equals $e \in \mathcal{E}$.\footnote{In Appendix \ref{S-sec:liquidity}, we discuss how to extend the analysis to allow the support of $E_i$ to be continuous by choosing a specific partition as in our baseline analysis. However, we focus on the discrete case in our application for computational reasons and hence focus on this case in the main text for simplicity.}\textsuperscript{,}\footnote{Note that we take the primitive to be the joint probability, rather than conditional probability of choosing $j$ given $e$ and the distribution of $e$. We do so as this ensures that our parameters and restrictions continue to be linear in the primitives, which our analysis heavily exploits. A limitation of this is that it precludes imposing shape or parametric restrictions directly on these finer primitives. We leave such an extension for future work, which may require insights from alternative tools in the literature on unobserved choice sets that introduce such richer primitives---see, e.g., \cite{barseghyan/etal:21} and references therein.} The proposition essentially follows by applying the arguments from Proposition \ref{prop:EB} for each value of $e \in \mathcal{E}$ and then summing across these values. To formally state it, as in Proposition \ref{prop:EB}, for each $e \in \mathcal{E}$, let $\tilde{\Delta}^{a,b}_{1}(e) \leq \ldots \leq \tilde{\Delta}^{a,b}_{|\mathcal{J}|}(e)$ denote the ordered values of $\{\tilde{p}_j(p^a,e) - \tilde{p}_j(p^b,e): j \in \mathcal{J}\}$, and let $\tilde{\mathcal{J}}_l^{a,b}(e) = \{j \in \mathcal{J} : \tilde{p}_j(p^a,e) - \tilde{p}_j(p^b,e) \geq \tilde{\Delta}^{a,b}_l(e)\}$ for $1 \leq l \leq |\mathcal{J}|$. Moreover, let $\tilde{q}(e) \equiv
 \sum_{j \in \mathcal{J}}\tilde{q}_j(p,e)$ denote the probability that $E_i = e \in \mathcal{E}$, and $\tilde{p}(p,e) = (\tilde{p}_j(p,e) : j \in \mathcal{J})$. Using this notation, we can state the expression in terms of $\tilde{q} \equiv (\tilde{q}_j : j \in \mathcal{J})$ as follows.

\begin{proposition}\label{prop:liquidity}
For each individual $i$, let Assumption \ref{ass:LC} hold and suppose that $\tilde{U}_{ij}$ is continuous and strictly increasing for each $j \in \mathcal{J}$. We then have that $B_i^{a,b}$ exists and is unique, and that
\begin{align}\label{eq:AB_t_q_liquidity}
 E[B_i^{a,b}] = \sum\limits_{e \in \mathcal{E}}\left(\tilde{\Delta}_1^{a,b}(e) \tilde{q}(e) + \sum_{l=1}^{|\mathcal{J}|-1} \sum\limits_{j \in \tilde{\mathcal{J}}_{l+1}^{a,b}(e)} \int\limits_{\tilde{\Delta}_{l}^{a,b}(e)}^{\tilde{\Delta}_{l+1}^{a,b}(e)} \tilde{q}_j\left(\text{min}\{\tilde{p}(p^a,e),\tilde{p}(p^b,e)+t\},e\right) dt \right)~.
\end{align}
\end{proposition}

In this case, given the above expression and that the relation between $q$ in \eqref{eq:q_def} and $\tilde{q}$ equals $q_j(p) = \sum_{e \in \mathcal{E}} \tilde{q}_j(\tilde{p}(p,e),e)$, 
our general parameter of interest in \eqref{eq:theta_q} can be stated in terms of $\tilde{q}$ as\footnote{As in Footnote \ref{foot:1}, when $g^{a,b}=0$, our analysis can be applied to the case of general price changes and not only $\tilde{p}(p^a,e)$ and $\tilde{p}(p^b,e)$, and also where the weights $g^a_j$ and $g^b_j$ depend on $e \in \mathcal{E}$.}
\begin{align}
 \tilde{\theta}(\tilde{q}) = &~g^{a,b} \sum\limits_{e \in \mathcal{E}}\left(\tilde{\Delta}_1^{a,b}(e)\tilde{q}(e) + \sum_{l=1}^{|\mathcal{J}|-1} \sum\limits_{j \in \tilde{\mathcal{J}}_{l+1}^{a,b}(e)} \int\limits_{\tilde{\Delta}_{l}^{a,b}(e)}^{\tilde{\Delta}_{l+1}^{a,b}(e)} \tilde{q}_j\left(\text{min}\{\tilde{p}(p^a,e),\tilde{p}(p^b,e)+t\},e\right) dt \right)   \nonumber \\
 &~+ \sum_{j \in J} g_j^a \sum_{e \in \mathcal{E}} \tilde{q}_j(\tilde{p}(p^a,e),e)  + g_j^b \sum_{e \in \mathcal{E}} \tilde{q}_j(\tilde{p}(p^b,e),e) ~. \label{eq:theta_q_liquidity}
\end{align}
In Appendix \ref{S-sec:liquidity}, we illustrate how to extend the analysis in Section \ref{sec:compute_identifiedset} to compute the identified set for this parameter when the restrictions in \eqref{eq:q_data} and \eqref{eq:q_logical1}-\eqref{eq:q_shape} and the additional ones from Assumption \ref{ass:LC} are written in terms of $\tilde{q}$. The main idea remains the same. Specifically, we can first construct a partition as in \eqref{eq:set_W} but for each value of $e \in \mathcal{E}$, and then, as in Proposition \ref{prop:identifiedset_base_1}, show that taking $\tilde{q}$ to be constant over such a partition leads to no loss of information and allows computing the identified set by solving two linear programs. Moreover, as in Sections \ref{sec:dim_red} and \ref{sec:parametric}, we can also extend this analysis to respectively allow for dimension reduction or parametric assumptions implemented again given each value of $e \in \mathcal{E}$.

\section{Evaluation of the DC Opportunity Scholarship Program}\label{sec:empirical}

\subsection{Background}\label{sec:background}

The DC Opportunity Scholarship Program (OSP) was a federally-funded school voucher program established by Congress in January 2004, and which started accepting students for the 2004-2005 school year. The OSP was structured similarly to other voucher programs that existed at the time \citep{epple/etal:17}. It was open to students residing in Washington, DC, and whose family income was no higher than 185\% of the federal poverty line (\$18,850 for a family of four in 2004).  It could be used only for K-12 education, and at the time of initial receipt was renewable for up to five years. It provided students a voucher worth \$7,500 that could be used to offset tuition, fees, and transportation at any private school of their choice participating in the program.

The law that established the program also mandated its evaluation, which culminated with a final report to Congress \citep{wolf/etal:10}. The report exploited the fact that the OSP randomly allocated vouchers to participating students. In particular, Congress expected the program to be oversubscribed, i.e. the number of applicants would exceed the number of available vouchers. As a result, it required that vouchers be randomly allocated to applicants through a lottery---see \cite{wolf/etal:10} for details on the lottery. \cite{wolf/etal:10} used this random allocation by comparing various outcomes of voucher recipients to non-recipients to experimentally evaluate the effect of voucher receipt on these outcomes. As summarized in their executive summary, they find mixed evidence on the effects of providing a voucher. Specifically, while the receipt of a voucher improved students' chances of graduating high school and raised parent's rating of school safety and satisfaction, they find no conclusive evidence of any significant effects on various outcomes corresponding to student achievement.

In what follows, we use the tools developed in the previous sections to complement these findings by analyzing the welfare effects of the price subsidy induced by the status quo voucher amount as well as counterfactual amounts. Our analysis is motivated by the fact that while the receipt of the voucher revealed mixed evidence on outcomes, parents may nonetheless value the voucher, potentially across dimensions not easily captured by the outcomes. Indeed, as highlighted below, the data reveals that a non-trivial proportion of voucher recipients used the voucher, which implies that they value receiving the voucher. Our analysis estimates these potential welfare benefits using data collected by the OSP.

\subsection{Data and Empirical Setting}\label{sec:setting}

We use the same data as that in \cite{wolf/etal:10}. It contains detailed information on the school setting for the first two years of the program, 2004 and 2005, and on a sample of 2,308 students who applied to the program in these years. Importantly, for our purposes, it includes the prices of private schools in the program and the enrollment choices of students. Across the two years, the composition of applicants and private schools in the program changed. To keep prices and the set of eligible schools the same for all students, we focus on the second year of the program, 2005, which contains around $80\%$ of the entire sample. In particular, it yields an analysis sample of 1,816 students. In addition, while these students were tracked for at least four years, we focus on their enrollment choices from the initial year to avoid complications from dynamics.

We start our analysis by describing how we translate the OSP setting in terms of our setup in Section \ref{sec:baseline}. For each student in the data, we take $D_i$ to denote their observed enrollment choice, and we let it take values in $\mathcal{J} = \mathcal{J}_v \cup \{g,n\}$, where $\mathcal{J}_v$ denotes the set of all private schools in the program, and $g$ and $n$ denote the alternatives of enrolling in any government school (which includes charter schools) and any private school not in the program, respectively.\footnote{We separately consider the alternatives of enrolling in a government school and a private school not in the program, rather than combining them into a single alternative, as we take the costs associated with them in \eqref{eq:AC} to be different.} In 2005, there were approximately 70 private schools in the program (out of a total of about 110 in Washington, DC), and thus we have that $|\mathcal{J}_v| \approx 70$ (rounded to the nearest ten for privacy purposes).

To define the support of the price vector $P_i$, note that the voucher affected only the prices (tuition) of private schools in the program, and hence there is no variation in the prices of government and private schools not in the program. The prices of the alternatives $g$ and $n$ are therefore normalized to zero. For the private schools in the program, the variation in prices is determined by the receipt of the status quo voucher. For each $j \in \mathcal{J}_v$, let $p^*_j \in \mathbf{R}_+$ denote the original price of the school recorded in the data, and we take $p_j(\tau) = \max\{p^*_j - \tau,0\}$ to denote its price under the application of a voucher of amount of $\tau \in \mathbf{R}_+$, as the voucher provided an amount of at most $\tau$ to cover tuition. Moreover, let $p(\tau) = (p_j(\tau) : j \in \mathcal{J})$ denote the vector of prices under a voucher amount of $\tau \in \mathbf{R}_+$, where note that $p_g(\tau) = p_n(\tau) = 0$ as their prices are normalized to 0. Denoting by $\tau_{\text{sq}}$ the status quo amount, the support of the observed prices is then given by $\mathcal{P}_{\text{obs}} = \{p(0),p(\tau_{\text{sq}})\}$, i.e. the prices with and without the status quo voucher. As the voucher was randomly assigned, we have that Assumption \ref{ass:E} is satisfied.

The objective of our empirical analysis is to learn the welfare effects of the price decrease induced by the voucher, i.e. \eqref{eq:E_B} when $p^a = p(0)$ and $p^b = p(\tau)$ given by
\begin{align}\label{eq:AB}
  AB(\tau) \equiv E[B_i^{a,b}]~.
\end{align}
Indeed, when $\tau = \tau_{\text{sq}}$, this corresponds to the effect of providing the status quo voucher amount, while when $\tau \neq \tau_{\text{sq}}$, it corresponds to that of a counterfactual voucher amount. To benchmark these benefits and perform a cost-benefit analysis, we also study additional parameters that measure the costs the government may face when individuals receive the voucher relative those when they do not receive it, which can be straightforwardly written as \eqref{eq:theta_q}. In particular, denoting by $\{c_j(\tau) : j \in \mathcal{J}\}$ the costs that the government associates with enrollment in the different alternatives under a voucher of amount $\tau$, we take the average cost of the voucher to be
\begin{align}\label{eq:AC}
  AC(\tau) = \sum_{j \in \mathcal{J}} c_j(\tau) q_j(p(\tau)) - c_j(0) q_j(p(0))~,
\end{align}
and the average surplus measuring benefit net of cost by
\begin{align}\label{eq:AS}
  AS(\tau) = AB(\tau) - AC(\tau)~.
\end{align}
We take $c_g(\tau) = c_g$ and $c_n(\tau) = 0$, i.e. the cost associated with government schools to be some known value $c_g$ and that with private schools not in the program to be zero; and $c_j(\tau) = \min\{p_{j}(0),\tau\} + \mu 1\{\tau > 0\}$, i.e. the cost associated with private schools in the program to be the voucher amount spent to cover the tuition reduction it induces plus some known administrative cost $\mu$ of operating the program (i.e. charged only when the voucher amount is positive). Importantly, making our cost-benefit analysis empirically interesting, observe that the average cost here captures the net cost of providing the voucher as it includes the costs of funding the voucher at eligible private schools as well as the potential cost savings from students moving out of government schools when offered a voucher. Indeed, absent the latter costs, the average surplus will otherwise be non-positive by construction as a student's willingness to pay for a voucher is necessarily less than the value of the price decrease it induces at the eligible private school they may enroll at and hence the cost of funding the voucher at that school.

For the known values in the costs, we take $c_g = \$$5,355, which corresponds to the educational expenditure reported by the \cite{us2005public}. This is lower than total per-pupil expenditure from the Census (\$12,979, which includes some fixed costs), or educational expenditure as measured in other sources (\$8,105, \cite{sable2006overview}). However, as our surplus parameter is increasing in $c_g$, we choose the smaller, more conservative value. On the other hand, we take $\mu = \$200$, which corresponds to cost of administration, adjudication and providing information to families for an alternative school voucher program reported in \cite{levin1997cost}---see Figure \ref{S-fig:AS_cost} for robustness to a range of other values of $c_g$ and $\mu$.\footnote{All cost values reported in this paragraph have been adjusted to 2004 dollars.}

In the OSP setting, as students were primarily from low-income families, it is likely that they may be liquidity constrained and unable to afford all the private schools, a feature commonly arising in education decisions \citep{lochner:12}. As highlighted in Section \ref{sec:liquidity}, this casts suspicion on the assumptions in our baseline setup and hence could bias the resulting estimates. To this end, in our analysis we also present results under the extension in Section \ref{sec:liquidity} that allows students to be liquidity constrained. When doing so, we take $r$ in Assumption \ref{ass:LC} to be a reasonably large value equal to $\$100,000$, which is an order of magnitude larger than the average price of private schools in the program and around five times the average income of families in the sample. This implies students prefer alternatives $g$ or $n$ to a private school in the program when its price is greater than $\$100,000$. Moreover, we take the discrete support of $E_i$ to equal $\mathcal{E} = \{0,~10,0000,~20,000,~30,000\}$. Given the prices of private schools in the program presented in Section \ref{sec:data_summ} below, this allows for types of students who either cannot afford any private school in the program or all of them absent the voucher, as well as intermediate types that can afford a small or large proportion of them.

\subsection{Descriptive Statistics}\label{sec:data_summ}

To better understand our results, we first present some descriptive statistics for the two main variables in the data our analysis exploits, namely the enrollment choices and original prices of the private schools that determines the price variation---see Appendices \ref{S-sec:data_construction}-\ref{S-sec:summstat_setting} for more details on the data and various statistics on the schools and the sample of individuals.

\begin{table}[!t]
\begin{center}
\caption{Enrollment shares across school type by voucher receipt}
\label{tab:shares_summ}
\scalebox{0.85}{
\def\arraystretch{0.9}
\makebox[\textwidth]{\begin{threeparttable}
 \begin{tabular}{ L{7.75cm}  C{3.25cm} C{3.25cm} C{3.25cm}  }  
 \toprule  
  & With voucher & Without Voucher & Difference \\ \midrule  
  Government schools             &  0.288 &  0.901 &  -0.613 \\ 
                                  & [0.453]   & [0.299]   & (0.018)  \\ 
   Private schools not in program &  0.014 &  0.020 &  -0.006 \\ 
                                  & [0.117]   & [0.140]   & (0.006)  \\ 
   Private schools in program     &  0.698 &  0.079 &  0.619 \\  
                                  & [0.459]   & [0.270]   & (0.018)  \\ \midrule 
   Observations & 1,090 & 730 &  \\  
  \bottomrule  
 \end{tabular}
\begin{tablenotes}[flushleft]
\setlength\labelsep{0pt}
\item \footnotesize Observations rounded to the nearest 10. Standard deviations in square brackets and robust standard errors in parentheses.
\item \footnotesize SOURCE:  \textit{Evaluation of the DC Opportunity Scholarship Program: Final Report (NCEE 2010-4018)}, U.S. Department of Education, National Center for Education Statistics previously unpublished tabulations.
\end{tablenotes}
\end{threeparttable}}}
\end{center}
\end{table}

Table \ref{tab:shares_summ} presents enrollment shares across the three types of schools, i.e. government schools and private schools in and not in the program, by voucher receipt. A relatively large proportion (69.8$\%$) choose to take up the voucher as revealed by those enrolled in private schools in the program. By revealed preference, this implies that recipients value the voucher. In addition, the voucher increases the proportion enrolling in voucher private schools by 61.9 percentage points, suggesting that prices play an important role in inducing private school enrollment. The voucher also produces a nearly symmetric decline in the proportion enrolled in government schools (-61.3 percentage points) implying that nearly all students induced into voucher schools would be in government ones absent the voucher.

\begin{figure}[!t]
\centering
\caption{Tuition prices across voucher private schools, and enrollment shares across them as well as government and non-voucher private schools.}
\makebox[\textwidth]{ \begin{tabular}{c c}
\subcaptionbox{Tuition of private schools in the program}{ ~~\includegraphics[width=2.75in]{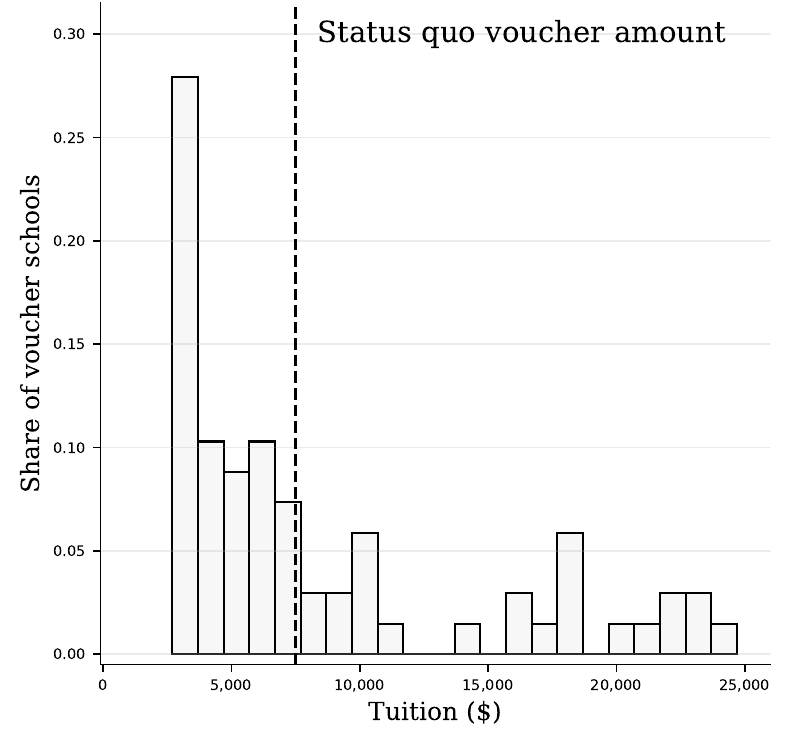}~~} &
\subcaptionbox{Enrollment shares}{~~\includegraphics[width=2.75in]{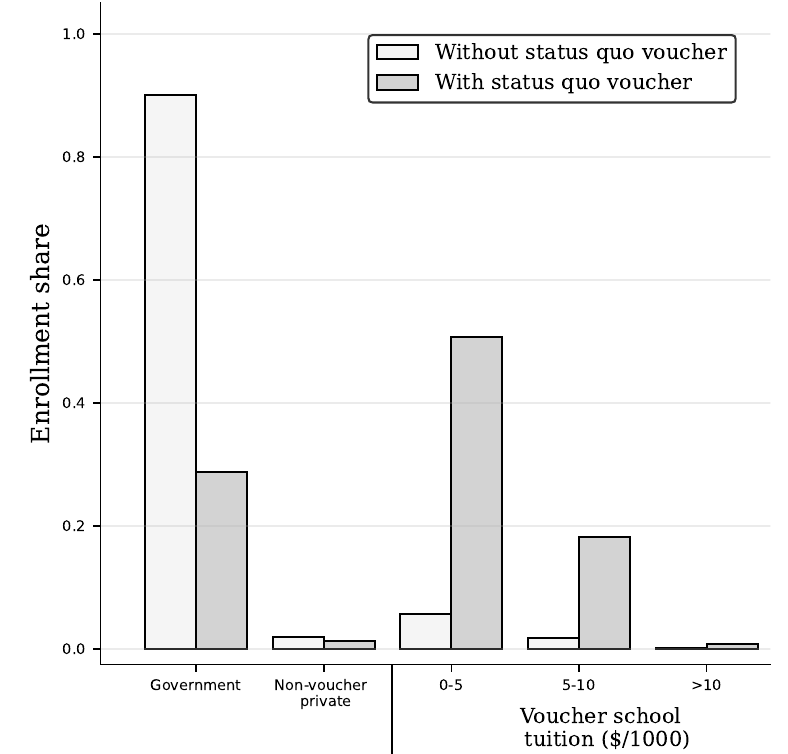}~~}
\end{tabular}}
\begin{tabular*}{0.99\textwidth}{c}
\multicolumn{1}{p{.99\hsize}}{\footnotesize SOURCE:  \textit{Evaluation of the DC Opportunity Scholarship Program: Final Report (NCEE 2010-4018)}, U.S. Department of Education, National Center for Education Statistics previously unpublished tabulations.}
\end{tabular*}
\label{fig:summ_stat}
\end{figure}

Figure \ref{fig:summ_stat} summarizes the variation in prices across the private schools in the program as well as the enrollment shares across various ranges of these prices. Figure \ref{fig:summ_stat}(a) reveals that a large number of voucher schools had low prices---around 80$\%$ had prices below the status quo voucher amount. Figure \ref{fig:summ_stat}(b) reveals that the voucher induced a significant proportion to enroll in these low-price schools---out of the 61.9 percentage point increase in the proportion attending a voucher private school, a full 59 percentage points (95\%) was into schools with prices less than the status quo voucher amount. Similarly, a large proportion of recipients (81$\%$) redeem the voucher at schools with prices below the cost of a government school. Given that the majority of these recipients would have enrolled in government schools absent the voucher as observed from Table \ref{tab:shares_summ}, this suggests that the government may face only small net costs or even savings from the provision of a voucher. Our estimates below make this point more precisely.

\subsection{Welfare Estimates}\label{sec:welfare_est}

We report below estimated lower and upper bounds for the parameters in \eqref{eq:AB}-\eqref{eq:AS} for both the status quo amount of $\tau=\tau_{\text{sq}}\equiv\$7,500$ as well as counterfactual amounts under various specifications of demand. Under the baseline specification, i.e. that in Section \ref{sec:assumptions} and that described in Appendix \ref{S-sec:liquidity} in the case of liquidity constraints, the dimensions of the linear programs to do so are intractable---see Table \ref{S-tab:dim}, which reports the dimensions for the various specification when $\tau=\tau_{\text{sq}}$. As discussed in Section \ref{sec:dim_red}, this is because these dimensions are extremely large when the number of alternatives is large, which is the case here as $|\mathcal{J}_v|$ is close to $70$. For our baseline specification, we therefore compute outer bounds using sub-programs as in Section \ref{sec:dim_red}, which have a substantially lower dimension as $|\mathcal{V}|$ and $|\mathcal{V}(e)|$ for $e \in \mathcal{E}$ in the case of liquidity constraints is small here. We also compute bounds under the separability restriction in Assumption \ref{ass:AS} as well as their parameterized versions in \eqref{eq:q_A_AS} for values of $K$ between 1 and 3, and their analogs under liquidity constraints in \eqref{S-eq:q_AS_liquidity} and \eqref{S-eq:q_para_liquidity}, respectively. Note that while the outer bounds in the baseline case may not always be sharp, comparing them to the sharp ones under the nonparametric separable case can reveal when they are equal and hence sharp.

When estimating the bounds by computing the corresponding linear programs under these various specifications, we note that we replace the enrollment shares $P(D_i = j | P_i = p)$ in the restriction in \eqref{eq:q_data} by their sample analogs that are estimated using the empirical distribution of the analysis sample.\footnote{We find that all the specifications exactly match the data and so the constraint sets for the programs are non-empty. If this was not the case, one could straightforwardly apply the procedure from \cite{mogstad/etal:18} that allows the specification to not exactly match the data due to sampling uncertainty.} To account for the subsequent sampling uncertainty, we then also report 90$\%$ confidence intervals, which we construct using a bootstrap procedure (using 1,000 draws) from \cite{bugni/etal:17} described in Appendix \ref{S-sec:inference}.

\begin{table}[!t]
\begin{center}
\caption{Estimated bounds and 90$\%$ confidence intervals on welfare effects for status quo amount}
\label{tab:welfare_status}
\scalebox{0.75}{
\def\arraystretch{0.9}
\makebox[\textwidth]{\begin{threeparttable}
 \begin{tabular}{ L{1.75cm} L{1.75cm}  R{1.25cm} R{1.25cm} C{0.1cm}  R{1.25cm} R{1.25cm} R{1.25cm} C{0.1cm}  R{1.25cm} R{1.25cm} C{0.1cm}  R{1.25cm} R{1.25cm} R{1.25cm} }  
 \toprule  
    & & \multicolumn{6}{c}{ Without Liquidity Constraints} & & \multicolumn{6}{c}{ With Liquidity Constraints}       \\  \cline{3-8} \cline{10-15}                
    & & \multicolumn{2}{c}{NP} &  & \multicolumn{3}{c}{P Separable, $K$} & & \multicolumn{2}{c}{NP} &  & \multicolumn{3}{c}{P Separable, $K$}       \\  \cline{3-4} \cline{6-8} \cline{10-11} \cline{13-15}                
    & & \multicolumn{1}{c}{Baseline} & \multicolumn{1}{c}{Separable} &  & 1 & 2 & 3 & & \multicolumn{1}{c}{Baseline} & \multicolumn{1}{c}{Separable} &  & 1 & 2 & 3  \\   
    & & (1) & (2) &  & (3) & (4) & (5) & & (6) & (7) &  & (8) & (9) & (10) \\  \midrule 
                                   & Lower CI   & \scriptsize 203 & \scriptsize 203 & & \scriptsize 1,598 & \scriptsize 1,039 & \scriptsize 763 & & \scriptsize 203 & \scriptsize 203 & & \scriptsize 1,598 & \scriptsize 1,039 & \scriptsize 763  \\  
 \multirow{2}{*}{$AB(\tau_{\text{sq}})$}  & LB & 362 & 362 & & 1,751 & 1,174 & 897 & & 362 & 362 & & 1,751 & 1,174 & 897 \\  
                                     & UB & 5,239 & 3,344 & & 1,853 & 2,449 & 2,747 & & 69,465 & 62,881 & & 31,560 & 44,065 & 50,333\\   
                                         & Upper CI & \scriptsize 5,511 & \scriptsize 3,559 & & \scriptsize 2,006 & \scriptsize 2,621 & \scriptsize 2,931 & & \scriptsize 73,143 & \scriptsize 67,553 & & \scriptsize 33,969 & \scriptsize 47,368 & \scriptsize 54,105  \\   
  \cline{2-15} 
                                    & Lower CI   & \scriptsize -121 & \scriptsize -121 & & \scriptsize -121 & \scriptsize -121 & \scriptsize -121 & & \scriptsize -121 & \scriptsize -121 & & \scriptsize -121 & \scriptsize -121 & \scriptsize -121  \\  
 \multirow{2}{*}{$AC(\tau_{\text{sq}})$}  & LB & \multirow{2}{*}{113} & \multirow{2}{*}{113} & & \multirow{2}{*}{113} & \multirow{2}{*}{113} & \multirow{2}{*}{113} & & \multirow{2}{*}{113} & \multirow{2}{*}{113} & & \multirow{2}{*}{113} & \multirow{2}{*}{113} & \multirow{2}{*}{113}  \\  
                                     & UB &                                                                         &                                                                         & &                                                                         &                                                                         &                                                                         &                                                                         &                                                                         \\   
                                         & Upper CI & \scriptsize 303 & \scriptsize 303 & & \scriptsize 303 & \scriptsize 303 & \scriptsize 303 & & \scriptsize 303 & \scriptsize 303 & & \scriptsize 303 & \scriptsize 303 & \scriptsize 303  \\   
  \cline{2-15} 
                                    & Lower CI   & \scriptsize 77 & \scriptsize 77 & & \scriptsize 1,453 & \scriptsize 883 & \scriptsize 619 & & \scriptsize 77 & \scriptsize 77 & & \scriptsize 1,453 & \scriptsize 883 & \scriptsize 619  \\  
 \multirow{2}{*}{$AS(\tau_{\text{sq}})$}  & LB & 249 & 249 & & 1,638 & 1,061 & 784 & & 249 & 249 & & 1,638 & 1,061 & 784 \\  
                                     & UB & 5,126 & 3,231 & & 1,740 & 2,336 & 2,634 & & 69,352 & 62,768 & & 31,447 & 43,952 & 50,220\\   
                                         & Upper CI & \scriptsize 5,441 & \scriptsize 3,465 & & \scriptsize 1,931 & \scriptsize 2,545 & \scriptsize 2,849 & & \scriptsize 72,980 & \scriptsize 67,665 & & \scriptsize 34,044 & \scriptsize 47,493 & \scriptsize 54,198  \\   
 \bottomrule  
 \end{tabular}
\begin{tablenotes}[flushleft]
\setlength\labelsep{0pt}
\item \footnotesize $AB(\tau_{\text{sq}})$, $AC(\tau_{\text{sq}})$, and $AS(\tau_{\text{sq}})$ denote average benefit, cost and surplus under status quo voucher amount of $\tau_{\text{sq}}$ as defined in \eqref{eq:AB}-\eqref{eq:AS}, respectively. NP denotes Nonparametric and P denotes Parametric. LB and UB denote lower and upper bound estimates, respectively, and Lower and Upper CI denote lower and upper values of the confidence interval, respectively. Lower and upper bound estimates are not repeated if they coincide.
\item \footnotesize SOURCE: \textit{Evaluation of the DC Opportunity Scholarship Program: Final Report (NCEE 2010-4018)}, U.S. Department of Education, National Center for Education Statistics previously unpublished tabulations.
\end{tablenotes}
\end{threeparttable}}}
\end{center}
\end{table}

Table \ref{tab:welfare_status} first presents our results for the status quo voucher amount. In the absence of liquidity constraints, the average benefit estimates reveal that a large range of benefits are credibly consistent with the data with the upper bounds under the nonparametric baseline and most flexible parametric specifications, given by $\$$5,239 and $\$$2,747, respectively, large relative to the entire voucher value of $\$$7,500. In contrast, the average cost, which is point identified as it is a function of demand at the two values of prices observed in the data, equals $\$113$ and is low compared to the voucher amount of $\$$7,500. As noted in Section \ref{sec:data_summ}, this is because a large proportion of recipients redeem the voucher at low-cost private schools relative to government schools they would have enrolled absent the voucher. Taking the benefits and costs together, the main finding is that the average surplus is positive, which we can observe is statistically significant and robustly holds under the baseline specification, and of a potentially large magnitude.

The second half of Table \ref{tab:welfare_status} accounts for liquidity constraints. We find that doing so the upper bounds on the average benefit grow dramatically. Even under the most restrictive parametric specification with $K=1$, the upper bound ($\$31,560$) is close to four times the value of the voucher. This means that not accounting for liquidity constraints could substantially downward bias the possible magnitude of benefits that are consistent with the data. There is no change in the lower bounds, however, meaning that our finding of positive net benefits is robust to accounting for liquidity constraints. We also note that under the baseline specification, the upper bound is only slightly smaller than the proportion of students choosing private schools in the program (69.8$\%$ as reported in Table \ref{tab:shares_summ}) times $\$$100,000. This captures that those who take up the voucher might all be choosing schools that were not previously affordable to them, and hence their willingness to pay is bounded above by the difference between their maximum willingness to pay and the price of their chosen school without the voucher. Given our choice of $r$ in Assumption \ref{ass:LC} and that most students choose relatively cheap schools absent the voucher, this is close to $\$$100,000. In this sense, the benefits can also be made arbitrarily large by taking a larger value of $r$. This is intuitive as unless we sufficiently restrict the tails of the valuation for the alternative, we can have no logical upper bound on the willingness to pay for it. This is consistent with related work in other settings and viewed as a transparent benefit of a nonparametric approach \citep[e.g.,][]{tebaldi/etal:19}.

\begin{figure}[!t]
\centering
\caption{Estimated bounds on welfare effects for a range of voucher amounts}
\makebox[\textwidth]{
\begin{tabular}{c c c}
\subcaptionbox{$AB(\tau)$}{ \includegraphics[width=2.1in]{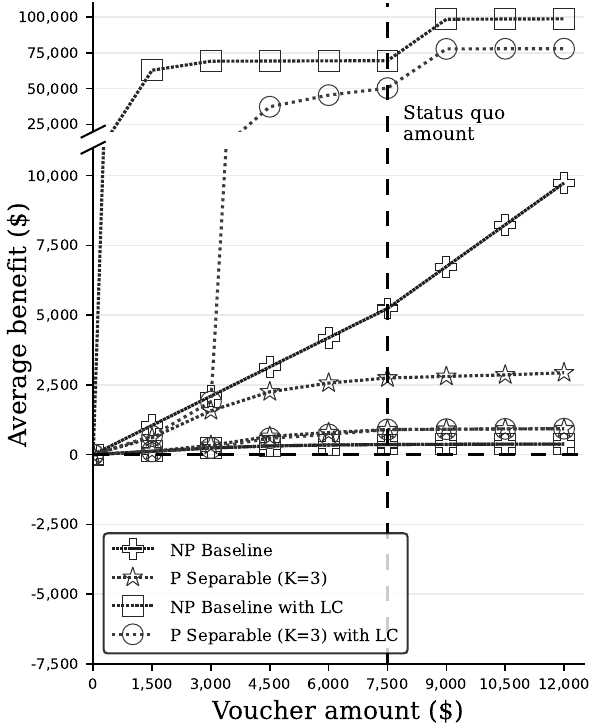}} & \subcaptionbox{$AC(\tau)$}{ \includegraphics[width=2.1in]{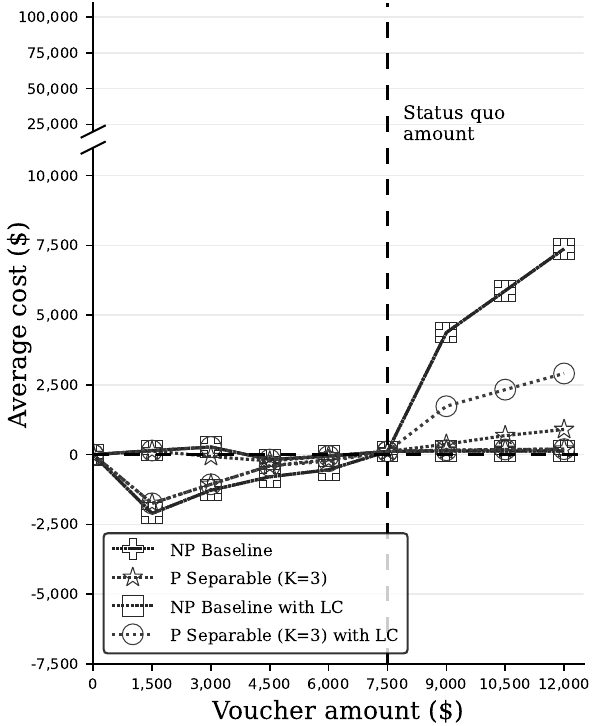}} &
\subcaptionbox{$AS(\tau)$}{\includegraphics[width=2.1in]{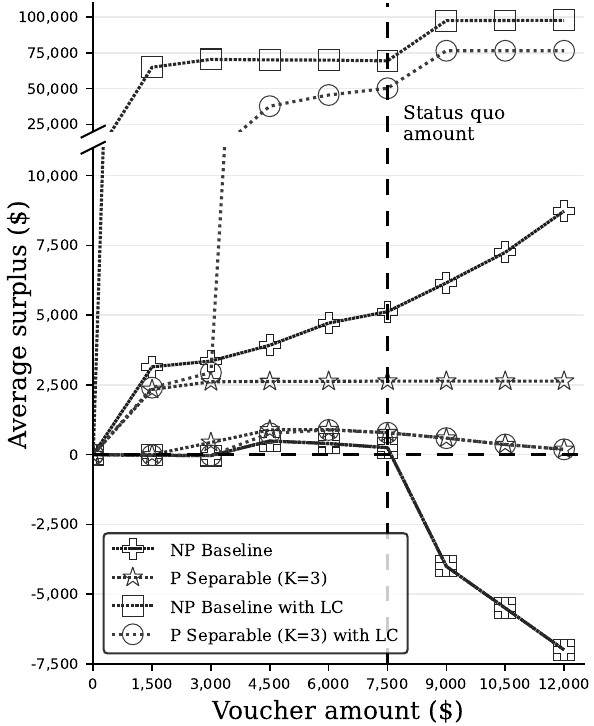}}
\end{tabular}}
\begin{tabular*}{0.99\textwidth}{c}
\multicolumn{1}{p{.99\hsize}}{\footnotesize $AB(\tau)$, $AC(\tau)$, and $AS(\tau)$ denote average benefit, cost and surplus under voucher amount $\tau$ as defined in \eqref{eq:AB}-\eqref{eq:AS}, respectively. NP denotes Nonparametric and P denotes Parametric, and LC denotes that the model allows students to be liquidity constrained.}\\
\multicolumn{1}{p{.99\hsize}}{\footnotesize SOURCE:  \textit{Evaluation of the DC Opportunity Scholarship Program: Final Report (NCEE 2010-4018)}, U.S. Department of Education, National Center for Education Statistics previously unpublished tabulations. }\\
\end{tabular*}
\label{fig:welfare_counter}
\end{figure}

Figure \ref{fig:welfare_counter} presents the results for a range of counterfactual amounts beyond the status quo, for the baseline nonparametric specification as well as the parametric one taking $K=3$. As in the case of the status quo, the lower bounds with and without liquidity constraints continue to remain the same, while the upper bounds in the presence of liquidity constraints are substantially larger. Moreover, for values below the status quo amount, we continue to robustly find positive, potentially large net benefits---Figure \ref{S-fig:AS_tau_ci} reveals that this is also generally statistically significant. In contrast, for those above the status quo amount, the conclusion of positive effects is dependent on the strength of the assumptions. Specifically, Figure \ref{fig:welfare_counter}(c) reveals that we will necessarily have positive net benefits for such amounts only if we impose parametric restrictions.

This can be explained by the underlying average benefits and costs in Figures \ref{fig:welfare_counter}(a)-(b), where the bounds for the nonparametric specification appear to be significantly tighter for voucher values below the status quo rather than those above it. Intuitively, this is because, unlike the parametric specification, the nonparametric one allows for unrestricted substitution patterns between schools and also, unlike values of the voucher below the status quo, there is no additional data at higher voucher amounts to provide information on the substitution patterns. In turn, the bounds  are wide, highlighting that a range of patterns are nonparametrically consistent with the data.

\subsection{The Role of Low-Tuition Schools}\label{sec:lowtuition}

We highlighted above that the positive effects arose in part due to the presence of low-tuition schools in the program that many recipients attend, but that have a small net cost to the government. We conclude this section by further exploring the importance of these schools in the program under the status quo voucher amount. Specifically, we analyze how our estimates change when we remove schools having prices less than a certain amount from the program. For a given $\kappa \in \mathbf{R}_+$, let $\mathcal{J}^{\kappa} = \{j \in \mathcal{J}_v : p_j(0) \leq \kappa\}$ denote the set of private schools in the program with prices no more than $\kappa$, and let $p^{\kappa}_j(\tau)$ be equal to $p_j(0)$ if $j \in \mathcal{J}^{\kappa}$ and $p_j(\tau)$ otherwise, i.e. the voucher amount is applied to only schools with prices above $\kappa$. We then study the parameter in \eqref{eq:E_B} when $p^a = p(0)$ and $p^b = p^{\kappa}(\tau)$,
\begin{align}\label{eq:AB_kappa}
AB^{\kappa}\left(\tau_{\text{sq}}\right) \equiv E[B_i^{a,b}]~,
\end{align}
as well as analogous versions of those in \eqref{eq:AC} and \eqref{eq:AS} given by
\begin{align}
AC^{\kappa}\left(\tau_{\text{sq}}\right) &= \sum_{j \in \mathcal{J}} c^{\kappa}_j(\tau_{\text{sq}}) \cdot q_j(p^{\kappa}(\tau_{\text{sq}})) - \sum_{j \in \mathcal{J}} c_j(0) \cdot q_j(p(0))~, \label{eq:AC_kappa} \\
AS^{\kappa}\left(\tau_{\text{sq}}\right) &= AB^{\kappa}\left(\tau_{\text{sq}}\right) - AC^{\kappa}\left(\tau_{\text{sq}}\right)~, \label{eq:AS_kappa}
\end{align}
where $c_j^{\kappa}(\tau_{\text{sq}}) = c_j(\tau_{\text{sq}})$ for $j \in \mathcal{J} \setminus \mathcal{J}^{^{\kappa}}$ and $c_j^{\kappa}(\tau_{\text{sq}}) = 0$ for $j \in \mathcal{J}^{\kappa}$, i.e. the same costs as in \eqref{eq:AC} except that the schools that are removed from the program are taken to have zero costs.

\begin{figure}[!t]
\centering
\caption{Estimated bounds on welfare effects when schools with tuition at most $\kappa$ are removed from the program}
\makebox[\textwidth]{\begin{tabular}{c c c}
\subcaptionbox{$AB^{\kappa}\left(\tau_{\text{sq}}\right)$}{ \includegraphics[width=2.1in]{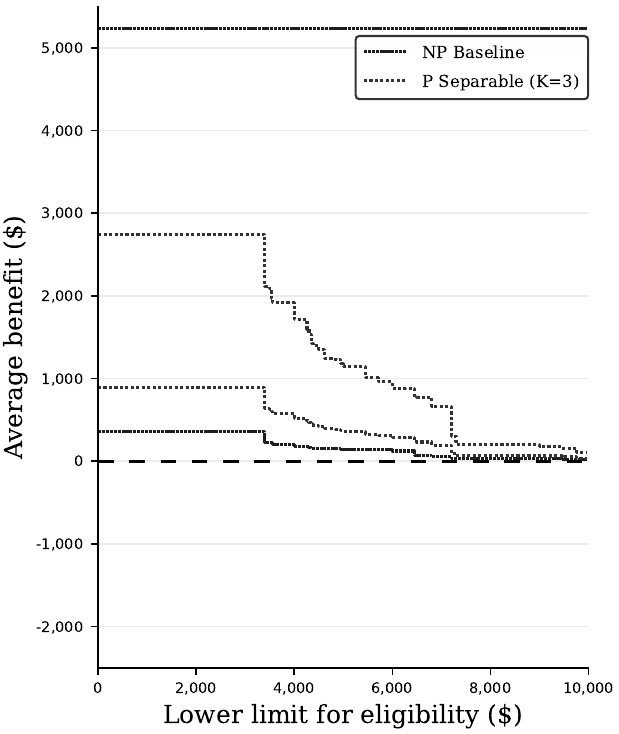}} & \subcaptionbox{$AC^{\kappa}\left(\tau_{\text{sq}}\right)$}{ \includegraphics[width=2.1in]{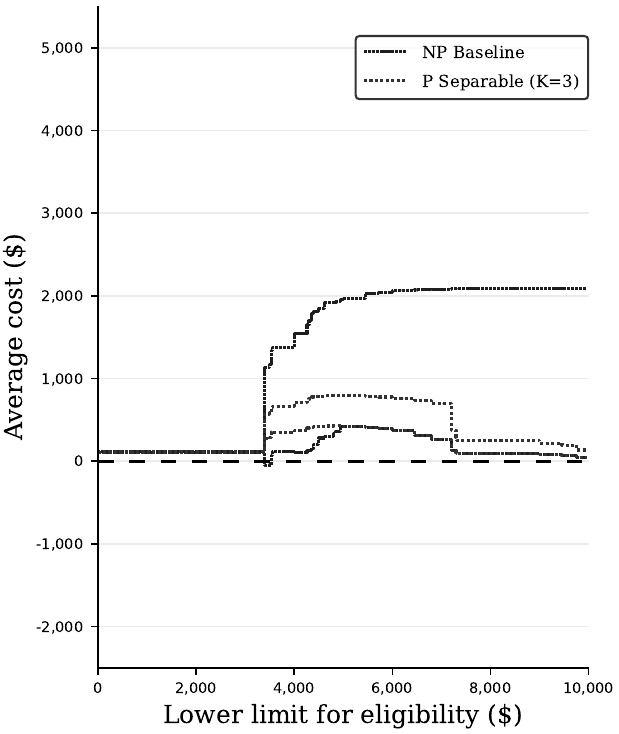}} &
\subcaptionbox{$AS^{\kappa}\left(\tau_{\text{sq}}\right)$}{\includegraphics[width=2.1in]{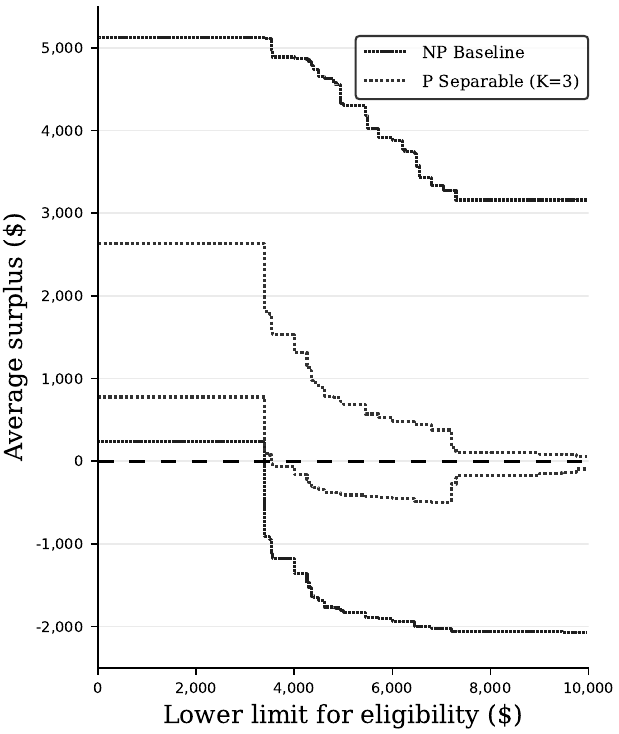}}
\end{tabular}}
\begin{tabular*}{0.99\textwidth}{c}
\multicolumn{1}{p{.99\hsize}}{\footnotesize $AB^\kappa(\tau_{\text{sq}})$, $AC^\kappa(\tau_{\text{sq}})$, and $AS^\kappa(\tau_{\text{sq}})$ denote average benefit, cost and surplus defined in \eqref{eq:AB_kappa}-\eqref{eq:AS_kappa}, respectively. NP denotes Nonparametric and P denotes Parametric. All models here take students to be not liquidity constrained.}\\
\multicolumn{1}{p{.99\hsize}}{\footnotesize SOURCE:  \textit{Evaluation of the DC Opportunity Scholarship Program: Final Report (NCEE 2010-4018)}, U.S. Department of Education, National Center for Education Statistics previously unpublished tabulations.}
\end{tabular*}
\label{fig:remove_low}
\end{figure}

Figure \ref{fig:remove_low} presents estimated bounds for a range of values of $\kappa$. Intuitively, as the baseline specification imposes no restrictions on substitution patterns and as the data exhibits no variation where the voucher is applied to only certain voucher schools, the bounds under the baseline specification can be wide. The bounds under the parametric specification in contrast can be substantially smaller. Across all specifications, Figure \ref{fig:remove_low}(c) suggests that the removal of low-tuition schools from the program generally results in the reduction of average surplus. Importantly, it reveals that removing schools with tuition of $\$$3,500 and lower could potentially cause it to have a negative surplus. A closer look at Figure \ref{fig:summ_stat}(a) reveals that nearly 30\% of schools in the program have tuition of at most this value. The estimates reveal that these low-tuition schools play a key role in explaining the positive net benefits that we find.

To provide some suggestive evidence on what features of these schools attract voucher recipients, Table \ref{S-tab:school-chars-tuition} compares average characteristics between schools charging above and below \$3,500 in tuition. Consistent with spending less money on instruction and having alternative sources of funding, the low-tuition schools have larger student-teacher ratios, are less likely to have individual tutors or programs for students with learning difficulties, and are more likely to be religious. Quantifying which of these features voucher recipients exactly value, however, requires a richer model that explicitly models their role in determining choice, which we leave for future work.

\section{Comparison to Traditional Parametric Methods}\label{sec:comparison}

In this section, we compare our empirical results and conclusions from the previous section to those we would obtain when applying traditional methods.

\subsection{Logit Specifications}\label{sec:logit_specs}

Recall from Section \ref{sec:identifiedset} that our identification problem requires imposing restrictions on how the demand functions vary with price. Traditional methods do so by imposing a parametric functional form on demand such that it is point identified by the variation in the data, which in turn point-identifies our parameters of interest. These parametrizations are commonly implied by imposing functional forms on the utilities and parametric distributions on the unobserved heterogeneity.

In our comparison, we consider versions of a standard logit parameterization of our empirical model in Section \ref{sec:setting} that begins by assuming
\begin{align}\label{eq:utility_para}
U_{ij}(Y_i - p_j) = \xi_{j} - \gamma_{i} p_{j} + \epsilon_{ij}
\end{align}
for $j \in \mathcal{J}$, i.e utility is linear in prices, with alternative-specific intercepts, individual-specific price coefficients, and individual and alternative-specific shocks.\footnote{Note that we do not include distance to school as commonly done in such parameterizations as our data does not contain information on such variables. Alternatively, we do not include school characteristics, such as those in Table \ref{S-tab:school-chars}, as they are subsumed in the alternative-specific intercepts, unless they are interacted with individual-specific coefficients. Given our focus is not on preferences for these characteristics \citep[e.g.,][]{carneiro/etal:19}, we do not introduce such interactions for simplicity.} Denoting by $X_i$ a vector of observed individual covariates, the two versions we consider impose different distributions on the unobserved heterogeneity $\gamma_i$ and $\epsilon_{ij}$ as follows:
\begin{enumerate}

\item[] \textbf{Mixed Logit (ML)}: $\gamma_{i} = \bar{\gamma}_{0} + \bar{\gamma}_{1}’ X_i + \nu_i$, where $\nu_i$ is normally distributed with mean 0 and variance $\sigma^2$; and $\epsilon_{ij}$ is distributed independently across $j$ as Type I extreme value.
 \item[] \textbf{Nested Logit (NL)}: $\gamma_{i} = \bar{\gamma}_{0} + \bar{\gamma}_{1}’ X_i$; and $\epsilon_{i} = (\epsilon_{ij} : j \in \mathcal{J})$ has a CDF evaluated at $\epsilon$ equal to $\text{exp}( -\sum\limits_{k \in \{1,2\}} ( \sum\limits_{j \in \mathcal{N}_k} e^{-\epsilon_j / \lambda_k} )^{\lambda_k} )$ for some $\lambda_1,\lambda_2 \in \mathbf{R}$, where $\mathcal{N}_1 = \mathcal{J}_v \cup \{n\}$, $\mathcal{N}_2 = \{g\}$.

\end{enumerate}
The ML specification introduces both observed and unobserved heterogeneity in the price coefficient, while NL introduces only observed heterogeneity, but allows dependence in the shocks across alternatives in the same nest, where there are two nests with one consisting of private schools and the other of government schools.\footnote{In unreported results, we also consider a standard Logit specification that has no unobserved heterogeneity and no dependence in the shocks. We find qualitatively similar results with respect to ML and NL, and hence focus on these richer specifications for brevity.} We take $X_i$ to be a vector of indicators for which bin the student's family income lies in, where we consider four bins determined by quartiles of its empirical distribution---see Table \ref{S-tab:student-chars} for descriptive statistics on the family income.

We also consider analogous parameterized versions of the liquidity constrained model in \eqref{eq:utlitymax_const}. Specifically, we take $\tilde{U}_{ij}$ to be as that in \eqref{eq:utility_para} with different specifications again imposing different assumptions on $\gamma_i$ and $\epsilon_{ij}$, and $C_{1i}(p)$ to be implied by a linear model on $E_i$ given by
\begin{align}
   E_i =  \tilde{\gamma}_{0} + \tilde{\gamma}_{1}’ X_i + \tilde{\nu}_i~,
 \end{align}
 where $\tilde{\nu}_i$ is normally distributed with mean 0 and variance $\tilde{\sigma}^2$, and statistically independent of $\epsilon_i$ and $\gamma_i$ conditional on $X_i$. We consider two specifications for $\gamma_i$ and $\epsilon_{ij}$, which correspond to ML and NL above, but with $\bar{\gamma}_1$ equal to a vector of zeros. In particular, the variation in covariates here is exploited to separate the role of the unobserved $E_i$ in determining choice rather than in allowing for observed heterogeneity in price responses---see, for e.g., \cite{arcidiacono/etal:16} for a further discussion who estimate such a model in a related school voucher setup.

Table \ref{S-tab:logit} reports maximum likelihood estimates for the parameters in the above specifications using the same analysis sample as in Section \ref{sec:empirical}.\footnote{The likelihood is simply given by $\prod_{i=1}^N q_{D_i}(P_i)$, where $N$ denotes sample size and $q$ corresponds to the implied demand function under the given specification, whose expressions are given in Table \ref{S-tab:logit_exp}.} Note that the flexibility in the above parameterizations is carefully chosen such that these parameters are point identified given the binary variation in prices induced by the voucher, after imposing the usual location and scale normalizations \citep[e.g.,][Chapter 2.5]{train:09}. As the underlying parameters are point identified, it follows that the corresponding demand functions are also point identified by plugging in the point identified parameter values in the expressions for demand, described in Table \ref{S-tab:logit_exp}, implied by the various specifications. Similarly, as we do in what follows, we can estimate demand by plugging the parameter estimates from Table \ref{S-tab:logit} into these demand expressions. We then estimate our parameters of interest by plugging estimated demand into \eqref{eq:theta_q} or, in the case of liquidity constraints that account for the fact that $E_i$ is continuous, the version of \eqref{eq:theta_q_liquidity} described in Table \ref{S-tab:logit_exp}.

\subsection{Observed and Counterfactual Demand Estimates}\label{sec:logit_fit}

\begin{figure}[!t]
\centering
\caption{Demand estimates for various specification under observed and counterfactual voucher amounts}
\makebox[\textwidth]{\begin{tabular}{c c c}
\subcaptionbox{Enrollment shares}{ \includegraphics[width=2.1in]{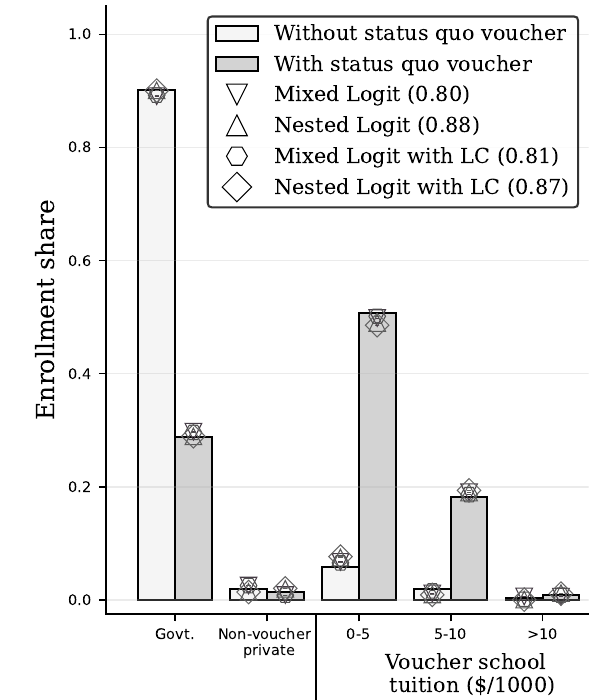}} & \subcaptionbox{Probability of voucher takeup}{ \includegraphics[width=2.1in]{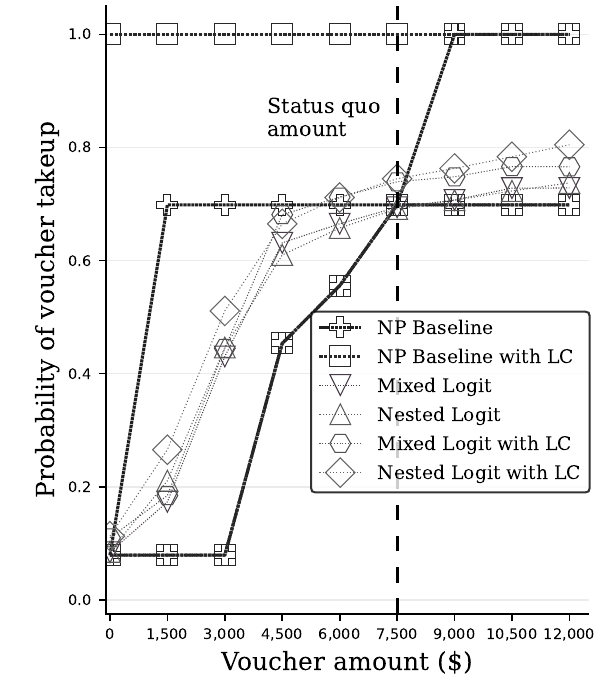}} &
\subcaptionbox{Probability of takeup at schools unaffordable without voucher}{\includegraphics[width=2.1in]{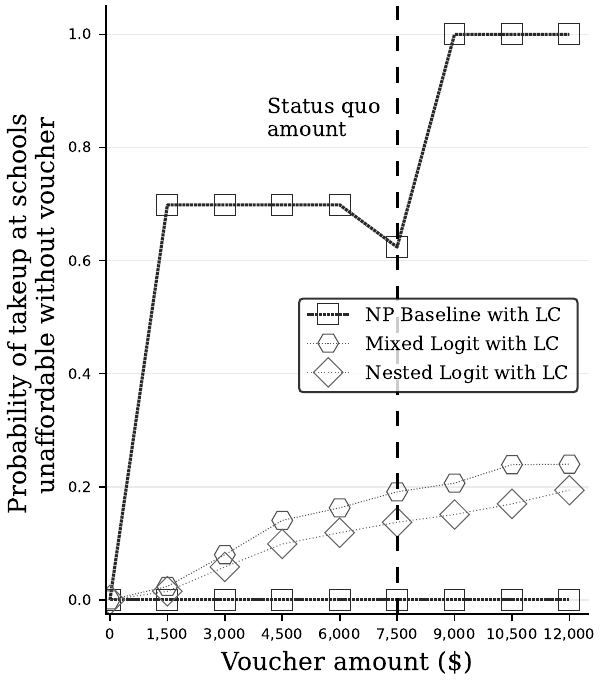}}
\end{tabular}}
\begin{tabular*}{0.99\textwidth}{c}
\multicolumn{1}{p{.99\hsize}}{\footnotesize NP denotes Nonparametric, and LC denotes that the model allows students to be liquidity constrained. In Panel (a), the numbers in brackets denote bootstrap-based $p$-values using 1,000 draws for specification tests that implied shares equal the observed ones.} \\
\multicolumn{1}{p{.99\hsize}}{\footnotesize SOURCE:  \textit{Evaluation of the DC Opportunity Scholarship Program: Final Report (NCEE 2010-4018)}, U.S. Department of Education, National Center for Education Statistics previously unpublished tabulations.}
\end{tabular*}
\label{fig:logit_fit}
\end{figure}

Before proceeding to the welfare estimates under the different logit specifications, we present various estimates of the demand functions implied by these specifications. In Figure \ref{fig:logit_fit}(a), we first analyze how well these demand functions match the observed shares by plotting them over the empirical enrollment shares in Figure \ref{fig:summ_stat}(b). Unlike our specifications that exactly match these shares, we can observe that this is not the case with the logit specifications. Nonetheless, the discrepancies are small. Heuristically, this is because there is only binary variation in the data, which is not too demanding to match relative to the flexibility of the logit models. We also statistically test the null hypothesis of no discrepancies by bootstrapping (using 1,000 draws) a test statistic based on the sum of squared difference between the implied and observed shares. The $p$-values reported in Figure \ref{fig:logit_fit}(a) reveal that the discrepancies are not statistically significant.

Next, to analyze the implied demand functions at prices beyond those observed in the data, Figure \ref{fig:logit_fit}(b) presents demand for the voucher, i.e. probability of choosing $j \in \mathcal{J}_v$, at various voucher amounts. Here, in the case of liquidity constraints, we plot demand taking all schools to be affordable, i.e. $\text{Prob}[\max_{j \in \mathcal{J}} \tilde{U}_{ij}(Y_i - p_j(\tau)) \in \mathcal{J}_v]$, to analyze how demand responds solely to price without relaxing liquidity constraints. We can observe that the logit models do not capture the range of demand functions credibly consistent with the data as revealed by our bounds, but instead all imply demand estimates that lie close to our lower bounds, where demand for the voucher is the lowest---Figure \ref{S-fig:parameter_logit_ci} reveals that this holds even when accounting for statistical uncertainty. This feature is even more pronounced in the case students are allowed to be liquidity constrained, where demand stays almost the same to that without liquidity constraints, while our upper bounds transparently reveal that demand for the voucher can in fact equal one---this is simply because allowing for a flexible liquidity constraint model, the data allows students who are not taking up the voucher to do so because they are liquidity constrained.

Finally, to better understand the magnitude of how liquidity constraints affect choices, Figure \ref{fig:logit_fit}(c) plots for the liquidity constraint models the probability of voucher take up at schools that are unaffordable absent the voucher, i.e. $\text{Prob}[\max_{j \in C_i(p(\tau))} \tilde{U}_{ij}(Y_i - p_j(\tau)) \in \tilde{C}_{i}]$ with $\tilde{C}_i = \{j \in \mathcal{J}_v : p_j(0) > E_i\}$. We can observe that while a positive proportion of students are impacted by the relaxation of liquidity constraints under the logit models, the proportion is relatively small and close to the lower bounds under the nonparametric specification. In contrast, the upper bounds reveal that large magnitudes are potentially consistent with the data. Similar to above, this is because under a flexible liquidity constrained model, the data allows all students who are taking up  the voucher to do so due to a relaxation of liquidity constraints.

\subsection{Welfare Estimates}\label{sec:logit_welfare}

\begin{figure}[!t]
\centering
\caption{Welfare estimates for status quo and counterfactual amounts under various specifications}
\makebox[\textwidth]{\begin{tabular}{c c c}
\subcaptionbox{Status quo amount}{ ~~\includegraphics[width=2.75in]{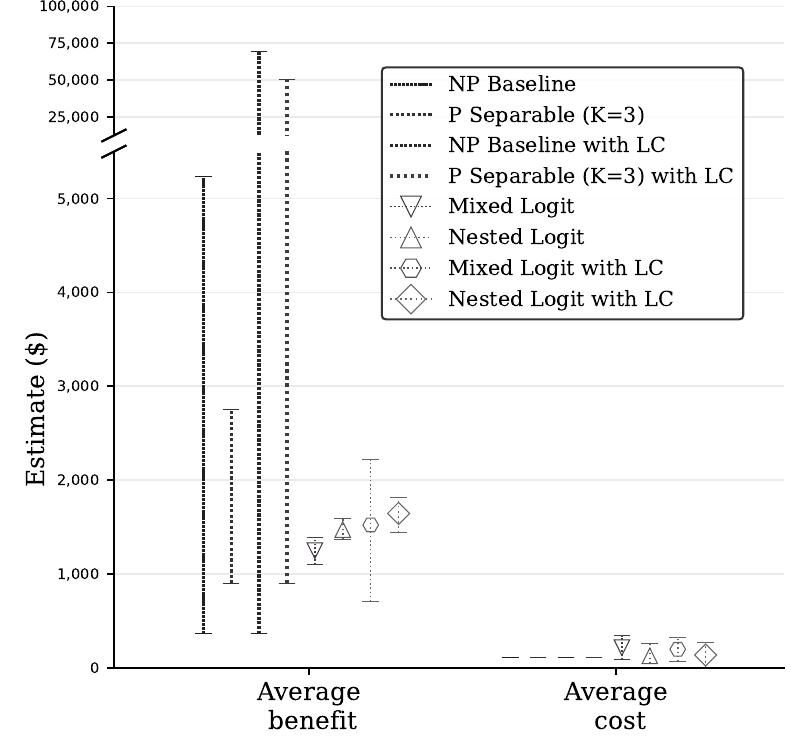}~~} & &
\subcaptionbox{$AS(\tau)$}{~~\includegraphics[width=2.75in]{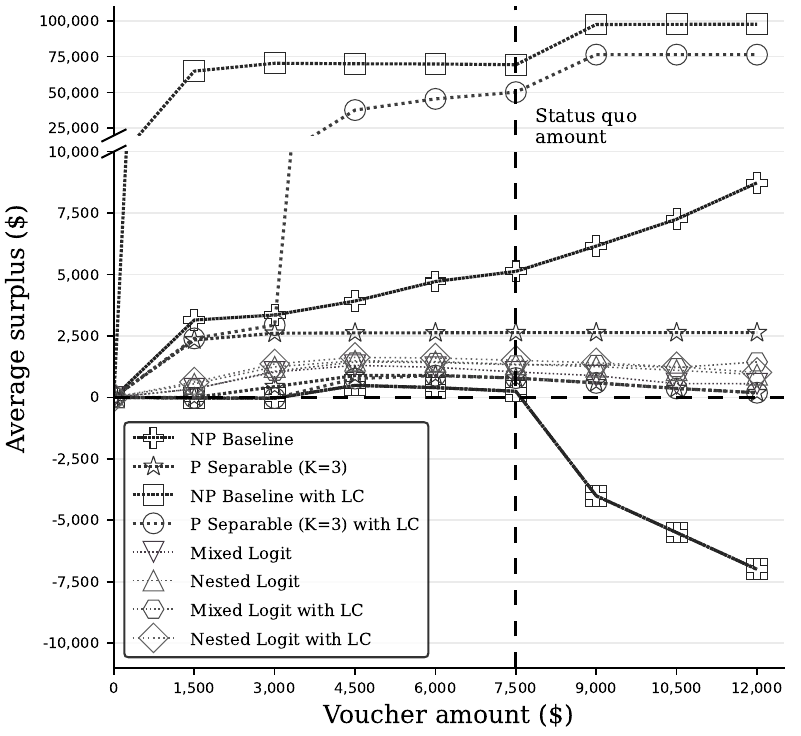}~~}
\end{tabular}}
\begin{tabular*}{0.99\textwidth}{c}
\multicolumn{1}{p{.99\hsize}}{\footnotesize NP denotes Nonparametric, P denotes Parametric, and LC denotes that the model allows students to be liquidity constrained. $AS(\tau)$ denotes average surplus under voucher amount $\tau$ as defined in \eqref{eq:AS}. For the nonparametric baseline and parametric separable specifications in Panel (a), the intervals denote the estimated lower and upper bounds, and, for the logit specifications, the markers denote the point estimates and the dashed intervals denote 90$\%$ confidence intervals computed using the percentile bootstrap with 1,000 draws. } \\
\multicolumn{1}{p{.99\hsize}}{\footnotesize SOURCE:  \textit{Evaluation of the DC Opportunity Scholarship Program: Final Report (NCEE 2010-4018)}, U.S. Department of Education, National Center for Education Statistics previously unpublished tabulations.}
\end{tabular*}
\label{fig:logit_welfare}
\end{figure}

As the logit models systematically limit attention to demand functions where demand for the voucher is lowest, it raises the concern that it may similarly lead us to conclude that the welfare gains are small. We conclude our analysis by assessing this concern relative to the conclusions we obtain in Section \ref{sec:welfare_est} from our robust estimates. To do so, Figure \ref{fig:logit_welfare}(a) presents the estimates for the average benefit and cost for the status quo voucher amount under the logit specifications along with estimated bounds under several of our specifications, and Figure \ref{fig:logit_welfare}(b) presents the resulting average surplus estimates in this case as well as under various counterfactual voucher amounts.

We can observe that the logit specifications all generate estimates that lie within our bounds. Intuitively, this is because the implied demand functions approximately fall in the estimated version of $\mathbf{Q}_B$ as they match well the data, as observed in Figure \ref{fig:logit_fit}(a), and satisfy the shape restrictions in \eqref{eq:q_shape} as $U_{ij}$ is generally increasing given that the estimated price coefficients are negative. In turn, as in our estimated bounds, they allow us to conclude that there is a positive net benefit for the voucher---though the conclusion for amounts larger than the status quo is more definitive here as our estimates caution that positive effects do not hold under the nonparametric specifications.

However, as the logit estimates all lie within a specific area of our bounds, they reveal a different picture on the magnitudes of the benefits. In particular, while our bounds conclude that there can potentially be large benefits, all the logit demand functions imply only small benefits. For example, from Figure \ref{fig:logit_welfare}(a), we can observe that our upper bounds suggest benefits could be up to half of the value of the status quo voucher in the case without liquidity constraints, but the logit estimates imply they will be no more than a fourth of its value. 

Moreover, as the logit specifications accounting for liquidity constraints imply only a small proportion of individuals are impacted by the relaxation of such constraints, the attenuation is even starker when accounting for liquidity constraints. While our bounds transparently reveal the benefits can become arbitrarily large, they stay almost the same in this case to those without liquidity constraints, and do not capture that there may exist a large downward bias from not accounting for such constraints. In summary, as this reveals that these low magnitudes of benefits are not driven by the data or shape restrictions, we conclude that the demand parameterizations implied by the logit specifications can indeed affect the conclusions one draws.

\section{Conclusion}\label{sec:conclusion}

In this paper, we develop new discrete choice tools to robustly learn about the average willing to pay for a price subsidy and its effects on demand given exogenous, discrete variation in prices. Our tools show how to characterize what we can learn when demand is allowed to be nonparametric as well as flexibly parameterized, both of which imply that our parameters are generally partially identified. We use our tools to perform a welfare analysis of the price subsidy provided by school vouchers in the DC Opportunity Scholarship Program. We also compare our empirical results to those one would obtain under standard logit parameterizations of demand and highlight how they can affect the empirical conclusions one draws.


\newpage
\bibliography{references.bib}

\end{document}


\author{
Vishal Kamat \\
Toulouse School of Economics\\
University of Toulouse Capitole\\
\url{vishal.kamat@tse-fr.eu}
\and
Samuel Norris\\
Department of Economics\\
University of British Columbia\\
\url{sam.norris@ubc.ca} \vspace{0.5cm}
}

\title{Supplementary Appendix to ``Estimating Welfare Effects in a Nonparametric Choice Model: The Case of School Vouchers''}

\maketitle

\vspace{-0.5cm}
\begin{abstract}
This document presents proofs and additional details pertinent to the main analysis for the authors' paper titled ``Estimating Welfare Effects in a Nonparametric Choice Model: The Case of School Vouchers.'' Section \ref{sec:procedure_V} describes how to construct the partition in our baseline analysis. Section \ref{sec:extensions} provides details on the various extensions to the baseline analysis. Section \ref{sec:inference} describes the procedure used to perform statistical inference in the empirical analysis. Section \ref{sec:add_details_emp} presents additional details relevant to the empirical analysis. Section \ref{sec:proof} presents proofs of all results. Section \ref{sec:add_figtab} presents additional tables and figures.
\end{abstract}

\thispagestyle{empty}

\newpage
\setcounter{page}{1}

\appendix
\renewcommand{\theequation}{S.\arabic{equation}}
\renewcommand{\thetable}{S.\arabic{table}}
\renewcommand{\thefigure}{S.\arabic{figure}}
\renewcommand{\thesection}{S.\arabic{section}}
\renewcommand{\theproposition}{S.\arabic{proposition}}
\renewcommand{\thelemma}{S.\arabic{lemma}}

\section{Constructing $\mathcal{V}$}\label{sec:procedure_V}

In this section, we show how to obtain a collection of sets $\mathcal{V}$ that partitions $\mathcal{P}^* = \bigcup\limits_{l=1}^L \mathcal{P}^*_l$ satisfying Definition \ref{O-def:V}.

To this end, observe first that the union of the sets in $\mathcal{V}^{a,b}$ and $\{\{p\} : p \in \mathcal{P}_{\text{obs}}\}$ equals $\mathcal{P}^*$, where $\mathcal{V}^{a,b} = \left\{ \mathcal{P}_l^{a,b}, \ldots, \mathcal{P}_{|\mathcal{J}|-1}^{a,b}, \{p^a\}, \{p^b\} \right\}$, and further that $\mathcal{V}^{a,b}$ and $\{\{p\} : p \in \mathcal{P}_{\text{obs}}\}$ each correspond to a partition of the union of the sets in it that satisfies Definition \ref{O-def:V}(i)-(iii) with respect to its union. However, for a given $v \in \mathcal{V}^{a,b}$ and $v' \in \{\{p\} : p \in \mathcal{P}_{\text{obs}}\}$, it may be that Definition \ref{O-def:V}(iii) is not satisfied. In particular, it may be the case that there exists $p \in \mathcal{P}_{\text{obs}}$ that intersects some $v \in \mathcal{V}^{a,b}$ in the $j$th dimension, i.e. $\{p_j\} \subset v_{[j]}$ for some $j \in \mathcal{J}$, and hence we have $\{p_j\} \neq v_{[j]}$ and $\{p_j\} \cap v_{[j]} \neq \emptyset$.

To obtain our partition, we therefore further partition the elements in $\mathcal{V}^{a,b}$ to obtain $\tilde{\mathcal{V}}^{a,b}$ such that Definition \ref{O-def:V}(iii) is satisfied between the elements of $\tilde{\mathcal{V}}^{a,b}$ and $\{\{p\} : p \in \mathcal{P}_{\text{obs}}\}$ by accounting for whether $p \in \mathcal{P}_{\text{obs}}$ and $v \in \mathcal{V}^{a,b}$ intersect in each of the $j \in \mathcal{J}$ dimensions. In particular, let $\mathcal{T} = \{\Delta^{a,b}_{l} : 1 \leq l \leq |\mathcal{J}|-1\} \cup \{\max\{\min\{p_j,p^a_j\},p^b_j+\Delta^{a,b}_1\} - p^b_j : j \in \mathcal{J},~p \in \mathcal{P}_{\text{obs}}\}\}$ denote the set of points such that $\{\min\{p^a_j,p^b_j + t\} : t \in \mathcal{T}\}$ correspond to the end-points of the sets in $\mathcal{P}_l^{a,b}$, $1 \leq l \leq |\mathcal{J}| - 1$, as well as where $p \in \mathcal{P}_{\text{obs}}$ may intersect the sets $\mathcal{V}^{a,b}$ in each of the $j \in \mathcal{J}$ dimensions---note here that the minimums and maximums ensure that we only include the intersection points if they are in $[p^b_j+\Delta^{a,b}_1,p^a_j]$. Denoting by $t_1 < \ldots < t_{|\mathcal{T}|}$ the ordered values of $\mathcal{T}$, then take $\tilde{\mathcal{V}}^{a,b} = \left\{ \tilde{\mathcal{P}}_l^{a,b}, \ldots, \tilde{\mathcal{P}}_{|\mathcal{T}|-1}^{a,b}, \{p^a\}, \{p^b\} \right\}$, where $\tilde{\mathcal{P}}^{a,b}_l = \left\{ p \in \mathcal{P} : p_j = \min \{p_j^a, p_j^b + t\},~t \in (t_l,t_{l+1}) \text{ for } j \in \mathcal{J} \right\}$ for $1 \leq l \leq |\mathcal{T}|-1$. As before, observe that $\tilde{\mathcal{V}}^{a,b}$ and $\{\{p\} : p \in \mathcal{P}_{\text{obs}}\}$ each correspond to a partition of the union of the sets in it that satisfy Definition \ref{O-def:V}(i)-(iii) with respect to its union. However, now by construction, we also have that Definition \ref{O-def:V}(iii) is satisfied for each $v \in \tilde{\mathcal{V}}^{a,b}$ and $v' \in \mathcal{P}^{\text{obs}}$. To this end, we take
\begin{align*}
\mathcal{V} = \tilde{\mathcal{V}}^{a,b} \bigcup \{\{p\} : p \in \mathcal{P}_{\text{obs}}\}~,
\end{align*}
which corresponds to a partition of $\mathcal{P}^*$ satisfying Definition \ref{O-def:V}.

Note that the above constructed $\mathcal{V}$ corresponds to the coarsest partition of $\mathcal{P}^*$ satisfying Definition \ref{O-def:V} in the sense for any other partition $\mathcal{V}'$ satisfying Definition \ref{O-def:V} we have that for all $v \in \mathcal{V}$ there exists $\mathcal{V}'' \subseteq \mathcal{V}'$ such that $\cup \{v' : v' \in \mathcal{V}''\} = v$, i.e. any other partition is finer and can be used to construct our choice of $\mathcal{V}$. To see this, suppose that this wasn't the case. This means that for some partition $\mathcal{V}'$ satisfying Definition \ref{O-def:V} there exists $v \in \mathcal{V}$ and $v' \in \mathcal{V}'$ such that $v' \not\subseteq v$ and $v \cap v' \neq \emptyset$, i.e. $v'$ is not contained in $v$ and that they overlap. In this case, it must be that $v = \left\{ p \in \mathcal{P} : p_j = \min \{p_j^a, p_j^b + t\},~t \in (t_l,t_{l+1}) \text{ for } j \in \mathcal{J} \right\}$ for some $1 \leq l \leq |\mathcal{T}|-1$ and $v' = \left\{ p \in \mathcal{P} : p_j = \min \{p_j^a, p_j^b + t\},~t \in (t',t'') \text{ for } j \in \mathcal{J} \right\}$ for some values of $t',t''$, where $t_{l+1} > t'$ or $t'' > t_l$. However, as the endpoints of $v$ equal those of sets in $\mathcal{P}^*$ in some dimensions, we have that $v'$ is not contained in and overlaps with these sets in $\mathcal{P}^*$ in these dimensions. In turn, $\mathcal{V}'$ cannot satisfy Definition \ref{O-def:V}(i), which is a contradiction.

\section{Additional Details on Extensions in Section \ref{O-sec:extensions}}\label{sec:extensions}

\subsection{Example of a Set $\mathbf{B}^{\text{r}}$}\label{sec:B_r}

In this section, we describe a set $\mathbf{B}^{\text{r}}$ such that $\phi(\mathbf{B}) \subseteq \mathbf{B}^{\text{r}}$ that we use in our empirical analysis when implementing the linear programs in \eqref{O-eq:b_r_opt}. To this end, it is first useful to consider an equivalent representation of the restrictions in \eqref{eq:b_shape} written in terms of pairs $w,w' \in \mathcal{W}$.

\begin{proposition}\label{prop:b_shape_alt}
$\beta$ satisfies \eqref{eq:b_shape} if and only if \begin{align}\label{eq:b_shape_alt}
\sum_{j \in \mathcal{J}^{>}_{w',w} \cup \mathcal{J}^{\dagger}} \beta_{j}(w) &\geq \sum_{j \in \mathcal{J}^{>}_{w',w} \cup \mathcal{J}^{\dagger}} \beta_{j}(w')
\end{align}
for each $\mathcal{J}^{\dagger} \subseteq \mathcal{J}^{=}_{w,w'}$ and $w,w' \in \mathcal{W}$, where
\begin{align*}
\mathcal{J}^{>}_{w,w'} &= \left\{ j \in \mathcal{J} : t > t' \text{ for all } t \in w_{[j]}~,~t' \in w'_{[j]} \right\}~, \\
\mathcal{J}^{=}_{w,w'} &= \mathcal{J} \setminus \left(  \mathcal{J}^{>}_{w,w'}  \cup  \mathcal{J}^{>}_{w',w}  \right)~.
\end{align*}
\end{proposition}

Given the equivalence between the restrictions in \eqref{eq:b_shape} and \eqref{eq:b_shape_alt}, we can alternatively write $\mathbf{B}$ in \eqref{eq:B_set} as
\begin{align*}
\mathbf{B} = \left\{ \beta \in \mathbf{R}^{d_{\beta}} : \beta \text{ satisfies } \eqref{eq:b_logical1}-\eqref{eq:b_logical2},~\eqref{eq:b_shape_alt}, \text{ and } \eqref{eq:b_data}  \right\}~.
\end{align*}
In this set, observe that each restriction on $\beta$ is for a given $w \in \mathcal{W}$ or for a pair of $w,w' \in \mathcal{W}$. Our choice of $\mathbf{B}^{\text{r}}$ corresponds to the subset of these restrictions on $\beta$ for $w \in \mathcal{W}^{\text{r}}$ or for pairs of $w,w' \in \mathcal{W}^{\text{r}}$, i.e. the subset of restrictions that directly correspond those that are in terms of $\beta^{\text{r}}$. More specifically, these restrictions correspond to the following
\begin{align}
\beta^{\text{r}}_{j}(w) &\geq 0~\text{ for each $j \in \mathcal{J}$ and $w \in \mathcal{W}^r$}~, \label{eq:b_r_1} \\
\sum_{j \in \mathcal{J}} \beta^{\text{r}}_{j}(w) &= 1~\text{ for each $w \in \mathcal{W}^r$}~, \\
\sum_{j \in \mathcal{J}^{>}_{w',w} \cup \mathcal{J}^{\dagger}} \beta^{\text{r}}_{j}(w) &\geq \sum_{j \in \mathcal{J}^{>}_{w',w} \cup \mathcal{J}^{\dagger}} \beta^{\text{r}}_{j}(w')~\text{ for each }~\mathcal{J}^{\dagger} \subseteq \mathcal{J}^{=}_{w,w'}~\text{ and }~w,w' \in \mathcal{W}^{\text{r}}~, \\
 \beta^{\text{r}}_{j}(\{p\})     &= \text{Prob}[D_i=j|P_i=p]~\text{ for each $j \in \mathcal{J}$, $p \in \mathcal{P}_{\text{obs}}$}~. \label{eq:b_r_end}
\end{align}
Then, denoting by $d_{\beta^{\text{r}}}$ the dimension of $\beta^{\text{r}}$, the set we consider is given by
\begin{align*}
\mathbf{B}^{\text{r}} = \left\{ \beta^{\text{r}} \in \mathbf{R}^{d_{\beta^{\text{r}}}} : \beta^{\text{r}} \text{ satisfies } \eqref{eq:b_r_1} - \eqref{eq:b_r_end} \right\}~.
\end{align*}

\subsection{Extension to Separability Assumptions}\label{sec:separability}

We noted in Section \ref{O-sec:dim_red} that we can additionally impose separability assumptions on demand to reduce the dimension of the linear programs to compute the identified set under the baseline nonparametric specification. In this section, we show how to extend the arguments in Section \ref{O-sec:compute_identifiedset} to compute the identified set under a general class of such separability assumptions.

For each $p \in \mathcal{P}$ and $\mathcal{J}' \subseteq \mathcal{J}$, let
\begin{align*}
p_{[\mathcal{J}']} = (p_j : j \in \mathcal{J}') \in \prod_{j \in \mathcal{J}'} [\underline{p}_j,\bar{p}_j]
\end{align*}
denote the sub-vector of $p$ with indices in $\mathcal{J}'$. Using this notation, the general separability assumption we consider can be stated as follows:
\begin{assumption2}{GS}{(General Separability)}\label{ass:S}~
For each $j \in \mathcal{J}$,
\begin{align}\label{eq:q_sep}
q_j(p) = \sum_{l \in \mathcal{L}_j} h_{jl}\left(p_{[\mathcal{J}_{jl}]}\right)~,
\end{align}
where $\mathcal{L}_j$ and $\mathcal{J}_{jl}$ are pre-specified subsets of $\mathcal{J}$, and $h_{jl}$ are some unknown functions.
\end{assumption2}
\noindent Assumption \ref{ass:S} imposes demand to be a sum of functions that are of lower dimension than demand. Observe that Assumption \ref{O-ass:AS} is a special case of this assumption, where $\mathcal{L}_j = \mathcal{J}$ and $\mathcal{J}_{jl} = \{l\}$. We highlight that it also allows for the case where no dimension reduction is imposed and demand is as that in baseline non-separable case, by taking $\mathcal{L}_j$ to be a singleton set such as $\{j\}$ and taking $\mathcal{J}_{jl} = \mathcal{J}$---see proof of Proposition \ref{O-prop:identifiedset_base_1} for more details. Under this assumption, our admissible set of demand functions is given by
\begin{align}\label{eq:Q_S}
\mathbf{Q}_S = \left\{q \in \mathbf{F} : q \text{ satisfies } \eqref{O-eq:q_logical1}-\eqref{O-eq:q_logical2},~\eqref{O-eq:q_shape} \text{ and } \eqref{eq:q_sep}\right\}~,
\end{align}
i.e. $\mathbf{Q}_B$ in \eqref{O-eq:Q_B}, but with the additional separability restriction in \eqref{eq:q_sep}.

In order to describe how to compute the identified set $\theta(\mathbf{Q}_S)$, we show as in Proposition \ref{O-prop:identifiedset_base_1} that there is no loss of information in first replacing $\mathbf{Q}_S$ with a certain finite-dimensional space, given which the identified set can then be computed using linear programs. To define this finite dimensional space, let $\mathcal{W}$ denote the constructed partition of $\mathcal{P}$ from \eqref{O-eq:set_W}, and, for each $w \in \mathcal{W}$ and $\mathcal{J}' \subseteq \mathcal{J}$, let
\begin{align*}
w_{[\mathcal{J}']} = \left\{ p_{[\mathcal{J}']} : p \in w  \right\} \in \prod_{j \in \mathcal{J}'} \mathcal{V}_j \equiv \mathcal{W}_{[\mathcal{J}']}
\end{align*}
denote the set that includes the sub-vector of prices in $w$ with indices in $\mathcal{J}'$---note that this is with some abuse in notation as for a single index $j \in \mathcal{J}$, we simply use $w_{[j]}$ as previously introduced rather than $w_{[\{j\}]}$. The finite dimensional space we consider is then given by
\begin{align}\label{eq:Q_S_fd}
\mathbf{Q}_S^{\text{fd}} = \left\{ q \in \mathbf{Q}_B : q_j(p) = \sum_{w \in \mathcal{W}} 1_w(p) \sum_{l \in \mathcal{L}_j}  \psi_{jl}(w_{[\mathcal{J}_{jl}]}) \text{ for some } \{\psi_{jl}(w_{[\mathcal{J}_{jl}]})\}_{\substack{w_{[\mathcal{J}_{jl}]} \in \mathcal{W}_{[\mathcal{J}_{jl}]}\\ l \in \mathcal{L}_j}},~ j \in \mathcal{J}  \right\}~,
\end{align}
i.e. the same as \eqref{O-eq:Q_B_fd} but with the additional restriction that the constant valued functions satisfy \eqref{eq:q_sep}. Let $\psi$ capture finite-dimensional variables $\{\psi_{jl}(w_{[\mathcal{J}_{jl}]}):w_{[\mathcal{J}_{jl}]} \in \mathcal{W}_{[\mathcal{J}_{jl}]},~l \in \mathcal{L}_j,~j \in \mathcal{J}\}$ in vector form. Observe that the dimension of this variable is given by $\sum\limits_{j \in \mathcal{J}}\sum\limits_{l \in \mathcal{L}_j} \prod\limits_{m \in \mathcal{J}_{[jl]}}|\mathcal{V}_m|$, while that of $\beta$ in \eqref{O-eq:Q_B_fd} is $|\mathcal{J}| \prod\limits_{m \in \mathcal{J}}|\mathcal{V}_m|$.

As in Section \ref{O-sec:equi_charac}, to show that $\theta(\mathbf{Q}_S)=\theta(\mathbf{Q}^{\text{fd}}_S)$ and how to compute $\theta(\mathbf{Q}_S^{\text{fd}})$, it is next useful to rewrite $\theta(\mathbf{Q}_S^{\text{fd}})$ in term of $\psi$. Let $\theta_S$ denote the continuous function that rewrites the parameter of interest in terms of $\psi$ in the sense that $\theta(q) = \theta_S(\psi)$.  Similarly, let
\begin{align}
\mathbf{S} = \left\{ \psi \in \mathbf{R}^{d_{\psi}} : \left( \sum_{w \in \mathcal{W}} 1_{w} \sum_{l \in \mathcal{L}_j}  \psi_{jl}(w_{[\mathcal{J}_{jl}]}) : j \in \mathcal{J} \right) \in \mathbf{Q}_B \right\}~,
\end{align}
where $d_{\psi}$ denotes the dimension of $\psi$, capture the restrictions on demand in terms of $\psi$. We can then write $\theta(\mathbf{Q}^{\text{fd}}_{S})$ in terms of $\psi$ by
\begin{align}\label{eq:s_identifiedset}
\Theta_S \equiv \left\{\theta_0 \in \mathbf{R} : \theta_S(\psi) = \theta_0 \text{ for some } \psi \in \mathbf{S} \right\}~.
\end{align}
As in Proposition \ref{O-prop:identifiedset_base_1}, we show in the following proposition that $\theta(\mathbf{Q}_S)$ is equal to $\Theta_S$, and that it can be compute by solving two optimization problem. We again highlight in its proof that $\theta_S$ is linear and that $\mathbf{S}$ is linear, which imply that the optimization problems are linear programs.

\begin{proposition}\label{prop:identifiedset_sep}
Suppose that $\mathbf{Q} = \mathbf{Q}_S$. Then, the identified set in \eqref{O-eq:theta_identifiedset} is equal to that in \eqref{eq:s_identifiedset}, i.e. $\Theta = \Theta_S$. In addition, if $\mathbf{S}$ is empty then by definition $\Theta_S$ is empty; whereas, if $\mathbf{S}$ is non-empty then the closure of $\Theta_S$ is given by $[\underline{\theta}_{S}, \bar{\theta}_{S}]$, where
\begin{align}\label{eq:bounds_opt_S}
\underline{\theta}_{S} = \inf_{\psi \in \mathbf{S}}~ \theta_S(\psi) ~\text{ and }~ \bar{\theta}_{S} = \sup_{\psi \in \mathbf{S}} ~\theta_S(\psi)~.
\end{align}
\end{proposition}

\subsection{Extension to Auxiliary Parametric Assumptions}\label{sec:parametric_ext}

In this section, we show to compute the identified set when Assumption \ref{O-ass:A} from Section \ref{O-sec:parametric} is additionally imposed. The admissible set of functions in this case is given by
\begin{align}\label{eq:Q_A}
\mathbf{Q}_A = \left\{q \in \bar{\mathbf{F}} : q \text{ satisfies } \eqref{O-eq:q_data}, \eqref{O-eq:q_logical1}-\eqref{O-eq:q_shape} \text{ and } \eqref{O-eq:q_para} \right\}~.
\end{align}
i.e. $\mathbf{Q}_B$ along with the restriction in \eqref{O-eq:q_para}, and the identified set by $\theta(\mathbf{Q}_A)$. Unlike the nonparametric analysis in Section \ref{O-sec:compute_identifiedset}, note that computing this identified set is a finite-dimensional problem as $\mathbf{Q}_A$ is a finite-dimensional parameterized space given that the demand functions are parameterized by the variable $\alpha$. In turn, $\theta(\mathbf{Q}_A)$ can be directly characterized by searching over $q$ in $\mathbf{Q}_A$ and then taking their image under the function $\theta$.

To show how to do this, it is useful, as in Section \ref{O-sec:equi_charac}, to rewrite $\theta(\mathbf{Q}_A)$ in terms of $\alpha$, the variable parameterizing $q \in \mathbf{Q}_A$. As $\theta$ is continuous in $q$ and $q$ is continuous in $\alpha$, let $\theta_A$ be the continuous function of $\alpha$ such that $\theta(q) = \theta_A(\alpha)$. Moreover, $\mathbf{Q}_A$ can be written in terms of $\alpha$ by
\begin{align}
\mathbf{A} = \left\{ \alpha \in \mathbf{R}^{d_{\alpha}}  : \left( \sum_{k=0}^{K_j} \alpha_{jk} \cdot b_{jk} : j \in \mathcal{J} \right) \in \mathbf{Q}_B  \right\}~.
\end{align}
where $d_{\alpha}$ denotes the dimension of $\alpha$, i.e. the set of values of $\alpha$ that ensure that the corresponding $q$ is in $\mathbf{Q}_B$. We can write $\theta(\mathbf{Q}_A)$ in terms of $\alpha$ by
\begin{align}\label{eq:A_identifiedset}
\theta_A(\mathbf{A}) = \left\{\theta_0 \in \mathbf{R} : \theta_A(\alpha) = \theta_0 \text{ for some } \alpha \in \mathbf{A} \right\} \equiv \Theta_A~.
\end{align}
In the following proposition, we show that when $\mathbf{A}$ is connected and non-empty, the closure of $\Theta_A$ is equal to an interval, where the endpoints can be characterized as solutions to two finite-dimensional optimization problems.
\begin{proposition}\label{prop:identifiedset_aux}
If $\mathbf{A}$ is empty then by definition $\Theta_A$ is empty; whereas, if $\mathbf{A}$ is connected and non-empty, then the closure of $\Theta_A$ is given by $[\underline{\theta}_{A}, \bar{\theta}_{A}]$, where
\begin{align}\label{eq:aux_optimization}
 \underline{\theta}_{A} = \inf_{\alpha \in \mathbf{A}}~ \theta_A(\alpha) ~\text{ and }~ \bar{\theta}_{A} = \sup_{\alpha \in \mathbf{A}} ~\theta_A(\alpha)~.
\end{align}
\end{proposition}

Proposition \ref{prop:identifiedset_aux} shows how to characterize the identified set under a general class of parametric restrictions. While the optimization problems in \eqref{eq:aux_optimization} are finite-dimensional, their tractability depends on the structure of the objective $\theta_A$ and the constraint set $\mathbf{A}$. Given the linear structure of $\theta$ and $\mathbf{Q}_B$ along with \eqref{O-eq:q_para}, we in fact have that $\theta_A$ is linear and $\mathbf{A}$ is determined by linear restrictions. For example, under the specific parameterization we consider in \eqref{O-eq:q_A_AS},  observe that \eqref{O-eq:theta_q} can be written in terms of $\alpha$ as follows
\begin{align}
\theta_{A}(\alpha) &\equiv g^{a,b} \Delta^{a,b}_1 + g^{a,b} \sum_{l=1}^{|\mathcal{J}|-1} \sum_{j \in \mathcal{J}^{a,b}_{l+1}} \int\limits_{\Delta_l^{a,b}}^{\Delta_{l+1}^{a,b}} \left( \sum_{m \in \mathcal{J}^{a,b}_{l+1}} \sum_{k=0}^K \alpha_{jmk} \cdot (p_m^b)^k   + \sum_{m \in \mathcal{J} \setminus \mathcal{J}^{a,b}_{l+1}} \sum_{k=0}^K \alpha_{jmk} \cdot (p_m^b + t)^k \right) dt \nonumber \\
&~~~+ \sum_{j \in \mathcal{J}} \sum_{m \in \mathcal{J}} \sum_{k=0}^K \alpha_{jmk} \cdot \left( g_j^a \cdot (p_m^a)^k + g_j^b \cdot (p_m^b)^k \right) \nonumber \\
& = g^{a,b} \Delta^{a,b}_1 + g^{a,b} \sum_{l=1}^{|\mathcal{J}|-1} \sum_{j \in \mathcal{J}^{a,b}_{l+1}} \left( \sum_{m \in \mathcal{J}^{a,b}_{l+1}} \sum_{k=0}^K \alpha_{jmk} \cdot (p_m^a)^k \cdot (\Delta_{l+1}^{a,b} - \Delta_{l}^{a,b}) ~+  \right. \nonumber  \\
& ~~~~~~~~~~~~~~~~~~~~~~~~~~~~~~~~~~~~~~\left. \sum_{m \in \mathcal{J} \setminus \mathcal{J}^{a,b}_{l+1}} \sum_{k=0}^K \alpha_{jmk} \cdot \left(\left. \frac{(p_m^b + t)^{k+1}}{k+1} \right|^{\Delta_{l+1}^{a,b}}_{\Delta_{l}^{a,b}} \right)  \right)~ \nonumber \\
&~~~+ \sum_{j \in \mathcal{J}} \sum_{m \in \mathcal{J}} \sum_{k=0}^K \alpha_{jmk} \cdot \left( g_j^a \cdot (p_m^a)^k + g_j^b \cdot (p_m^b)^k \right) \label{eq:AS_AB}
\end{align}
where the first line follows from directly substituting the relation between $q$ and $\alpha$ from \eqref{O-eq:q_A_AS} in \eqref{O-eq:theta_q}, and the second line from evaluating the integrals in the first part of the expression. Similarly, to see the linear restrictions that determine $\mathbf{A}$ here, observe that by substituting the relation between $q$ and $\alpha$ into the various restrictions in \eqref{O-eq:q_logical1}-\eqref{O-eq:q_logical2}, \eqref{O-eq:q_shape} and \eqref{O-eq:q_data} we obtain
\begin{align}
\sum_{m \in \mathcal{J}} \sum_{k=0}^K \alpha_{jmk} \cdot p_m^k &\geq 0 ~\text{ for each } j \in \mathcal{J}~, \label{eq:AS_logical1}\\
\sum_{j \in \mathcal{J}}  \sum_{m \in \mathcal{J}} \sum_{k=0}^K \alpha_{jmk} \cdot p_m^k &= 1
\end{align}
for all $p \in \mathcal{P}$,
\begin{align}\label{eq:AS_shape}
\sum_{m \in \mathcal{J}} \sum_{k=0}^K \alpha_{jmk} \cdot \left((p_m)^k - (p'_m)^k \right) \geq 0
\end{align}
for each $j \in \mathcal{J} \setminus \mathcal{J}'$ and $p,p' \in \mathcal{P}$ such that $p_j > p'_j$ for $j \in \mathcal{J}' \subseteq \mathcal{J}$ and $p_j = p'_j$ for $j \in \mathcal{J} \setminus \mathcal{J}'$, and
\begin{align}\label{eq:AS_data}
\sum_{m=1}^J \sum_{k=0}^K \alpha_{jmk} \cdot (p_m)^k &= \text{Prob}[D=j|P_i =p]~,
\end{align}
for each $j \in \mathcal{J}$ and $p \in \mathcal{P}_{\text{obs}}$, given which we then have that $\mathbf{A} = \{ \alpha \in \mathbf{R}^{d_{\alpha}} : \alpha \text{ satisfies } \eqref{eq:AS_logical1} - \eqref{eq:AS_data} \}$.

However, note that the restrictions in \eqref{O-eq:q_logical1}-\eqref{O-eq:q_shape} are evaluated at every price in the continuous space $\mathcal{P}$, which implies that the resulting optimization problems can be generally difficult to compute. In practice, we therefore propose to consider the following alternative optimization problems
\begin{align}\label{eq:aux_optimization_r}
\underline{\theta}_{A}^{\text{r}} = \min_{\alpha \in \mathbf{A}^{\text{r}}} \theta_{A}(\alpha) ~ \text{ and } ~ \bar{\theta}_{A}^{\text{r}} = \max_{\alpha \in \mathbf{A}^{\text{r}}} \theta_{A}(\alpha)~,
\end{align}
where $\mathbf{A}^{\text{r}}$ corresponds to a subset of $\mathbf{A}$ with the restrictions in \eqref{O-eq:q_logical1}-\eqref{O-eq:q_shape} evaluated on a discrete set of prices $\mathcal{P}^{\text{r}} = \{p_0, \ldots, p_L\}$. For example, under the parameterization in \eqref{O-eq:q_A_AS}, we have by simplifying and removing some jointly redundant restrictions using some algebra that the set of restrictions in \eqref{eq:AS_logical1}-\eqref{eq:AS_data} when evaluated on $\mathcal{P}^{\text{r}}$ can be explicitly given by
\begin{align}
\sum_{k=0}^K \alpha_{jjk} \cdot p_{l,j}^k + \sum_{m \in \mathcal{J} \setminus \{j\}} \sum_{k=0}^K \alpha_{jmk} \cdot p_{0,m}^k &\geq 0~ \text{ for each } j \in \mathcal{J},~ 0 \leq l \leq L~, \label{eq:ASr_logical1} \\
\sum_{j \in \mathcal{J}} \sum_{m \in \mathcal{J}} \sum_{k=0}^K \alpha_{jmk} \cdot p^k_{0,m} &= 1~, \\
 \sum_{j \in \mathcal{J}} \sum_{k=0}^K \alpha_{jmk} \cdot \Delta p_{l+1,m}^k &= 0~ \text{ for } m \in \mathcal{J},~ 0 \leq l \leq L - 1~, \\
 \sum_{k=0}^K \alpha_{jmk} \cdot  \Delta p_{l+1,m}^k  &\geq 0~ \text{ for each } j \in \mathcal{J},~m \neq j \in \mathcal{J}, ~ 0 \leq l \leq L - 1~, \  \label{eq:ASr_shape}
\end{align}
and \eqref{eq:AS_data}, given which $\mathbf{A}^{\text{r}} = \{ \alpha \in \mathbf{R}^{d_{\alpha}} : \alpha \text{ satisfies } \eqref{eq:ASr_logical1} - \eqref{eq:ASr_shape} \text{ and } \eqref{eq:AS_data}\}$, where $\{p_{0,j}, \ldots, p_{L,j}\}$ denotes the set of ordered values of $\mathcal{P}_j^{\text{r}}$ for each $j \in \mathcal{J}$ and $\Delta p_{l,j}^k = p_{l,j}^k - p_{l-1,j}^k$ denotes the difference in two consecutive prices in this set with each of these two prices raised to the power of $k$. As the objective and the finite number of restrictions determining the constraint sets of these problems are linear in $\alpha$, the problems in \eqref{eq:aux_optimization_r} are linear programs and hence generally computationally tractable. But, since $\mathbf{A} \subseteq \mathbf{A}^{\text{r}}$, note that these problems only provide an outer set for $\Theta_A$, i.e. $\Theta_A \subseteq [\underline{\theta}_{A}^{\text{r}}, \bar{\theta}_{A}^{\text{r}}]$, similar in spirit to those in \eqref{O-eq:b_r_opt} with respect to $\Theta_B$.

To implement the above, we need to choose the discrete set of prices $\mathcal{P}^{\text{r}}$. In our application in Section \ref{O-sec:empirical}, we take  $\mathcal{P}^{\text{r}} = \prod_{j=1}^J \mathcal{P}_j^{\text{r}}$, where $\mathcal{P}_j^{\text{r}} = \{\underline{p}_j + \frac{l}{L} (\bar{p}_j - \underline{p}_j): 0 \leq l \leq \bar{L}\} \cup \left\{p_j : p \in \{p^a,p^b\} \cup \mathcal{P}_{\text{obs}}\right\}$ for some pre-specified value of $\bar{L}$ and with $\underline{p}_j$ and $\bar{p}_j$ equal to the minimum and maximum values of $j$th dimension of the prices in $\{p^a,p^b\} \cup \mathcal{P}_{\text{obs}}$, i.e. a set of $(\bar{L}+1)$ equidistant points between the minimum and maximum support of prices used in the definition of the parameter and observed in the data. The empirical results take $\bar{L} = 6$. In unreported results, we find that increasing the value of $\bar{L}$ generally tightens the bounds in a gradual manner, but at the cost of increased computational time, specifically for the confidence intervals.

\subsection{Extension to Liquidity Constraints}\label{sec:liquidity}

In this section, we illustrate how to compute the identified set in the presence of liquidity constraints as modeled in Section \ref{O-sec:liquidity}. To this end, it is useful to explicitly state the analogs of various restrictions in \eqref{O-eq:q_data} and \eqref{O-eq:q_logical1}-\eqref{O-eq:q_shape}, and the additional ones from Assumption \ref{O-ass:LC} in terms of the richer primitives $\tilde{q}$ and the resulting identified set. Exploiting the relation between $\tilde{q}$ and $q$ given by $q_j(p) = \sum_{e \in \mathcal{E}} \tilde{q}_j(\tilde{p}(p,e),e)$, the data restrictions in \eqref{O-eq:q_data} correspond to
\begin{align}
 \sum_{e \in \mathcal{E}} \tilde{q}_j(\tilde{p}(p,e),e) & = P(D = j | P = p)~, \label{eq:q_tilde_data}
\end{align}
for $j \in \mathcal{J},~p \in \mathcal{P}_{\text{obs}}$; the logical restrictions that the richer version of demand must be positive and integrate out to one as in \eqref{O-eq:q_logical1}-\eqref{O-eq:q_logical2} correspond to
\begin{align}
 \tilde{q}(p,e) \geq 0~ \label{eq:q_tilde_ligical_1}
\end{align}
for $j \in \mathcal{J}$, $e \in \mathcal{E}$, $p \in \mathcal{P}$, and
\begin{align}
 \sum_{e \in \mathcal{E}} \sum_{j \in \mathcal{J}} \tilde{q}_j(p_e,e) =  1 \label{eq:q_tilde_logical_2}
\end{align}
where $p_e \in \mathcal{P}$ for each $e \in \mathcal{E}$, respectively; the shape restrictions capturing $\tilde{U}_{ij}$ being increasing as in \eqref{O-eq:q_shape} correspond to
\begin{align}
 \tilde{q}_j(p,e) \geq \tilde{q}_j(p',e) \label{eq:q_tilde_shape_1}
\end{align}
for $e \in \mathcal{E}$, $p,p' \in \mathcal{P}$ and $j \in \mathcal{J} \setminus \mathcal{J}'$ such that $p_j \geq p'_j$ for $j \in \mathcal{J}'$ and $p_j = p'_j$ for $j \in \mathcal{J} \setminus \mathcal{J}'$; and, finally, the additional restrictions introduced when invoking Assumption \ref{O-ass:LC} that takes demand for certain alternatives to be zero when their price is greater than $r$, which we take to be given by
\begin{align}
 \tilde{q}_j(p,e) = 0 \label{eq:q_tilde_shape_2}
\end{align}
for $p \in \mathcal{P}$ such that $p_j \geq r$, and $j \in \mathcal{J}_1$ and $e \in \mathcal{E}$ such that $p'_j > e$ for some $p' \in \mathcal{P}_{\text{obs}} \cup \{p^a,p^b\}$, i.e. when $j \notin C_i(p')$ for some $p' \in \mathcal{P}_{\text{obs}} \cup \{p^a,p^b\}$, which observe is when Assumption \ref{O-ass:LC} is invoked in our analysis to replace unaffordability in terms of prices by taking it to be equal to $r$ and obtain the expression in \eqref{O-eq:theta_q_liquidity} and the data restrictions in \eqref{eq:q_tilde_data}.\footnote{Note that we do not take \eqref{eq:q_tilde_shape_2} to hold for all $j \in \mathcal{J}_1$, but only if the alternative is unaffordable at certain relevant prices to ensure that no additional restrictions are imposed if it is affordable and so that the analysis in this case remains equivalent to our analysis without liquidity constraints.} The admissible space of demand functions in this case then corresponds to
\begin{align}
 \tilde{\mathbf{Q}}_B = \left\{ \tilde{q} \in \tilde{\bar{\mathbf{F}}} : \tilde{q} \text{ satisfies }   \eqref{eq:q_tilde_data} - \eqref{eq:q_tilde_shape_2} \right\}
\end{align}
where $\tilde{\bar{\mathbf{F}}}$ denotes the set of all functions from $\mathcal{P} \times \mathcal{E}$ to $\mathbf{R}^{|\mathcal{J}|}$, and in turn, as in \eqref{O-eq:theta_identifiedset}, the identified set is given by $\tilde{\theta}(\tilde{\mathbf{Q}}_B)$.

To compute the identified set, the main idea remains the same as that in Section \ref{O-sec:compute_identifiedset}, but performs the analysis conditional on each $e \in \mathcal{E}$ to capture the fact that we have a richer primitive $\tilde{q}$ that accounts for the individual's available income $e$. Specifically, for each $e \in \mathcal{E}$, we can construct a partition $\tilde{\mathcal{W}}(e)$ as in \eqref{O-eq:set_W} using the set of prices in \eqref{O-eq:set_P} given $e \in \mathcal{E}$, i.e.
 
\begin{align}\label{eq:set_P_tilde}
\left\{\tilde{\mathcal{P}}_l^{a,b}(e) : 1 \leq l \leq |\mathcal{J}|-1 \right\} \cup \left\{\{\tilde{p}(p,e)\} : p \in \left\{p^a,p^b\right\} \cup \mathcal{P}_{\text{obs}} \right\}
\end{align}
with $\tilde{\mathcal{P}}_l^{a,b}(e) = \{ p \in \mathcal{P} : p_j = \min \{\tilde{p}_j(p^a,e), \tilde{p}_j(p^b,e) + t\} \text{ for } t \in (\tilde{\Delta}_l^{a,b}(e),\tilde{\Delta}_{l+1}^{a,b}(e)),~ j \in \mathcal{J}  \}$, and then consider the finite-dimensional space of functions given by
\begin{align*}
 \tilde{\mathbf{Q}}_B^{\text{fd}} = \left\{q \in \tilde{\mathbf{Q}}_B : \tilde{q}_j(p,e) = \sum_{w \in \tilde{\mathcal{W}}(e)} 1_w(p) \cdot  \tilde{\beta}_{j}(w,e) \text{ for some } \{\tilde{\beta}_{j}(w,e)\}_{w \in \tilde{\mathcal{W}}} \text{ for each } j \in \mathcal{J},~e \in \mathcal{E} \right\}~,
\end{align*}
where $\{\tilde{\beta}_j(w,e) : w \in \tilde{\mathcal{W}}(e),~e \in \mathcal{E},~j \in \mathcal{J}\}$ are unknown parameters that parameterizes the space here. As in Section \ref{O-sec:equi_charac}, it can be similarly shown that this space leads to no loss of information, i.e. $\tilde{\theta}(\tilde{\mathbf{Q}}_B) = \tilde{\theta}(\tilde{\mathbf{Q}}_B^{\text{fd}})$, and that the identified set can be computed by solving two analogous versions of the finite-dimensional linear programs in \eqref{O-eq:bounds_opt}, where the objective and restrictions correspond to \eqref{O-eq:theta_q_liquidity} and \eqref{eq:q_tilde_data}-\eqref{eq:q_tilde_shape_2} written in terms of $\tilde{\beta}$, respectively. 

Moreover, as in Section \ref{O-sec:dim_red}, we can also consider sub-programs based on sub-vectors of $\tilde{\beta}$ that are defined over subsets of $\tilde{\mathcal{W}}(e)$ used to define the parameter for each $e \in \mathcal{E}$, i.e.
\begin{align*}
\tilde{\mathcal{W}}^{\text{r}}(e) = \left\{ w \in \tilde{\mathcal{W}}(e) : w = \prod_{j \in \mathcal{J}} v_{[j]} \text{ for some } v \in \tilde{\mathcal{V}}(e) \right\}~,
\end{align*}
where $\tilde{\mathcal{V}}(e)$ denotes the partition of the union of the sets in \eqref{eq:set_P_tilde} satisfying Definition \ref{O-def:V}, and separability assumptions such as
\begin{align}\label{eq:q_AS_liquidity}
 \tilde{q}_j(p,e) = \sum_{m \in \mathcal{J}} \tilde{h}_{j}(p_m,e)
\end{align}
for each $j \in \mathcal{J}$ and $e \in \mathcal{E}$ for some unknown functions $\{\tilde{h}_{jm} : m \in \mathcal{J}\}$, i.e. requiring demand to be additively separable in prices of all alternatives. As in Section \ref{O-sec:parametric}, we can also impose parametric restrictions such as
\begin{align}\label{eq:q_para_liquidity}
 \tilde{q}_j(p,e) = \sum_{k=0}^{K_j} \tilde{\alpha}_{jk}(e) \cdot b_{jk}(p)
\end{align}
for each $j \in \mathcal{J}$ and $e \in \mathcal{E}$ for some unknown parameters $\{\tilde{\alpha}_{jk}(e) : 0 \leq k \leq K_j\}$ and known functions $\{b_{jk} : 0 \leq k \leq K_j\}$, i.e. require demand to be a linear function of some basis of prices. In particular, observe that this corresponds to the analyses in Sections \ref{sec:separability} and \ref{sec:parametric_ext} but for each value of $e \in \mathcal{E}$. In turn, we can analogously apply the arguments from these sections for each value of $e \in \mathcal{E}$.

To conclude this section, note that the above analysis assumed that the support of available income, $\mathcal{E}$, was discrete. Our analysis, however, can be extended to also allow for it to be continuous. In particular, we can consider a finite partition $\mathcal{F}$ of $\mathcal{E}$ such that the prices $\tilde{p}(e,p)$ for $p \in \{p^a,p^b\} \cup \mathcal{P}_{\text{obs}}$ stay constant for values of $e \in f$ for each $f \in \mathcal{F}$. For example, denoting by $e^{a,b}_{1} \leq \ldots \leq e^{a,b}_{\tilde{M}}$ the order values of $\{p_j : p \in \{p^a,p^b\} \cup \mathcal{P}_{\text{obs}},~j\in \mathcal{J}\} \cup \{-\infty,\infty\}$, we can consider
\begin{align*}
\mathcal{F} = \{(e^{a,b}_{1}, e^{a,b}_{2}), [e^{a,b}_{2} , e^{a,b}_{3}) , \ldots,  [e^{a,b}_{\tilde{M}-1},e^{a,b}_{\tilde{M}})\}~.
\end{align*}
As the value of income in elements of this partition face the same price, it is rich enough to equivalently define the data restrictions as well as the parameter of interest and in turn, similar to the arguments in Section \ref{O-sec:compute_identifiedset}, capture the same information as that of the underlying continuous income variable. We can then proceed as above by performing the analysis for each value of $f \in \mathcal{F}$. While this allows for a continuous support, note that in practice the cardinality of $\mathcal{F}$ can be quite large and hence make the resulting optimization problem very high dimensional and difficult to compute---this arises in our application as $|\mathcal{J}|$ is large. This is why in our empirical analysis we instead focus on a discretized version of $\mathcal{E}$, where we can choose it to have a low cardinality so that the optimization problems are tractable.

\section{Statistical Inference}\label{sec:inference}

In this section, we describe how we construct confidence intervals for our parameters in our empirical analysis in Section \ref{O-sec:empirical} using the bootstrap procedure from \cite{bugni/etal:17}. To this end, let
\begin{align}\label{eq:data}
\{ (D_i,P_i) : 1 \leq i \leq N\}
\end{align}
denote our sample of $N$ observations, assumed to be independently and identically distributed, on which our statistical tests are based.

Recall that our parameters of interest are generally bounded across our various specifications, where the lower and upper bounds are given by minimization and maximization problems, respectively. We construct confidence intervals such that each point in these bounds lies in the interval with probability at least $(1-\alpha)$ for some pre-specified value of $\alpha \in (0,1)$. In order to describe the common procedure that we use across all the parameters and specifications, it is useful to first define the common structure present in all these cases. To this end, note that each point in the bounds across these cases can be written as $c’\pi$ for some vector $\pi \in \mathbf{R}^{d_\pi}$ of dimension $d_\pi$ that satisfies
\begin{align}
A_1\pi &= b_1~, \label{eq:xR1}\\
A_2\pi &\leq b_2~, \label{eq:xR2}
\end{align}
where $\pi$ corresponds to the optimizing variable in the minimization and maximization problems, $c$ corresponds to the vector defining the objective in these problems that depends on the choice of parameter, \eqref{eq:xR1} capture the restrictions imposed on the optimizing variable by the observed shares through \eqref{O-eq:q_data}, and \eqref{eq:xR2} capture the restrictions imposed by the shape restrictions in \eqref{O-eq:q_logical1}-\eqref{O-eq:q_shape}. Note that across these cases the values $c$, $A_1$, $A_2$ and $b_2$ are known and deterministic, and only $b_1$ needs to be estimated as it corresponds to the observed enrollment shares.

We construct confidence intervals for the various parameters across the various specifications by test inversion. In particular, we test the null hypothesis at level $\alpha$ that there exists a $\pi$ satisfying the restrictions in \eqref{eq:xR1} and \eqref{eq:xR2} such that $c’\pi = \theta_0$ for some given value of $\theta_0 \in \mathbf{R}$. Confidence intervals are then constructed by collecting the values of $\theta_0$ that are not rejected.

We test this null hypothesis using a bootstrap procedure from \cite{bugni/etal:17}, who show it can have several desirable theoretical properties that account for the fact that the parameter of interest is generally partially identified. Our above setup can be mapped into their general framework, given which the test procedure can be described in the following steps:
\begin{enumerate}
\item Compute the test statistic
\begin{align*}
 TS_N(\theta_0) = N \cdot \min_{\pi} (A_1\pi - \hat{b}_1)’\hat{\Sigma}^{-1} (A_1\pi - \hat{b}_1)
\end{align*}
subject to $\pi$ satisfying $c’\pi = \theta_0$ and the restrictions in \eqref{eq:xR2}, where $\hat{b}_1$ corresponds to the empirical counterpart of $b_1$ using the data in \eqref{eq:data} and $\hat{\Sigma}$ corresponds to a diagonal matrix consisting of the sample variances of the entries of $\hat{b}_1$.

\item For $l=1, \ldots, B$, compute the so-called minimum resampling bootstrap test statistics of the form
\begin{align*}
 TS^{MR}_{l,N}(\theta_0) = \min\left\{ TS_{l,N}^{DR}(\theta_0),TS_{l,N}^{PR}(\theta_0) \right\}
\end{align*}
where
\begin{align*}
 TS_{l,N}^{DR}(\theta_0) = N \cdot (\hat{b}_1 - \hat{b}_{l,1})' \hat{\Sigma}^{-1} (\hat{b}_1 - \hat{b}_{l,1})
\end{align*}
and
\begin{align*}
 TS_{l,N}^{PR}(\theta_0) = N \cdot \min_{\pi}\left(\hat{b}_1 - \hat{b}_{l,1} + \frac{1}{\kappa_N}(A_1\pi - \hat{b}_1)\right)' \hat{\Sigma}^{-1} \left(\hat{b}_1 - \hat{b}_{l,1}+ \frac{1}{\kappa_N}(A_1\pi - \hat{b}_1)\right)
\end{align*}
subject to $\pi$ satisfying $c’\pi = \theta_0$ and the restrictions in \eqref{eq:xR2}, and $\hat{b}_{l,1}$ corresponds to the analog of $\hat{b}_1$ using the $l$th boostrap sample drawn with replacement from the data in \eqref{eq:data}. Here note that $\kappa_N$ is a tuning parameter that satisfies $\kappa_N \to \infty$ and $\kappa_N / \sqrt{N} \to 0$ as $N \to \infty$. For our empirical results, following \cite{bugni/etal:17}, we take $\kappa_N = \sqrt{\text{ln} N}$.

\item Compute the critical value of the test $\hat{c}(1-\alpha,\theta_0)$ by taking the $(1-\alpha)$-quantile of the distribution of computed bootstrap test statistics. The test procedure is given by $\phi_N(\theta_0) = 1\{TS_N(\theta_0) > \hat{c}(1-\alpha,\theta_0)\}$, i.e we reject if the test statistic is greater than the critical value.

\end{enumerate}
Given the above test procedure for a given value of $\theta_0$, we can then use it construct confidence intervals by collecting the set all points we don’t reject, namely $\{\theta_0 \in \mathbf{R} : \phi_N(\theta_0) = 0 \}$.

\section{Additional Details on Empirical Analysis}\label{sec:add_details_emp}

\subsection{Data Construction}\label{sec:data_construction}

In this section, we describe how we construct the data used in our empirical analysis in Section \ref{O-sec:empirical}. The original data sample comes from the replication files for the evaluation of OSP, which are available from the US Department of Education \citep{wolf/etal:10}. Recall that our analysis focuses on the initial school choice for students who entered the experiment in 2005.  Beginning with this subsample, we make the following data-cleaning choices to reach our final analysis data.

Our analysis requires only the prices (as measured by the tuition) of the participating private schools and the school choices of the students (to compute their enrollment shares). For all participating private schools, we observe tuition in either the first or second year of the study, but not necessarily both. If we observe tuition only in the first year, we assume that it was unchanged between the first and second year (recall we use only the second year of data). For school choices, we have that they are missing for around 36\% of the students in our data. By a fortunate quirk of the research design, however, participating private schools reported all voucher students to the researchers. Unobserved school choices must therefore be either in non-participating private schools, or government-funded schools. For these students, we assume they enroll in these two groups at the same relative rate that students with observed choices enroll in non-participating private and government-funded schools. Once we obtain these school choices, we weight these observed choices using the baseline weights of the original evaluation---see \citet[Appendix A.7]{wolf/etal:10} for details on how these weights were constructed.

\subsection{Summary Statistics on School Setting}\label{sec:summstat_setting}

\begin{table}[!t]\centering 
\def\sym#1{\ifmmode^{#1}\else\(^{#1}\)\fi} 
\caption{Student and family characteristics by voucher receipt}\label{tab:student-chars} 
\scalebox{.85}{ 
\begin{tabular*}{.9\hsize}{@{\hskip\tabcolsep\extracolsep\fill}l*{1}{ccc}} 
\toprule 
                                &With voucher &  Without voucher &Difference         \\
\midrule
Mother married (=1)             &     0.52&     0.55&   -0.034         \\
Mother years education          &    12.20&    12.25&   -0.057         \\
Mother works full time (=1)     &     0.35&     0.38&   -0.028         \\
Mother works part time (=1)     &     0.11&     0.11&    0.007         \\
Family income (\$)              & 16,725& 17,372& -647         \\
HH receives govt transfers (=1) &     0.03&     0.01&    0.016 \\
Household size                  &     4.11&     4.14&   -0.030         \\
Black (=1)                      &     1.00&     1.00&   -0.001         \\
Male (=1)                       &     0.50&     0.47&    0.030         \\
Grade \(\leq\) 5 (=1)           &     0.65&     0.65&    0.000         \\
Grade 6-8 (=1)                  &     0.21&     0.21&   -0.000         \\
Grade \(\geq\) 9 (=1)           &     0.14&     0.14&    0.000         \\
Child learning disabilities (=1)&     0.09&     0.09&   -0.004         \\
\midrule
Observations                    &    1,090&      730&                  \\
\bottomrule \multicolumn{4}{p{.86\hsize}}{\footnotesize 
 Table shows mean student and family characteristics by treatment group, weighted using the baseline weights.  Observations rounded to the nearest 10.}\\ 
\multicolumn{4}{p{.88\hsize}}{\footnotesize SOURCE:  \textit{Evaluation of the DC Opportunity Scholarship Program: Final Report (NCEE 2010-4018)}, U.S. Department of Education, National Center for Education Statistics previously unpublished tabulations.  }\\ 
 \end{tabular*}} \end{table}

In this section, we describe additional information on the students and schools in our analysis data. Table \ref{tab:student-chars} reports mean characteristics of students and their families. Only families making less than 185\% of the federal poverty line were eligible for the program, and so unsurprisingly the students are relatively disadvantaged. Approximately 50\% of the students' mothers were married, and fewer than 50\% were employed at baseline. Family income was slightly less than \$17,000. Baseline achievement reflects both positive and negative selection: families chose to participate in the experiment, but they also had to be relatively poor to qualify. The table also reveals that voucher recipients and non-recipients are balanced in terms of the various predetermined characteristics. This suggests that the receipt of the voucher was random, in line with Assumption \ref{O-ass:E}.

\begin{table}[!t]\centering
\def\sym#1{\ifmmode^{#1}\else\(^{#1}\)\fi}
\caption{Characteristics of sample schools\label{tab:school-chars}}
\scalebox{.85}{ \begin{tabular*}{.87\hsize}{@{\hskip\tabcolsep\extracolsep\fill}l*{1}{ccc}}
\toprule
                                &  Private &   Government-funded & Difference         \\
\midrule
\multicolumn{4}{c}{\emph{Panel A: Unweighted characteristics}} \\
Share minority                  &     0.73&     0.96&   -0.227\\

School size                     &   222.97&   325.90& -102.924\\
Student/teacher ratio           &     8.92&    12.82&   -3.897\\
Catholic (=1)                   &     0.35&     0.00&    0.354\\
Other religious (=1)            &     0.20&     0.00&    0.200\\
Secular (=1)                    &     0.45&     1.00&   -0.554\\
Gifted program (=1)             &     0.35&     0.39&   -0.040         \\
Learning difficulties program (=1)&     0.48&     0.93&   -0.447\\
Individual tutors available (=1)&     0.64&     0.69&   -0.052         \\
Students tracked by ability (=1)&     0.79&     0.60&    0.192\\
Remedial classes available (=1) &     0.61&     0.68&   -0.070         \\
\addlinespace
\multicolumn{4}{c}{\emph{Panel B: Attendance-weighted characteristics}} \\
Share minority                  &     0.96&     0.98&   -0.017         \\
School size                     &   205.86&   419.46& -213.605\\
Student/teacher ratio           &    13.17&    13.72&   -0.551         \\
Catholic (=1)                   &     0.53&     0.00&    0.534\\
Other religious (=1)            &     0.25&     0.00&    0.252\\
Secular (=1)                    &     0.21&     1.00&   -0.786\\
Gifted program (=1)             &     0.34&     0.34&   -0.000         \\
Learning difficulties program (=1)&     0.45&     0.96&   -0.518 \\
Individual tutors available (=1)&     0.80&     0.77&    0.026         \\
Students tracked by ability (=1)&     0.70&     0.55&    0.157         \\
Remedial classes available (=1) &     0.68&     0.73&   -0.048         \\
\midrule
Observations                    &       60&      160&                  \\
\bottomrule \multicolumn{4}{p{.85\hsize}}{\footnotesize 
 Displays school characteristics for private and government-run schools. We do not break out the private schools by participation status because the non-participating schools almost never responded. Observations rounded to the nearest 10.}\\ 
\multicolumn{4}{p{.85\hsize}}{\footnotesize SOURCE:  \textit{Evaluation of the DC Opportunity Scholarship Program: Final Report (NCEE 2010-4018)}, U.S. Department of Education, National Center for Education Statistics previously unpublished tabulations.  }\\  \end{tabular*}} \end{table}

Table \ref{tab:school-chars} reports characteristics of the private and government-funded schools in the sample, both unweighted and weighted by attendance. Panel A reveals that the private schools are substantially whiter, have smaller student/teacher ratios, and are more likely to track students by ability. Most strikingly, many of the private schools are religious---35\% of them are Catholic, and an additional 20\% another religion. In addition, private schools tend to have lower share of minorities, lower share of student/teacher ratio, lower school sizes, have more students tracked by ability and have lower learning difficulties program. Comparing Panel A and Panel B reveals that among the schools that students actually attended (as reported in the attendance-weighted results in Panel B), there are smaller differences between private and government-funded schools. For example, while the average private school is only 73\% minority (relative to 96\% for the government schools), the average private school attended by a voucher student is 96\% minority.

\section{Proofs}\label{sec:proof}

\subsection{Proof of Proposition \ref{O-prop:EB}}\label{sec:proof_EB}

The proof of this proposition follows from \citet[][Proposition 1]{bhattacharya:18} and \citet[][Theorem 1]{bhattacharya:18}. We reproduce these proofs here in the context of our setup and notation for completeness. For convenience, we drop the $i$ sub-index here.

To see why the variable $B^{a,b}$ exists and is unique, note first that the right hand side of \eqref{O-eq:B_t} is continuous in $B^{a,b}$ as $U_j$ is a continuous function for each $j \in \mathcal{J}$. In addition, since $Y - p_j^a > Y - p_j^b - B^{a,b}$ and $Y - p_j^a  < Y - p_j^b - B^{a,b}$ for $j \in \mathcal{J}$ when $B^{a,b} > \Delta^{a,b}_{|\mathcal{J}|}$ and $B^{a,b} < \Delta^{a,b}_{1}$, respectively, note that it follows from the fact that $U_j$ is strictly increasing for each $j \in \mathcal{J}$ that if $B^{a,b} > \Delta^{a,b}_{|\mathcal{J}|}$ then the right hand side of \eqref{O-eq:B_t} is strictly smaller than its left hand side, whereas if $B^{a,b} < \Delta^{a,b}_{1}$ then the right hand side will be strictly greater than the left hand side. From these two points, it then follows by the intermediate value theorem that there exists a $B^{a,b} \in [\Delta^{a,b}_{1},\Delta^{a,b}_{|\mathcal{J}|}]$ such that the right hand side equals the left hand side, i.e. a solution to \eqref{O-eq:B_t} exists. Furthermore, given that $U_j$ is strictly increasing for each $j \in \mathcal{J}$, it follows that the solution must be unique.

To see why the average value of $B^{a,b}$ is given by \eqref{O-eq:AB_t_q}, note from above that $B^{a,b} \in [\Delta^{a,b}_{1},\Delta^{a,b}_{|\mathcal{J}|}]$ and hence that $\text{Prob}[B^{a,b} \leq t] = 0$ for $t < \Delta^{a,b}_{1}$ and $\text{Prob}[B^{a,b} \leq t] = 1$ for $t \geq \Delta^{a,b}_{|\mathcal{J}|}$. To calculate this probability for $t \in [\Delta^{a,b}_{1},\Delta^{a,b}_{|\mathcal{J}|})$, note that since $U_j$ is strictly increasing for each $j \in \mathcal{J}$, we have that $B^{a,b} \leq t$ is equivalent to
\begin{gather*}
\max_{j \in \mathcal{J}} U_{j}\left(Y - p_{j}^a\right) \geq \max_{j \in \mathcal{J}} U_{j}\left(Y - p_{j}^b - t\right) ~.
\end{gather*}
It follows that for $t \in [\Delta^{a,b}_l,\Delta^{a,b}_{l+1})$ for $l=1,\ldots,|\mathcal{J}|-1$, we have
\begin{align*}
\text{Prob}[B^{a,b} \leq t] =&  \sum_{j \in \mathcal{J}} \text{Prob}\left[ U_j(Y - p^a_j) \geq \max \left\{ \max_{m \in \mathcal{J} \setminus \{j\}}U_m(Y - p^{a}_m),~ \max_{m \in \mathcal{J}} U_m(Y - p^b_m - t) \right\}  \right]  \\
=&  \sum_{j \in \mathcal{J} \setminus \mathcal{J}^{a,b}_{l+1}} \text{Prob}\left[ U_j(Y - p^a_j) \geq \max \begin{Bmatrix}
\max\limits_{m \in \mathcal{J} \setminus \mathcal{J}^{a,b}_{l+1} \cup \{j\}}U_m(Y - p^{a}_m) , \\[0.25cm]
\max\limits_{m \in \mathcal{J}^{a,b}_{l+1}} U_m(Y - p^b_m - t)
\end{Bmatrix}  \right]  \\       
=& \sum\limits_{j \in \mathcal{J} \setminus \mathcal{J}^{a,b}_{l+1}} q_j\left(\min\{p^a,p^b + t\}\right)~,
\end{align*}
where the second equality follows from the fact that $U_j$ is strictly increasing for each $j \in \mathcal{J}$ along with $Y - p_j^b - t \geq Y - p_j^a$ for $j \in \mathcal{J}^{a,b}_{l+1}$ and $Y - p_j^b - t \leq Y - p_j^a$ for $j \in \mathcal{J} \setminus \mathcal{J}^{a,b}_{l+1}$, and the final equality follows from the definition of the average demand functions along with noting that $\min\{p^a_j,p^{b}_j +t\} = p^b_j + t$ for $j \in \mathcal{J}^{a,b}_{l+1}$ and $\min\{p^a_j,p^{b}_j +t\} = p^{a}_j$ for $j \in \mathcal{J} \setminus \mathcal{J}^{a,b}_{l}$. Finally, given that $B^{a,b}$ is a positive random variable, we have that its expectation is given by
\begin{align*}
E[B^{a,b}] = \int_{0}^{\infty} \left[1 - \text{Prob}[B^{a,b} \leq t] \right] dt~,
\end{align*}
it follows from the above characterization of $\text{Prob}[B^{a,b} \leq t]$ along with noting that
\begin{align*}
\sum\limits_{j \in \mathcal{J}_{l+1}^{a,b}} q_j\left(\min\{p^a,p^b + t\}\right) = 1 - \sum\limits_{j \in \mathcal{J} \setminus \mathcal{J}_{l+1}^{a,b}} q_j\left(\min\{p^a,p^b + t\}\right)~~,
\end{align*}
that the average value of $B^{a,b}$ is given by \eqref{O-eq:AB_t_q}. This concludes the proof.

\subsection{Proof of Proposition \ref{O-prop:identifiedset_base_1}}\label{sec:proof_base}

In order to prove the proposition, we need to show that $\Theta \subseteq \Theta_B$ and $\Theta_B \subseteq \Theta$, and that $\Theta_B = [\underline{\theta}_B,\bar{\theta}_B]$ if $\mathbf{B}$ is non-empty. The proof of the first two parts, i.e. $\Theta \subseteq \Theta_B$ and $\Theta_B \subseteq \Theta$, respectively follow directly from the first two parts of Proposition \ref{prop:identifiedset_sep} because $\mathbf{Q}_B$ corresponds to a special case of $\mathbf{Q}_S$ in \eqref{eq:Q_S} when taking $L_j$ to be a singleton set such as $\{j\}$ and $\mathcal{J}_{jl} = \mathcal{J}$ in \eqref{eq:q_sep} for each $j \in \mathcal{J}$. It remains to prove the third part, i.e. $\Theta_B = [\underline{\theta}_B,\bar{\theta}_B]$ if $\mathbf{B}$ is non-empty, as the third part of Proposition \ref{prop:identifiedset_sep} only shows that the closure of the set is a closed interval. Here we additionally show that $\mathbf{B}$ is not only convex but also compact, which implies that image of it through $\theta_B$ is a closed interval with end points given by \eqref{O-eq:bounds_opt}.

In order to show $\mathbf{B}$ is convex and compact, it is useful to rewrite \eqref{eq:s_logical1}-\eqref{eq:s_data} in terms of $\beta$ when taking $L_j$ to be a singleton set such as $\{j\}$ and $\mathcal{J}_{jl} = \mathcal{J}$. In this case, we have that \eqref{eq:s_logical1}-\eqref{eq:s_logical2} corresponds to
\begin{align}
 \beta_{j}(w) &\geq 0~\text{ for each $j \in \mathcal{J}$}~,  \label{eq:b_logical1} \\
 \sum_{j \in \mathcal{J}} \beta_{j}(w) &= 1~ \label{eq:b_logical2}
\end{align}
for each $w \in \mathcal{W}$; \eqref{eq:s_shape} corresponds to
\begin{align}\label{eq:b_shape}
 \beta_{j}(w) \geq \beta_{j}(w')
\end{align}
for each $w,w' \in \mathcal{W}$ and $j \in \mathcal{J} \setminus \mathcal{J}'$ such that $t > t'$ for all $t \in w_{[m]},~t' \in w'_{[m]}$ for $m \in \mathcal{J}' \subseteq \mathcal{J}$ and $w_{[m]} = w'_{[m]}$ for $m \in \mathcal{J} \setminus \mathcal{J}'$; and \eqref{eq:s_data} corresponds to
\begin{align}
\beta_{j}\left(\{p\}\right) = \text{Prob}[D_i = j | P_i = p]~,  \label{eq:b_data}
\end{align}
for each $j \in \mathcal{J}$ and $p \in \mathcal{P}_{\text{obs}}$. Hence, it follows that
\begin{align}\label{eq:B_set}
\mathbf{B} = \left\{ \beta \in \mathbf{R}^{d_{\beta}} : \beta \text{ satisfies } \eqref{eq:b_logical1}-\eqref{eq:b_data}  \right\}~.
\end{align}
Given that $\mathbf{B}$ is determined by linear inequality restrictions and further that each variable is bounded by the restrictions in \eqref{eq:b_logical1}-\eqref{eq:b_logical2}, it follows that it is bounded and compact, which concludes the proof.

\subsection{Proof of Proposition \ref{O-prop:liquidity}}\label{sec:proof_liquidity}

Given Assumption \ref{O-ass:LC}, it follows from \eqref{O-eq:relation_choiceset_price} that the willingness to pay in \eqref{O-eq:B_t_const_LC} can be defined as
\begin{align}\label{eq:B_t_const_LC}
\max_{j \in \mathcal{J}}\tilde{U}_{ij}(Y_{i} - \tilde{p}_j(p^a,E_i)) = \max_{j \in \mathcal{J}}\tilde{U}_{ij}(Y_{i} - \tilde{p}_j(p^b,E_i) - B_i^{a,b})~.
\end{align}
Conditional on $E_i = e \in \mathcal{E}$, it then follows from Proposition \ref{O-prop:EB} as $\tilde{U}_{ij}$ is continuous and strictly increasing for each $j \in \mathcal{J}$ than $B_i^{a,b}$ exists and is unique.

We can similarly apply Proposition \ref{O-prop:EB} to obtain the expression in \eqref{O-eq:AB_t_q_liquidity}. Specifically, for each $e \in \mathcal{E}$, let
\begin{align}\label{eq:q_tilde_cond}
 \tilde{q}_j(p|e) = \tilde{q}_j(p,e) / \tilde{q}(e)
\end{align}
denote the demand for alternative $j \in \mathcal{J}$ under prices $p \in \mathcal{P}$ conditional on individuals having a maximum affordable income of $e$. For each $e \in \mathcal{E}$, we can then apply Proposition \ref{O-prop:EB} to obtain the average willingness to pay conditional on $E_i = e$ given by
\begin{align}\label{eq:B_cond_e}
 E[B_i^{a,b}|E_i = e] = \tilde{\Delta}_1^{a,b}(e) + \sum_{l=1}^{|\mathcal{J}|-1} \sum\limits_{j \in \tilde{\mathcal{J}}_{l+1}^{a,b}(e)} \int\limits_{\tilde{\Delta}_{l}^{a,b}(e)}^{\tilde{\Delta}_{l+1}^{a,b}(e)} \tilde{q}_j\left(\text{min}\{\tilde{p}(p^a,e),\tilde{p}(p^b,e)+t\}|e\right) dt~.
\end{align}
Noting that
\begin{align*}
 E[B_i^{a,b}] = \sum_{e \in \mathcal{E}} \tilde{q}(e) \cdot E[B_i^{a,b}|E_i = e]
\end{align*}
and then plugging in the expression from \eqref{eq:B_cond_e} and using the relation in \eqref{eq:q_tilde_cond} gives the expression in \eqref{O-eq:AB_t_q_liquidity}, which completes the proof.

\subsection{Proof of Proposition \ref{prop:b_shape_alt}}

To see why the restrictions in \eqref{eq:b_shape_alt} imply those in \eqref{eq:b_shape}, consider $w,w' \in \mathcal{W}$ such that $t > t'$ for all $t \in w_{[j]},~t' \in w'_{[j]}$ for $j \in \mathcal{J}' \subseteq \mathcal{J}$ and $w_{[j]} = w'_{[j]}$ for $j \in \mathcal{J} \setminus \mathcal{J}'$. In this case, note that $\mathcal{J}^{>}_{w,w'} = \mathcal{J}'$, $\mathcal{J}^{>}_{w',w} = \emptyset$, and $\mathcal{J}^{=}_{w,w'} = \mathcal{J} \setminus \mathcal{J}'$. Then, taking $\mathcal{J}^{\dagger} = \{j\}$ for each $j \in \mathcal{J}^{=}_{w,w'}$ in \eqref{eq:b_shape_alt} implies that \eqref{eq:b_shape} holds for each $j \in \mathcal{J}'$. To see why the restrictions in \eqref{eq:b_shape} imply those in \eqref{eq:b_shape_alt}, consider $w,w' \in \mathcal{W}$ as well as a $w'' \in \mathcal{W}$ such that we have $w''_{[j]} = w'_{[j]} = w_{[j]}$ for $j \in \mathcal{J} \setminus \left(  \mathcal{J}^{>}_{w,w'}  \cup  \mathcal{J}^{>}_{w',w}  \right)$, $w''_{[j]} = w_{[j]}$ for $j \in  \mathcal{J}^{>}_{w,w'}$, and $w''_{[j]} = w'_{[j]}$ for $j \in  \mathcal{J}^{>}_{w',w}$. Since this implies that $t'' > t$ for all $t \in w_{[j]}$, $t'' \in w''_{[j]}$ for $j \in \mathcal{J}^{>}_{w',w}$ and $w_{[j]}=w''_{[j]}$ for $j \in \mathcal{J} \setminus  \mathcal{J}^{>}_{w’,w}$, it follows from \eqref{eq:b_shape} that
\begin{align}\label{eq:b_shape_alt_1}
\beta_j(w'') \geq \beta_j(w)
\end{align}
for each $j \in \mathcal{J}^{=}_{w,w'} \cup  \mathcal{J}^{>}_{w,w'}$. Similarly, since it also implies that $t'' > t'$ for all $t \in w'_{[j]}$, $t'' \in w''_{[j]}$ for $j \in \mathcal{J}^{>}_{w,w'}$ and $w'_{[j]}=w''_{[j]}$ for $j \in \mathcal{J} \setminus \mathcal{J}^{>}_{w,w’}$, it also follows from \eqref{eq:b_shape} that
\begin{align}\label{eq:b_shape_alt_2}
\beta_j(w'') \geq \beta_j(w')
\end{align}
for each $j \in \mathcal{J}^{=}_{w,w'} \cup  \mathcal{J}^{>}_{w',w}$. Then, for each $\mathcal{J}^{\dagger} \subseteq \mathcal{J}^{=}_{w,w'}$, this implies that \eqref{eq:b_shape_alt} holds as
\begin{align*}
\sum_{j \in \mathcal{J}^{>}_{w',w} \cup \mathcal{J}^{\dagger}} \beta_{j}(w') \leq \sum_{j \in \mathcal{J}^{>}_{w',w} \cup \mathcal{J}^{\dagger}} \beta_{j}(w'')
&= 1 - \sum_{j \in \mathcal{J} \setminus \left( \mathcal{J}^{>}_{w',w} \cup \mathcal{J}^{\dagger} \right)} \beta_{j}(w'') \\
&\leq 1 - \sum_{j \in \mathcal{J} \setminus \left( \mathcal{J}^{>}_{w',w} \cup \mathcal{J}^{\dagger} \right)} \beta_{j}(w)
= \sum_{j \in \mathcal{J}^{>}_{w',w} \cup \mathcal{J}^{\dagger}} \beta_{j}(w)
\end{align*}
where the first inequality follows from \eqref{eq:b_shape_alt_2}, the second equality follows from \eqref{eq:b_logical2}, the third inequality from \eqref{eq:b_shape_alt_1}, and the final equality from \eqref{eq:b_logical2}.

\subsection{Proof of Proposition \ref{prop:identifiedset_sep}}\label{sec:proof_sep}

In order to prove the proposition, we need to show that $\Theta \subseteq \Theta_S$ and $\Theta_S \subseteq \Theta$, and that $\text{closure}(\Theta_S) = [\underline{\theta}_S,\bar{\theta}_S]$ if $\mathbf{S}$ is non-empty. Below, we divide the proof into three parts respectively showing each of these statements. First, we show $\Theta \subseteq \Theta_S$, i.e for every $\theta_0 \in \Theta$ there exists a $\psi \in \mathbf{S}$ such that $\theta_S(\psi) = \theta_0$. Second, we show $\Theta_S \subseteq \Theta$, i.e. for every $\theta_0 \in \Theta_S$ there exists a $q \in \mathbf{Q}_S$ such that $\theta(q) = \theta_0$. Third, we show that if $\mathbf{S}$ is non-empty then it its closure is given by $\Theta_S = [\underline{\theta}_S,\bar{\theta}_S]$~.

Before proceeding, it is useful to first explicitly state the restrictions on $\psi$ that characterize $\mathbf{S}$ as well as the expression for $\theta_S$. To this end, note that $\mathbf{S}$ corresponds to all $\psi$ such that the corresponding $q$ determined by \eqref{eq:Q_S_fd} satisfy \eqref{O-eq:q_logical1}-\eqref{O-eq:q_logical2}, \eqref{O-eq:q_shape} and \eqref{O-eq:q_data}. Given that \eqref{eq:Q_S_fd} requires $q$ to be a constant valued function over $w \in \mathcal{W}$, it follows that \eqref{O-eq:q_logical1}-\eqref{O-eq:q_logical2}, \eqref{O-eq:q_shape} and \eqref{O-eq:q_data} need to be satisfied across only values in $\mathcal{W}$. In particular, observe that  \eqref{O-eq:q_logical1} and \eqref{O-eq:q_logical2} equivalently correspond to
\begin{align}
 \sum_{l \in \mathcal{L}_j} \psi_{jl}(w_{[\mathcal{J}_{jl}]}) &\geq 0~\text{ for each $j \in \mathcal{J}$}~,  \label{eq:s_logical1} \\
 \sum_{j \in \mathcal{J}} \sum_{l \in \mathcal{L}_j} \psi_{jl}(w_{[\mathcal{J}_{jl}]}) &= 1~ \label{eq:s_logical2}
\end{align}
for each $w \in \mathcal{W}$, and, given the way $\mathcal{W}$ was constructed, \eqref{O-eq:q_shape} corresponds to
\begin{align}\label{eq:s_shape}
 \sum_{l \in \mathcal{L}_j} \psi_{jl}(w_{[\mathcal{J}_{jl}]}) \geq \sum_{l \in \mathcal{L}_j} \psi_{jl}(w'_{[\mathcal{J}_{jl}]})
\end{align}
for each $w,w' \in \mathcal{W}$ and $j \in \mathcal{J} \setminus \mathcal{J}'$ such that $t > t'$ for all $t \in w_{[m]},~t' \in w'_{[m]}$ for $m \in \mathcal{J}' \subseteq \mathcal{J}$ and $w_{[m]} = w'_{[m]}$ for $m \in \mathcal{J} \setminus \mathcal{J}'$. Similarly, since $\{p\} \in \mathcal{W}$ for each $p \in \mathcal{P}_{\text{obs}}$ given the way $\mathcal{W}$ was constructed, observe that \eqref{O-eq:q_data} corresponds to
\begin{align}
\sum_{l \in \mathcal{L}_j} \psi_{jl}\left(\{p_{[\mathcal{J}_{jl}]}\}\right) = \text{Prob}\left[D_i = j | P_i = p\right]~, \label{eq:s_data}
\end{align}
for each $j \in \mathcal{J}$ and $p \in \mathcal{P}_{\text{obs}}$. Then, it follows $\mathbf{S}$ can be written as
\begin{align}\label{eq:S_set}
\mathbf{S} = \left\{ \psi \in \mathbf{R}^{d_{\psi}} : \psi \text{ satisfies } \eqref{eq:s_logical1}-\eqref{eq:s_data}  \right\}~.
\end{align}
For the parameter of interest, observe that \eqref{O-eq:AB_t_q} can be written in terms of $\psi$ as
\begin{align}
E[B_i^{a,b}] &= \Delta_1^{a,b} + \sum_{l=1}^{|\mathcal{J}|-1} \sum\limits_{j \in \mathcal{J}_{l+1}^{a,b}} \int\limits_{\Delta_{l}^{a,b}}^{\Delta_{l+1}^{a,b}} q_j\left(\text{min}\{p^a,p^b+t\}\right) dt~, \\
&= \Delta_1^{a,b} + \sum_{l=1}^{|\mathcal{J}|-1} \sum\limits_{j \in \mathcal{J}_{l+1}^{a,b}} \sum_{v \in \mathcal{V}^{a,b}_l} \int\limits_{\underline{t}_{v}}^{\bar{t}_{v}} q_j\left(\text{min}\{p^a,p^b+t\}\right) dt~, \label{eq:AB_t_2}\\
&= \Delta_1^{a,b} + \sum_{l=1}^{|\mathcal{J}|-1} \sum\limits_{j \in \mathcal{J}_{l+1}^{a,b}} \sum_{v \in \mathcal{V}^{a,b}_l} (\bar{t}_v - \underline{t}_v) \sum_{l \in \mathcal{L}_j} \psi_{jl}(w(v)_{[\mathcal{J}_{jl}]})
\end{align}
where $\mathcal{V}_l^{a,b} \subseteq \mathcal{V}$ is such that $\bigcup\limits_{v \in \mathcal{V}_l^{a,b}} v = \mathcal{P}_l^{a,b}$, which exists given Definition \ref{O-def:V}(i), and $w(v) \equiv \prod\limits_{j \in \mathcal{J}}v_{[j]} \in \mathcal{W}$ given $v \in \mathcal{V}$. In particular, the first line simply recalls \eqref{O-eq:AB_t_q}, the second line follows from rewriting it in terms of the collection of sets $\mathcal{V}$ and exploiting the fact that given the form of $\mathcal{P}_l^{a,b}$, we have that, for $v \in \mathcal{V}_l^{a,b}$, $v = \{p \in \mathcal{P} : p = \min\{p^a,p^a + t\} \text{ for } t \in (\underline{t}_v,\bar{t}_v) \text{ or } t \in [\underline{t}_v,\bar{t}_v) \text{ or } t \in (\underline{t}_v,\bar{t}_v]\}$ for some values of $\underline{t}_v$ and $\bar{t}_v$, and the third line then directly follows from substituting in the equation from \eqref{eq:Q_S_fd}. In turn, along with the fact that $\{p^b\},\{p^a\} \in \mathcal{W}$ given how $\mathcal{W}$ was constructed, observe that we have that \eqref{O-eq:theta_q} can be written in terms of $\psi$ as follows
\begin{align}
\theta_S(\psi) = &~g^{a,b} \Delta^{a,b}_1 + g^{a,b}\sum_{l=1}^{|\mathcal{J}|-1} \sum\limits_{j \in \mathcal{J}_{l+1}^{a,b}} \sum_{v \in \mathcal{V}^{a,b}_l} (\bar{t}_v - \underline{t}_v) \sum_{l \in \mathcal{L}_j} \psi_{jl}(w(v)_{[\mathcal{J}_{jl}]}) \nonumber \\
&~+ \sum_{j \in \mathcal{J}} \left(g^a_j \sum_{l \in \mathcal{L}_j} \psi_{jl}\left(\{p^a_{[\mathcal{J}_{jl}]}\}\right)+ g^b_j \sum_{l \in \mathcal{L}_j} \psi_{jl}\left(\{p^b_{[\mathcal{J}_{jl}]}\}\right) \right)~. \label{eq:theta_S}
\end{align}

Given the explicit characterizations of $\mathbf{S}$ and $\theta_S$, we now proceed to presenting the proofs of each of the three parts.

\textbf{Part 1:} Since $\theta_0 \in \Theta$, there exists by definition a $q \in \mathbf{Q}$ such that $\theta(q) = \theta_0$. Using this $q$, we construct a $\psi^{\dagger}$ such that $\psi^{\dagger} \in \mathbf{S}$ and $\theta_S(\psi^{\dagger}) = \theta_0$. In particular, given that $q$ satisfies \eqref{eq:q_sep}, we take $\psi^{\dagger}$ to be such that
\begin{align}\label{eq:s_constr}
\psi^{\dagger}_{jl}(w_{[\mathcal{J}_{jl}]}) = \int\limits_{0}^{1} h_{jl}\left( p_{[\mathcal{J}_{jl}]}(t,w) \right) dt
\end{align}
for each $w \in \mathcal{W}$, $j \in \mathcal{J}$ and $l \in \mathcal{L}_j$, where, for $t \in (0,1)$, $p(t,w) = \left( p_j(t,w) : j \in \mathcal{J}  \right)$ with $p_j(t,w) = \underline{w}_{[j]} + (\bar{w}_{[j]} - \underline{w}_{[j]}) t$, and $\bar{w}_{[j]} = \sup\{t : t \in w_{[j]}\}$ and $\underline{w}_{[j]} = \inf\{t : t \in w_{[j]}\}$.

In order to show $\psi^{\dagger} \in \mathbf{S}$, we need to show it satisfies the restrictions in \eqref{eq:s_logical1}-\eqref{eq:s_data}. The restriction in \eqref{eq:s_logical1} is satisfied for each $w \in \mathcal{W}$ and $j \in \mathcal{J}$ as
\begin{align*}
\sum_{l \in \mathcal{L}_j} \psi^{\dagger}_{jl}(w_{[\mathcal{J}_{jl}]}) = \sum_{l \in \mathcal{L}_j} \int\limits_{0}^{1} h_{jl}\left( p_{[\mathcal{J}_{jl}]}(t,w) \right) dt = \int\limits_{0}^{1} q_j\left( p(t,w) \right) dt \geq 0~,
\end{align*}
where the equalities follow from \eqref{eq:s_constr} and \eqref{eq:q_sep}, respectively, and the inequality from \eqref{O-eq:q_logical1}. Similarly, the restriction in \eqref{eq:s_logical2} is satisfied for each $w \in \mathcal{W}$ and $j \in \mathcal{J}$ as
\begin{align*}
\sum_{j \in \mathcal{J}} \sum_{l \in \mathcal{L}_j} \psi^{\dagger}_{jl}(w_{[\mathcal{J}_{jl}]}) = \int\limits_{0}^{1} \sum_{l \in \mathcal{L}_j} h_{jl}\left( p_{[\mathcal{J}_{jl}]}(t,w) \right) dt = 1~,
\end{align*}
where the equalities follows from \eqref{eq:s_constr} and \eqref{O-eq:q_logical2}, respectively. To see why \eqref{eq:s_shape} is satisfied, take $w,w' \in \mathcal{W}$ such that $t > t'$ for all $t \in w_{[j]},~t' \in w'_{[j=]}$ for $j \in \mathcal{J}' \subseteq \mathcal{J}$ and $w_{[j]} = w'_{[j]}$ for $j \in \mathcal{J} \setminus \mathcal{J}'$. For $t \in (0,1)$, observe that $p_j(t,w) > p_j(t,w)$ for $j \in \mathcal{J}'$ and $p_j(t,w) = p_j(t,w')$ for $j \in \mathcal{J} \setminus \mathcal{J}'$. In turn, it follows from \eqref{O-eq:q_shape} that
\begin{align*}
q_j(t,w) \geq q_j(t,w')
\end{align*}
for each $t \in (0,1)$ and $j \in \mathcal{J} \setminus \mathcal{J}'$. Taking the integral over $t \in (0,1)$ and then rewriting using \eqref{eq:q_sep} and \eqref{eq:s_constr}, it directly follows that \eqref{eq:s_shape} is satisfied. Finally, for the restriction in \eqref{eq:s_data}, observe that it is satisfied as
\begin{align*}
\sum_{l \in \mathcal{L}_j} \psi^{\dagger}_{jl}\left(\{p_{[\mathcal{J}_{jl}]}\}\right) = \sum_{l \in \mathcal{L}_j} h_{jl}\left(p_{[\mathcal{J}_{jl}]}\right) = q_j(p) = P_{j|0}~,
\end{align*}
for each $j \in \mathcal{J}$ and $p \in \mathcal{P}_{\text{obs}}$, where the equalities follow from \eqref{eq:s_constr}, \eqref{eq:q_sep}, and \eqref{O-eq:q_data}, respectively.

In order to show $\psi^{\dagger}$ satisfies $\theta_S(\psi^{\dagger}) = \theta_0$, note that it is sufficient to show $\theta_S(\psi^{\dagger}) = \theta(q)$ as $\theta(q) = \theta_0$. To this end, observe that the various components in \eqref{eq:AB_t_2} can be written in terms of $\psi^{\dagger}$ as
\begin{align*}
\int\limits^{\bar{t}_v}_{\underline{t}_v} q_j\left(\min\{p^a, p^b + t\} \right) dt &= (\bar{t}_v - \underline{t}_v)\int\limits_{0}^{1} q_j\left( p(t,w(v)) \right) dt~,\\
&=  (\bar{t}_v - \underline{t}_v) \int\limits_{0}^{1} \sum_{l \in \mathcal{L}_j} h_{jl}\left( p_{[\mathcal{J}_{jl}]}(t,w(v)) \right) dt~,\\
&=  (\bar{t}_v - \underline{t}_v) \sum_{l \in \mathcal{L}_j} \psi^{\dagger}_{jl}\left( w_{[\mathcal{J}_{jl}]}(v) \right) dt
\end{align*}
for each $v \in \mathcal{V}^{a,b}_l$, $1 \leq l \leq |\mathcal{J}|-1$, where the first equality follows from the change of variables $t = \underline{t}_v + (\bar{t}_v - \underline{t}_v) \cdot t'$ for $t' \in [0,1]$ along with the above definition of $p(t,w(v))$, and the second and third equalities follow from \eqref{eq:s_constr} and \eqref{eq:q_sep}, respectively; and that
\begin{align*}
\sum_{j \in J} g_j^a q_j(p^a) + g_j^b q_j(p^b) &=  \sum_{j \in J} \left( g_j^a \sum_{l \in \mathcal{L}_j} h_{jl}\left(p^a_{[\mathcal{J}_{jl}]}\right) + g_j^b \sum_{l \in \mathcal{L}_j} h_{jl}\left(p^b_{[\mathcal{J}_{jl}]}\right) \right) \\
&= \sum_{j \in J} \left( g_j^a \sum_{l \in \mathcal{L}_j} \psi^{\dagger}_{jl}\left(\{p^a_{[\mathcal{J}_{jl}]}\}\right) + g_j^b \sum_{l \in \mathcal{L}_j} \psi^{\dagger}_{jl}\left(\{p^b_{[\mathcal{J}_{jl}]}\}\right) \right)
\end{align*}
where the equalities follow from \eqref{eq:q_sep} and \eqref{eq:s_constr}, respectively. Substituting these terms in \eqref{O-eq:theta_q}, we obtain the expression in \eqref{eq:theta_S} from which it follows that $\theta(q) = \theta_S(\psi^{\dagger})$. This completes the first part of the proof.

\textbf{Part 2:} Since $\theta_0 \in \Theta_S$, there exist by definition a $\psi \in \mathbf{S}$ such that $\theta_S(\psi) = \theta_0$ and, in turn, by how $\Theta_S$ and $\theta_S$ were constructed, a $q^{\dagger} \in \mathbf{Q}_S^{\text{fd}}$ that is related to $\psi$ by the equation in \eqref{eq:Q_S_fd} such that $\theta(q^{\dagger}) = \theta_S(\psi) = \theta_0$. Since it holds that $\mathbf{Q}^{\text{fd}}_S \subseteq \mathbf{Q}_S$, it follows that $q \in \mathbf{Q}$. This completes the second part of the proof.

\textbf{Part 3}: Given the various linear restrictions that define $\mathbf{S}$ in \eqref{eq:S_set}, observe that $\mathbf{S}$ is a convex set. In addition, it is also a non-empty set by assumption. Then, since $\theta_S$ is a continuous real-valued scalar function, it follows that the image of this function over $\mathbf{S}$ given by $\Theta_S$ is a convex and non-empty set on the real line, i.e. an interval whose closure has endpoints given by \eqref{O-eq:bounds_opt}. This completes the final part of the proof.

\subsection{Proof of Proposition \ref{prop:identifiedset_aux}}\label{sec:proof_aux}

Note $\mathbf{A}$ is a connected and non-empty set. Then, since $\theta_A$ is a continuous real-valued scalar function, it follows that the image of this function over $\mathbf{A}$ given by $\Theta_A$ is a connected and non-empty set on the real line, i.e. an interval whose closure has endpoints given by \eqref{eq:aux_optimization}.


\section{Additional Table and Figures}\label{sec:add_figtab}

\begin{table}[H]
\begin{center}
\caption{Dimension of optimizing variable under different specifications for the programs estimating welfare effects under the status quo voucher amount}
\label{tab:dim}
\scalebox{0.75}{
\def\arraystretch{0.9}
\makebox[\textwidth]{\begin{threeparttable}
 \begin{tabular}{ R{2cm} R{2cm} R{2cm} C{0.1cm}  R{2cm} R{2cm} R{2cm}  }  
 \toprule  
    \multicolumn{7}{c}{Without Liquidity Constraints}       \\  \cline{1-7}                
    \multicolumn{3}{c}{Nonparametric} &  & \multicolumn{3}{c}{Parametric Separable, $K$}       \\  \cline{1-3} \cline{5-7}                
    \multicolumn{1}{c}{True Baseline} & \multicolumn{1}{c}{Outer Baseline} & \multicolumn{1}{c}{Separable} &  & 1 & 2 & 3    \\  \midrule  
       1.01e+69 & 1,856 & 65,774  & & 5,619 & 8,428 & 11,237    \\ \midrule  
    \multicolumn{7}{c}{With Liquidity Constraints}       \\  \cline{1-7}                
    \multicolumn{3}{c}{Nonparametric} &  & \multicolumn{3}{c}{Parametric Separable, $K$}       \\  \cline{1-3} \cline{5-7}                
    \multicolumn{1}{c}{True Baseline} & \multicolumn{1}{c}{Outer Baseline} & \multicolumn{1}{c}{Separable} &  & 1 & 2 & 3    \\  \midrule  
       2.26e+71 & 6,417 & 200,609  & & 22,476 & 33,712 & 44,948    \\  
 \bottomrule  
 \end{tabular}
\begin{tablenotes}[flushleft]
\setlength\labelsep{0pt}
\item \footnotesize True Baseline true denotes the programs under baseline specification in \eqref{O-eq:bounds_opt}, and Outer  Baseline outer denotes the programs under the baseline specification in \eqref{O-eq:b_r_opt}. \item \footnotesize SOURCE:  \textit{Evaluation of the DC Opportunity Scholarship Program: Final Report (NCEE 2010-4018)}, U.S. Department of Education, National Center for Education Statistics previously unpublished tabulations.
\end{tablenotes}
\end{threeparttable}}}
\end{center}
\end{table}

\begin{table}[H]
\begin{center}
\caption{Average school characteristics by tuition level  for voucher private schools}
\label{tab:school-chars-tuition}
\scalebox{0.75}{
\def\arraystretch{0.9}
\makebox[\textwidth]{\begin{threeparttable}
 \begin{tabular}{ L{7cm} R{3cm} R{3cm} R{3cm}  } 
\toprule 
                                & Tuition $\leq$ $\$$3,500 & Tuition $>$ $\$$3,500 & Difference \\
\midrule
School size                     &  196.103&  281.843&  -85.740\\
Student/teacher ratio           &   13.000&    9.860&    3.140\\
Catholic (=1)                   &    0.829&    0.159&    0.671\\
Other religious (=1)            &    0.003&    0.256&   -0.253\\
Gifted program (=1)             &    0.243&    0.397&   -0.155\\
Learning difficulties program (=1)&    0.446&    0.547&   -0.101\\
Individual tutors available (=1)&    0.609&    0.774&   -0.165\\
Students tracked by ability (=1)&    0.763&    0.729&    0.034\\
Remedial classes available (=1) &    0.646&    0.619&    0.027\\
\midrule
Number of schools                    &       20&       50&         \\
\bottomrule  
 \end{tabular}
\begin{tablenotes}[flushleft]
\setlength\labelsep{0pt}
\item \footnotesize SOURCE:  \textit{Evaluation of the DC Opportunity Scholarship Program: Final Report (NCEE 2010-4018)}, U.S. Department of Education, National Center for Education Statistics previously unpublished tabulations. Observations rounded to the nearest 10.
\end{tablenotes}
\end{threeparttable}}}
\end{center}
\end{table}

\begin{table}[H]
\begin{center}
\caption{Estimates of the underlying parameters for the various logit specifications}
\label{tab:logit}
\scalebox{0.75}{
\def\arraystretch{0.9}
\makebox[\textwidth]{\begin{threeparttable}
 \begin{tabular}{ L{2cm} C{2.5cm} C{2.5cm} C{0.5cm} C{2.5cm} C{2.5cm}  }  
 \toprule  
   & \multicolumn{5}{c}{Specification}   \\  \cline{2-6}  
   & \multicolumn{2}{c}{With LC} & & \multicolumn{2}{c}{Without LC}   \\  \cline{2-3} \cline{5-6}  
   & Mixed Logit & Nested Logit & & Mixed Logit & Nested Logit  \\  \midrule  
  $\bar{\gamma}_0$     &    12.18      &    8.45      &  &    13.82      &    7.45      \\   
                        &   [2.05]      &   [0.89]      &  &   [8.98]  &   [0.74] \\   
  $\bar{\gamma}_{11}$  &    -0.88      &    -0.24      &  &  &   \\    
                        &   [1.15]      &   [1.10]      &  &  &  \\    
  $\bar{\gamma}_{12}$  &    -0.88      &    0.01      &  &  &  \\    
                        &   [1.05]      &   [1.12]      &  &  &  \\    
  $\bar{\gamma}_{13}$  &    -3.09      &    -2.25      &  &  &  \\    
                        &   [1.23]      &   [0.96]      &  &  &  \\    
  $\tilde{\gamma}_{0}$ &                                             &                                             &  &    5,094.68  &    6,342.16      \\    
                        &                                             &                                             &  &   [1,453.52]  &   [4,033.85]  \\    
  $\tilde{\gamma}_{1}$ &                                             &                                             &  &    1,310.06  &    966.52      \\    
                        &                                             &                                             &  &   [2,095.44]  &   [4,509.57]  \\    
  $\tilde{\gamma}_{2}$ &                                             &                                             &  &    7,431.60  &    1,586.91        \\    
                        &                                             &                                             &  &   [4,963.80]  &   [3,904.02]  \\    
  $\tilde{\gamma}_{3}$ &                                             &                                             &  &    3,944.48  &    25,918.62        \\    
                        &                                             &                                             &  &   [1,257.73]  &   [10,586.12]  \\    
  $\sigma$              &    4.37      &                                             &  &    7.51  &       \\    
                        &    [1.31]     &                                             &  &   [7.41]  &  \\    
  $\lambda_1$           &                                             &    1.87      &  &                                         &    2.00      \\    
                        &                                             &   [0.47]      &  &                                         &   [0.46]  \\    
  $\tilde{\sigma}$     &                                             &                                             &  &    3,944.48  &    5,166.43        \\    
                        &                                             &                                             &  &   [1,257.73]  &   [2,969.95]  \\    
 \bottomrule  
 \end{tabular}
\begin{tablenotes}[flushleft]
\setlength\labelsep{0pt}
\item \footnotesize $\bar{\gamma}_{0}$, $\bar{\gamma}_{11}$, $\bar{\gamma}_{12}$, $\bar{\gamma}_{13}$, and $\sigma$ are multiplied by $10^{4}$ so that they can be visually presented in the same scale as the remaining parameters. Standard errors reported in square brackets computed using bootstrap with 1,000 draws. $\bar{\gamma}_{1l}$ corresponds to the coefficient on the $l$th income bin, i.e. incomes between the $l$ and $l+1$ quartiles of the empirical distribution. The coefficient on the fourth income bin is normalized to 0 due to avoid perfect multicollinearity with the respect to the constant term. $\lambda_2$ is normalized to 1 as there is a single alternative in $\mathcal{N}_2$. LC denotes that the model allows students to be liquidity constrained.
\item \footnotesize SOURCE:  \textit{Evaluation of the DC Opportunity Scholarship Program: Final Report (NCEE 2010-4018)}, U.S. Department of Education, National Center for Education Statistics previously unpublished tabulations.
\end{tablenotes}
\end{threeparttable}}}
\end{center}
\end{table}

\begin{table}[H]
\begin{center}
\caption{Expressions for demand with and without liquidity constraints, and the parameter of interest with liquidity constraints under the various logit specifications}
\label{tab:logit_exp}
\scalebox{0.725}{
\def\arraystretch{0.9}
\makebox[\textwidth]{\begin{threeparttable}
\begin{tabular}{L{26cm}}  
\toprule  
  \underline{Demand expressions without LC}: For each $j \in \mathcal{J}$, we have
  $${\displaystyle q_j(p) = \sum_{x \in \mathcal{X}} \left( \int\limits \dfrac{e^{\xi_j - (\bar{\gamma}_0 + \bar{\gamma}_1'x + \nu) p_j}}{ \sum\limits_{l \in \mathcal{J}} e^{\xi_l - (\bar{\gamma}_0 + \bar{\gamma}_1'x + \nu) p_l}} \phi\left(\frac{\nu}{\sigma}\right) d\nu \right) \text{Prob}[X_i = x]}$$
  under specification ML, and
  $$\displaystyle q_j(p) = \sum_{x \in \mathcal{X}} \dfrac{e^{\left(\xi_j - (\bar{\gamma}_0 + \bar{\gamma}_1'x) p_j\right) / \lambda_k} \left( \sum\limits_{l \in \mathcal{N}'} e^{ \left(\xi_l - (\bar{\gamma}_0 + \bar{\gamma}_l'x) p_l\right) / \lambda' }  \right)^{\lambda' - 1}}{ \sum\limits_{k \in \{1,2\}}  \left( \sum\limits_{l \in \mathcal{N}_k} e^{ \left(\xi_l - (\bar{\gamma}_0 + \bar{\gamma}_l'x) p_l\right) / \lambda_k }  \right)^{\lambda_k}} \text{Prob}[X_i = x] $$
  where $\mathcal{N}' = \mathcal{N}_1$ and $\lambda' = \lambda_1$ if $j \in \mathcal{N}_1$ and $\mathcal{N}' = \mathcal{N}_2$ and $\lambda' = \lambda_2$ otherwise under specification NL.\\ \hline
  \underline{Demand expressions with LC}: Denoting by $e_1(p) \leq \ldots \leq e_{M^\dagger}(p)$ denote the ordered values of $\{p_j : j \in \mathcal{J}\} \cup \{-\infty,\infty\}$ for $p \in \mathcal{P}$, we have for each $j \in \mathcal{J}$ that
  $${\displaystyle q_j(p) =  \sum_{m=1}^{M^\dagger-1}  \int\limits \dfrac{1_{j,m} e^{\xi_j - (\bar{\gamma}_0 + \nu) p_j}}{\sum\limits_{l \in \mathcal{J} } 1_{j,m} e^{\xi_l - (\bar{\gamma}_0 + \nu) p_l}} \phi\left(\frac{\nu}{\sigma}\right) d\nu \left( \sum_{x \in \mathcal{X}}\left(\Phi\left(\frac{e_m(p) - \tilde{\gamma}_0 - \tilde{\gamma}_1'x}{\tilde{\sigma}}\right) - \Phi\left(\frac{e_{m-1}(p) - \tilde{\gamma}_0 - \tilde{\gamma}_1'x}{\tilde{\sigma}}\right) \right) \text{Prob}[X_i = x] \right)}~,$$\\ 
 where $1_{j,m} = 1$ if $j \in \{g,n\}$ and $1_{j,m} = 1\{p_j < e_{m+1}(p)\}$ if $j \in \mathcal{J} \setminus \{g,n\}$ under specification ML, and
  $$\displaystyle q_j(p) =  \sum_{m=1}^{M^\dagger-1} \dfrac{1_{j,m}e^{\left(\xi_j - \bar{\gamma}_0 p_j\right) / \lambda_k} \left( \sum\limits_{l \in \mathcal{N}'} 1_{l,m} e^{ \left(\xi_l - \bar{\gamma}_0 p_l\right) / \lambda' }  \right)^{\lambda' - 1}}{ \sum\limits_{k \in \{1,2\}}  \left( \sum\limits_{l \in \mathcal{N}_k} 1_{l,m} e^{ \left(\xi_l - \bar{\gamma}_0 p_l\right) / \lambda_k }  \right)^{\lambda_k}} \left( \sum_{x \in \mathcal{X}} \left(\Phi\left(\frac{e_m(p) - \tilde{\gamma}_0 - \tilde{\gamma}_1'x}{\tilde{\sigma}}\right) - \Phi\left(\frac{e_{m-1}(p) - \tilde{\gamma}_0 - \tilde{\gamma}_1'x}{\tilde{\sigma}}\right) \right) \text{Prob}[X_i = x] \right)~, $$
where $1_{j,m} = 1$ if $j \in \{g,n\}$ and $1_{j,m} = 1\{p_j < e_{m+1}(p)\}$ if $j \in \mathcal{J} \setminus \{g,n\}$, where $\mathcal{N}' = \mathcal{N}_1$ and $\lambda' = \lambda_1$ if $j \in \mathcal{N}_1$ and $\mathcal{N}' = \mathcal{N}_2$ and $\lambda' = \lambda_2$ otherwise under specification NL. \\
\hline
  \underline{Parameter of interest with LC}: Denoting by $e^{a,b}_1 \leq \ldots \leq e^{a,b}_{\tilde{M}}$ the ordered values of $\{p_j : j \in \mathcal{J},~p \in \{p^a,p^b\}\} \cup \{-\infty,\infty\}$, the parameter in \eqref{O-eq:theta_q_liquidity} corresponds to
  \begin{align*} 
  \tilde{\theta}(\tilde{q}) = \sum_{m=1}^{\tilde{M}-1} \text{Prob}[e \in (e^{a,b}_{m},e^{a,b}_{m+1})] \left(g^{a,b} \left(\tilde{\Delta}_1^{a,b}(e^{a,b}_m) + \sum_{l=1}^{|\mathcal{J}|-1} \sum\limits_{j \in \tilde{\mathcal{J}}_{l+1}^{a,b}(e^{a,b}_m)} \int\limits_{\tilde{\Delta}_{l}^{a,b}(e^{a,b}_m)}^{\tilde{\Delta}_{l+1}^{a,b}(e^{a,b}_m)} \tilde{q}_j\left(\text{min}\{\tilde{p}(p^a,e^{a,b}_m),\tilde{p}(p^b,e^{a,b}_m)+t\}\right) dt \right) + \sum_{j \in J} g_j^a \tilde{q}_j(\tilde{p}(p^a,e^{a,b}_m))  + g_j^b \tilde{q}_j(\tilde{p}(p^b,e^{a,b}_m)) \right)~,
    \end{align*}
  where
  $$\displaystyle \text{Prob}[e \in (e^{a,b}_{m},e^{a,b}_{m+1})] = \sum_{x \in \mathcal{X}}\left(\Phi\left(\frac{e^{a,b}_{m+1} - \tilde{\gamma}_0 - \tilde{\gamma}_1'x}{\tilde{\sigma}}\right) - \Phi\left(\frac{e^{a,b}_{m} - \tilde{\gamma}_0 - \tilde{\gamma}_1'x}{\tilde{\sigma}}\right) \right)\text{Prob}[X_i = x]~, $$
  and
  $$\displaystyle \tilde{q}_j(p) =  \int\limits \dfrac{e^{\xi_j - (\bar{\gamma}_0 + \nu) p_j}}{ \sum\limits_{l \in \mathcal{J}} e^{\xi_l - (\bar{\gamma}_0 + \nu) p_l}} \phi\left(\frac{\nu}{\sigma}\right) d\nu $$
  under specification ML and
  $$\displaystyle \tilde{q}_j(p) = \dfrac{e^{\left(\xi_j - (\bar{\gamma}_0 + \bar{\gamma}_1'x) p_j\right) / \lambda_k} \left( \sum\limits_{l \in \mathcal{N}'} e^{ \left(\xi_l - (\bar{\gamma}_0 + \bar{\gamma}_l'x) p_l\right) / \lambda' }  \right)^{\lambda' - 1}}{ \sum\limits_{k \in \{1,2\}}  \left( \sum\limits_{l \in \mathcal{N}_k} e^{ \left(\xi_l - (\bar{\gamma}_0 + \bar{\gamma}_l'x) p_l\right) / \lambda_k }  \right)^{\lambda_k}}$$
  with $\mathcal{N}' = \mathcal{N}_1$ and $\lambda' = \lambda_1$ if $j \in \mathcal{N}_1$ and $\mathcal{N}' = \mathcal{N}_2$ and $\lambda' = \lambda_2$ otherwise under specification NL. \\
\bottomrule  
\end{tabular}

\begin{tablenotes}[flushleft]
\setlength\labelsep{0pt}
\item \footnotesize $\mathcal{X}$ denotes the support of $X_i$, $\phi$ and $\Phi$ are the density and cumulative distribution function of a standard normal distribution, respectively. LC denotes that the model allows students to be liquidity constrained. Note that for specifications ML and NL without LC, Assumption \ref{O-ass:LC} is satisfied taking $r \to \infty$ and we hence take $r \to \infty$ when implementing the function $\tilde{p}$, defined in Section \ref{O-sec:liquidity}, in the above expression.
\end{tablenotes}
\end{threeparttable}}}
\end{center}
\end{table}

\begin{figure}[H]
\centering
\caption{Estimated bounds on average surplus under nonparametric baseline specification for alternative values of government school costs ($c_g$) and administrative costs ($\mu$)}
\makebox[\textwidth]{ \begin{tabular}{c c c}
\subcaptionbox{Government school costs}{ \includegraphics[width=2.5in]{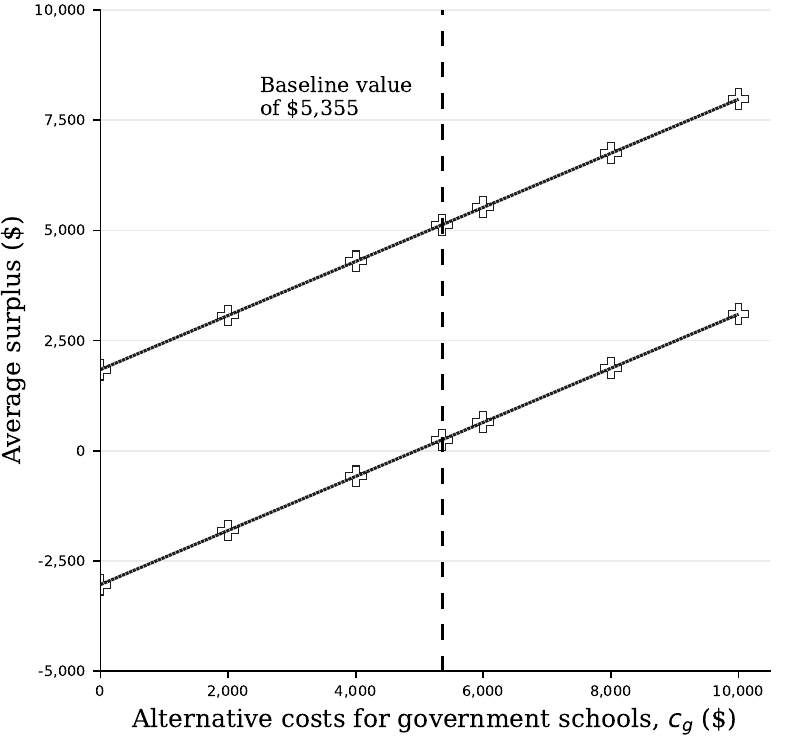}} & ~~~~ &
\subcaptionbox{Administrative costs}{\includegraphics[width=2.5in]{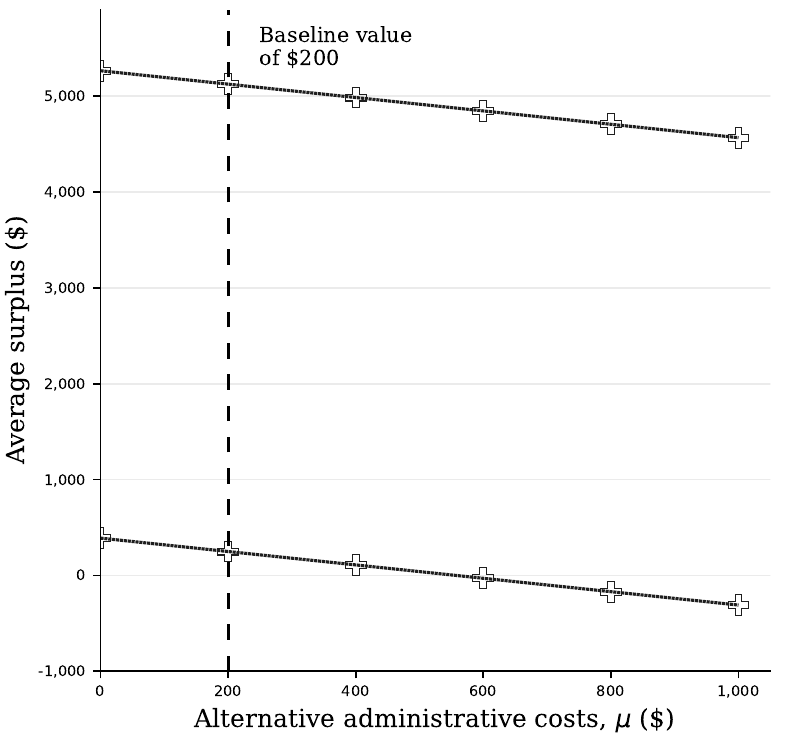}}
\end{tabular}}
\begin{tabular*}{0.99\textwidth}{c}
\multicolumn{1}{p{.99\hsize}}{\footnotesize Observe that as long as we assume that $c_g$ is at least around $\$$5,000, and that $\gamma$ is at most around $\$$500, we continue to robustly find positive net average benefits.}\\
\multicolumn{1}{p{.99\hsize}}{\footnotesize SOURCE:  \textit{Evaluation of the DC Opportunity Scholarship Program: Final Report (NCEE 2010-4018)}, U.S. Department of Education, National Center for Education Statistics previously unpublished tabulations.}
\end{tabular*}
\label{fig:AS_cost}
\end{figure}

\begin{figure}[H]
\centering
\caption{90$\%$ confidence intervals under nonparametric baseline specification on average surplus for a range of voucher amounts}
\makebox[\textwidth]{ \includegraphics[width=2.5in]{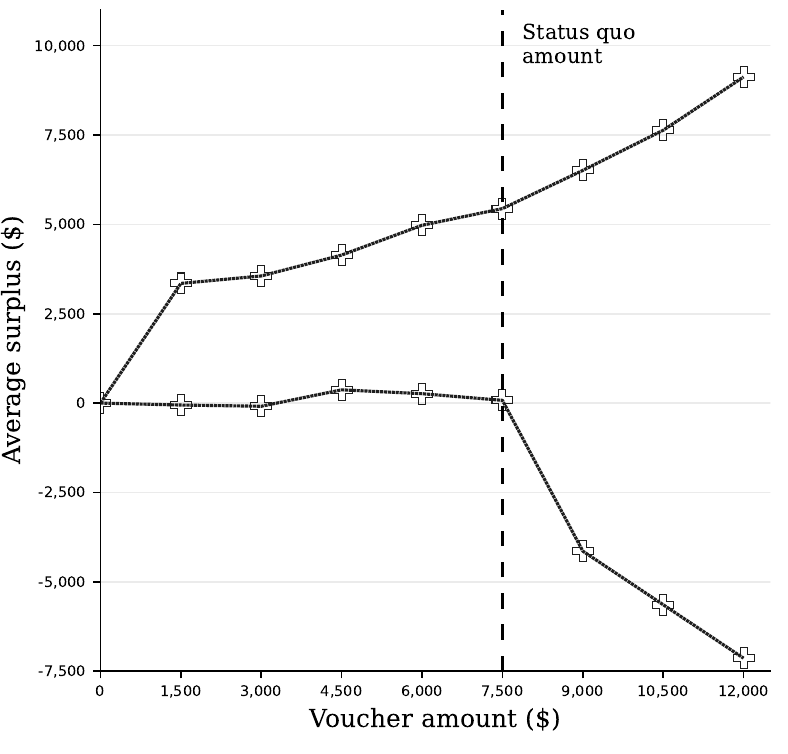}}
\begin{tabular*}{0.99\textwidth}{c}
\multicolumn{1}{p{.99\hsize}}{\footnotesize SOURCE:  \textit{Evaluation of the DC Opportunity Scholarship Program: Final Report (NCEE 2010-4018)}, U.S. Department of Education, National Center for Education Statistics previously unpublished tabulations.}
\end{tabular*}
\label{fig:AS_tau_ci}
\end{figure}

\begin{figure}[H]
\centering
\caption{Estimates along with 90$\%$ confidence intervals for the logit specifications for various demand and welfare parameters at various counterfactual voucher amounts}
\makebox[\textwidth]{
\begin{tabular}{c c c}
\subcaptionbox{Probability of voucher takeup}{ \includegraphics[width=2.25in]{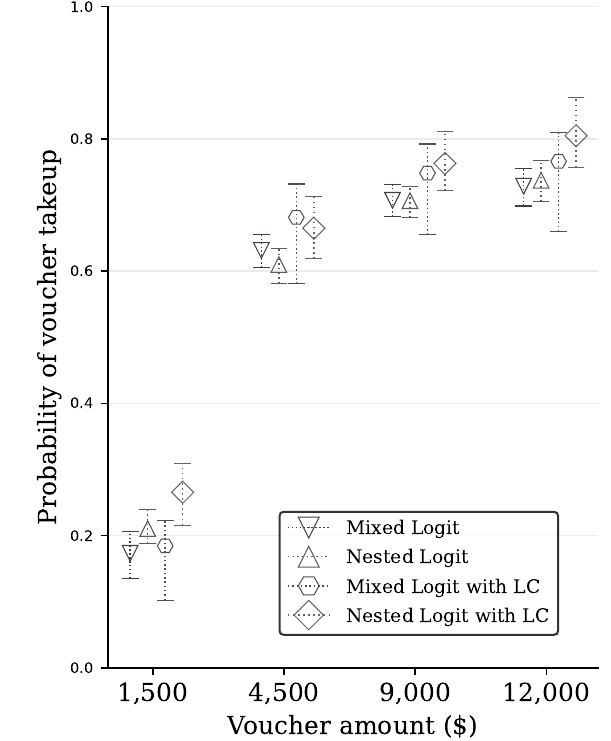}} & \subcaptionbox{Probability of takeup at schools with tuition at least voucher amount}{ \includegraphics[width=2.25in]{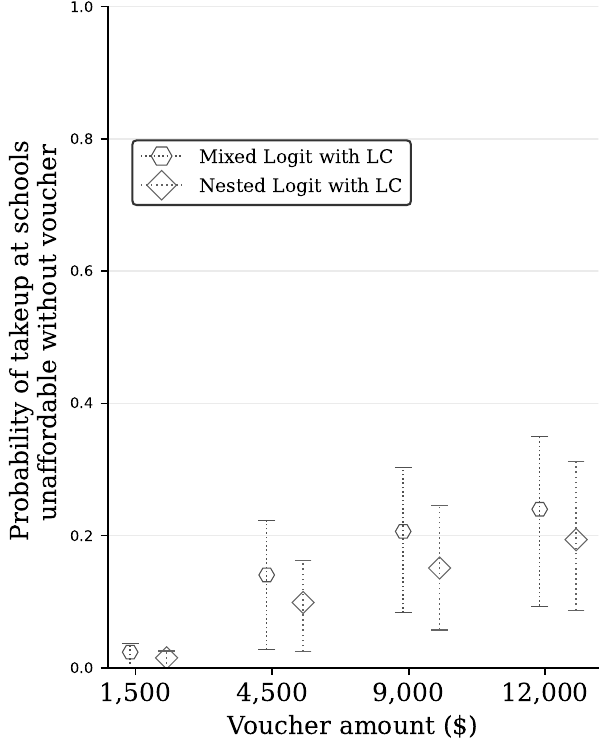}} &
\subcaptionbox{Average surplus}{\includegraphics[width=2.25in]{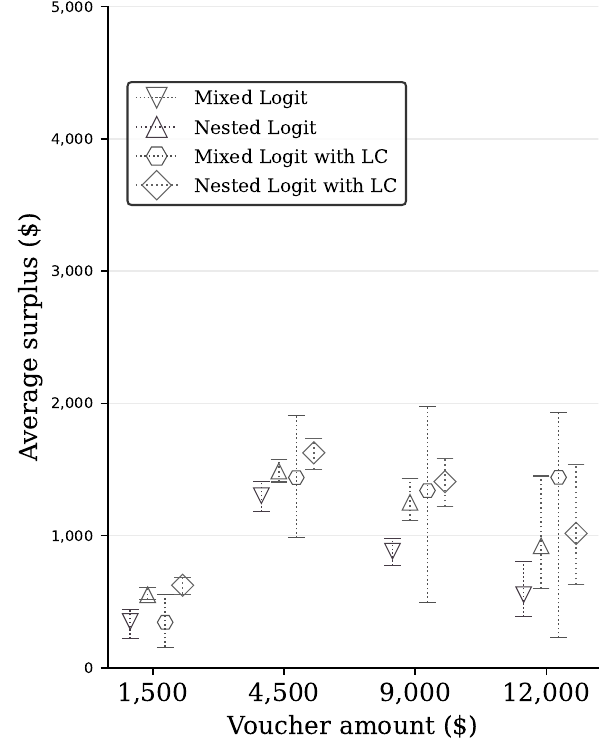}}
\end{tabular}}
\begin{tabular*}{0.99\textwidth}{c}
\multicolumn{1}{p{.99\hsize}}{\footnotesize For the various logit specifications, the markers denote the point estimates and the dashed intervals denote 90$\%$ confidence intervals computed using the percentile bootstrap with 1,000 draws. LC denotes that the model allows students to be liquidity constrained.} \\
\multicolumn{1}{p{.99\hsize}}{\footnotesize SOURCE:  \textit{Evaluation of the DC Opportunity Scholarship Program: Final Report (NCEE 2010-4018)}, U.S. Department of Education, National Center for Education Statistics previously unpublished tabulations. }
\end{tabular*}
\label{fig:parameter_logit_ci}
\end{figure}

\newpage

\bibliography{references.bib}